\documentclass[fleqn,usenatbib]{mnras}

\usepackage{newtxtext,newtxmath}
\usepackage[T1]{fontenc}
\usepackage[dvipsnames]{xcolor} 

\usepackage{xspace}
\def\m87{M\,87\xspace}
\newcommand{\ra}{\textit{RadioAstron}\xspace} 

\usepackage{graphicx} 
\usepackage{amsmath}  
\usepackage{orcidlink}
\usepackage[normalem]{ulem} 

\defcitealias{2023ApJ...952...34K}{Paper~I}

\graphicspath{{./figs/}}

\title[RadioAstron observations of M87 at 4.8\,GHz]{\textit{RadioAstron} space-VLBI imaging of the jet in M87: II. The parsec-scale structure at 4.8\,GHz}

\author[E. V. Kravchenko et al.]{E. V. Kravchenko$^{\orcidlink{0000-0003-4540-4095}}$,$^{1}$\thanks{E-mail: evgenia.v.kravchenko@gmail.com (EVK)}
T. Savolainen$^{\orcidlink{0000-0001-6214-1085}}$,$^{2,3,4}$ 
A. S. Nikonov$^{\orcidlink{0000-0002-4009-9186}}$,$^{4}$
I. N. Pashchenko$^{\orcidlink{0000-0002-9404-7023}}$,$^{1,5}$
E. E. Nokhrina$^{\orcidlink{0000-0002-3897-2417}}$,$^{1,5}$
\newauthor
P. A. Voitsik$^{\orcidlink{0000-0002-1290-1629}}$,$^{1}$
L. I. Gurvits$^{\orcidlink{0000-0002-0694-2459}}$,$^{6,7}$
Y. Y. Kovalev$^{\orcidlink{0000-0001-9303-3263}}$,$^{4}$
J.-Y. Kim$^{\orcidlink{0000-0001-8229-7183}}$,$^{8}$
M. M. Lisakov$^{\orcidlink{0000-0001-6088-3819}}$,$^{9}$
A. P. Lobanov$^{\orcidlink{0000-0003-1622-1484}}$,$^{4}$
\newauthor
K. V. Sokolovsky$^{\orcidlink{0000-0001-5991-6863}}$$^{10,11}$
\\
$^{1}$Lebedev Physical Institute, Leninsky prospekt 53, 119991 Moscow, Russia\\ 
$^{2}$Aalto University Department of Electronic and Nanoengineering, PL15500, FI-00076 Aalto, Finland\\ 
$^{3}$Aalto University Mets\"{a}hovi Radio Observatory, Mets\"{a}hovintie 114, FI-02540 Kylm\"{a}l\"{a}, Finland\\ 
$^{4}$Max-Planck-Institut f\"{u}r Radioastronomie, Auf dem H\"{u}gel 69, Bonn, 53121, Germany\\ 
$^{5}$Moscow Center for Advanced Studies, Kulakova str. 20, Moscow 123592, Russia\\
$^{6}$Joint Institute for VLBI ERIC, Oude Hoogeveensedijk 4, 7991PD Dwingeloo, The Netherlands\\ 
$^{7}$Faculty of Aerospace Engineering, Delft University of Technology, Kluyverweg 1, 2629HS Delft, The Netherlands\\ 
$^{8}$Department of Physics, Ulsan National Institute of Science and Technology (UNIST), 50 UNIST-gil, Eonyang-eup, Ulju-gun, Ulsan 44919, Republic of Korea\\ 
$^{9}$ Instituto de F\'{i}sica, Pontificia Universidad Cat\'{o}lica de Valpara\'{i}so, Casilla 4059, Valpara\'{i}so, Chile \\ 
$^{10}$Department of Astronomy, University of Illinois at Urbana-Champaign, 1002 W. Green Street, Urbana, IL 61801, USA \\ 
$^{11}$Department of Physics and Astronomy, Texas Tech University, Lubbock, TX 79409, USA \\ 
}

\date{
Accepted 2026 September 15. Received 2026 September 15; in original form 2026 February 4
}

\pubyear{\the\year{}}

\begin{document}
\label{firstpage}
\pagerange{\pageref{firstpage}--\pageref{lastpage}}
\maketitle

\begin{abstract}
We present {\it RadioAstron} observations of the M87 jet obtained on 2014 February 4-5 at 4.8\,GHz ($\lambda=6.2$\,cm). 
The correlated signals between the 10-m space antenna and ground-based radio telescopes were detected up to projected baselines of $\thicksim3$\,Earth diameters (0.65~G$\lambda$). 
The orientation of the jet downstream of the core shifts northward, consistent with the long-term oscillating trend reported in other studies.
An elliptical body mode of the Kelvin-Helmholtz instability can create the observed helical shape of the jet.
The estimated wavelength of this instability mode suggests that its origin may be related to some processes in the accretion disc. 
We present 1.7 - 4.8\,GHz and 4.8 - 15.4\,GHz spectral index maps showing flattening of the spectrum along the outer streamlines and along the jet spine. 
Simulations and analysis show that this spectral flattening is partially artificial and is a common systematic effect of 
the \texttt{CLEAN} procedure of VLBI imaging. The spectral index of the M87 jet is consistent with the optically thin emission, while its steepening can be produced by the propagation of the instability modes causing higher pressure regions near the jet boundary and in its interior.
\end{abstract}

\begin{keywords}
galaxies: active -- galaxies: jets -- galaxies: individual: M87 -- radio continuum: galaxies -- techniques: interferometric
\end{keywords}




\section{Introduction} \label{sec:intro}

Messier~87 (also known as NGC\,4486, Virgo~A, 1228$+$126, 3C~274B; hereafter -- M87) is one of the most studied extragalactic sources in astronomy. 
It is a super-giant elliptical galaxy at a distance of $(16.8 \pm 0.8)$~Mpc \citep{2010A&A...524A..71B, 2019ApJ...875L...4E}.
It is one of the nearest galaxies hosting an active galactic nucleus (AGN).
Given its high brightness across the electromagnetic spectrum, the prominent jet of M87 has been extensively monitored since its optical discovery more than a century ago \citep{1918PLicO..13....9C}. 
Owing to the Very Long Baseline Interferometry (VLBI) technique \citep{1965R&QE....8..461M}, the jet of M87 has been well-studied at scales from a few Schwarzschild radii $R_{\rm s}$ \citep[$100R_{\rm s}\approx0.06$~pc; e.g.][]{2023Natur.616..686L} up to $10^7$ $R_{\rm s}$ \citep[e.g.][]{2007ApJ...668L..27K, 2016ApJ...833...56A,2018ApJ...855..128W}, and even beyond \citep[e.g.][]{2012A&A...547A..56D,2025ApJ...988...28W}.
Direct imaging of the black hole shadow by the Event Horizon Telescope (EHT) collaboration in total and polarized light \citep{2019ApJ...875L...1E, 2021ApJ...910L..12E} confirmed the existence of the supermassive black hole (SMBH) and its critical role in jet launching.
Following the transformative EHT results \citep{2019ApJ...875L...4E}, the inner structure of the central source of M87 has been closely scrutinised by all available high-resolution facilities in spatial, time and frequency domains. 
These observations revealed that extended and time-variable emission structures fill the region between the SMBH photon rings and what is conventionally called the jet core \citep{2022NatAs.tmp..261A}.
The recent high-resolution images from the Global mm-VLBI array show that the edge-brightened jet connects to the accretion flow of the black hole \citep{2023Natur.616..686L}. 
Notably, the jet-launching region near the black hole exhibits a broader emission profile than predicted for a purely black-hole-driven jet, indicating the possible presence of an accretion disc wind.
This echoes \ra{} imaging results of another nearby radio galaxy, 3C~84 \citep[62~Mpc;][]{2013AJ....146...86T},
which suggested that its jet may be too wide to be launched from the black hole ergosphere \citep{2018NatAs...2..472G}.
Thanks to its proximity, M87 offers a unique opportunity for detailed studies of structural patterns in AGN jets \citep[e.g.][and references therein]{Nokhrina+2019MNRAS}.

The parsec-scale VLBI studies show that the M87 jet gradually accelerates over the distance range of $10^6$\,$R_{\rm s}$ \citep[$\thicksim240$ pc; ][]{2019ApJ...887..147P}, which coincides with the jet collimation zone \citep{2012ApJ...745L..28A, 2018ApJ...868..146N}. This is consistent with the gradual conversion of Poynting flux to kinetic energy flux, as predicted by the magnetohydrodynamic (MHD) acceleration model. 
\citet{2018ApJ...855..128W} analysed images of the M87 jet obtained with the Very Long Baseline Array (VLBA) over two decades and found that the position angle of the innermost section of the jet oscillates. This oscillation can be traced down to 140\,$R_{\rm s}$ \citep{2018AA...616A.188K}. 
From the VLBI observations obtained between 2000 and 2022 at 22-24 and 43\,GHz, the variations in the position angle of the jet show a quasi-periodic behaviour, which is well described by a model of a precessing jet with a period $T_\mathrm{prec}= (11.24\pm0.47)$~years \citep{2023Natur.621..711C}. 
Quasi-periodic jet position angle variations can be produced, for example, by a system of a spinning black hole and a misaligned accretion disc that induces the Lense-Thirring precession.

Mapping the synchrotron spectrum may constrain the particle and magnetic field distribution in the jet at $10-10^4$\,$R_{\rm s}$. 
Low-sensitivity spectral index\footnote{We define the spectral index $\alpha$ as $S(\nu)\propto\nu^{\alpha}$, where $S$ is the flux density observed at frequency $\nu$.} images show a steepening of the spectra within 10\,mas off the VLBI core down to $\alpha\approx -2.5$ \citep{2007ApJ...660..200L, 2014AJ....147..143H, 2016ApJ...833...56A, 2019Galax...7...86Z, 2020A&A...637L...6K, 2023A&A...673A.159R}. 
However, our recent high-dynamic range VLBA observations \citep{2023MNRAS.526.5949N} show a median spectral index along the jet $\alpha = -0.5\pm0.2$ at the same distances [inner 50~mas] from the core. In turn, the spectra were reported to flatten in some regions \citep{2023MNRAS.526.5949N}, which can indicate plasma stratification \citep{2023MNRAS.523..887F}. With high-quality, high-resolution images of the M87 jet presented in this paper, we aim to resolve this ambiguity. 

\begin{figure}
\centering
\includegraphics[width=0.95\columnwidth]{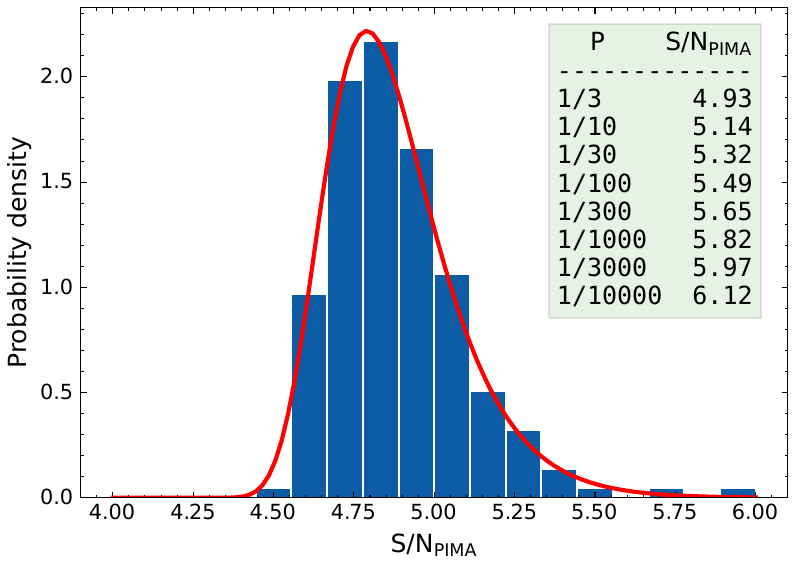}\\
\includegraphics[angle=270,width=0.95\columnwidth]{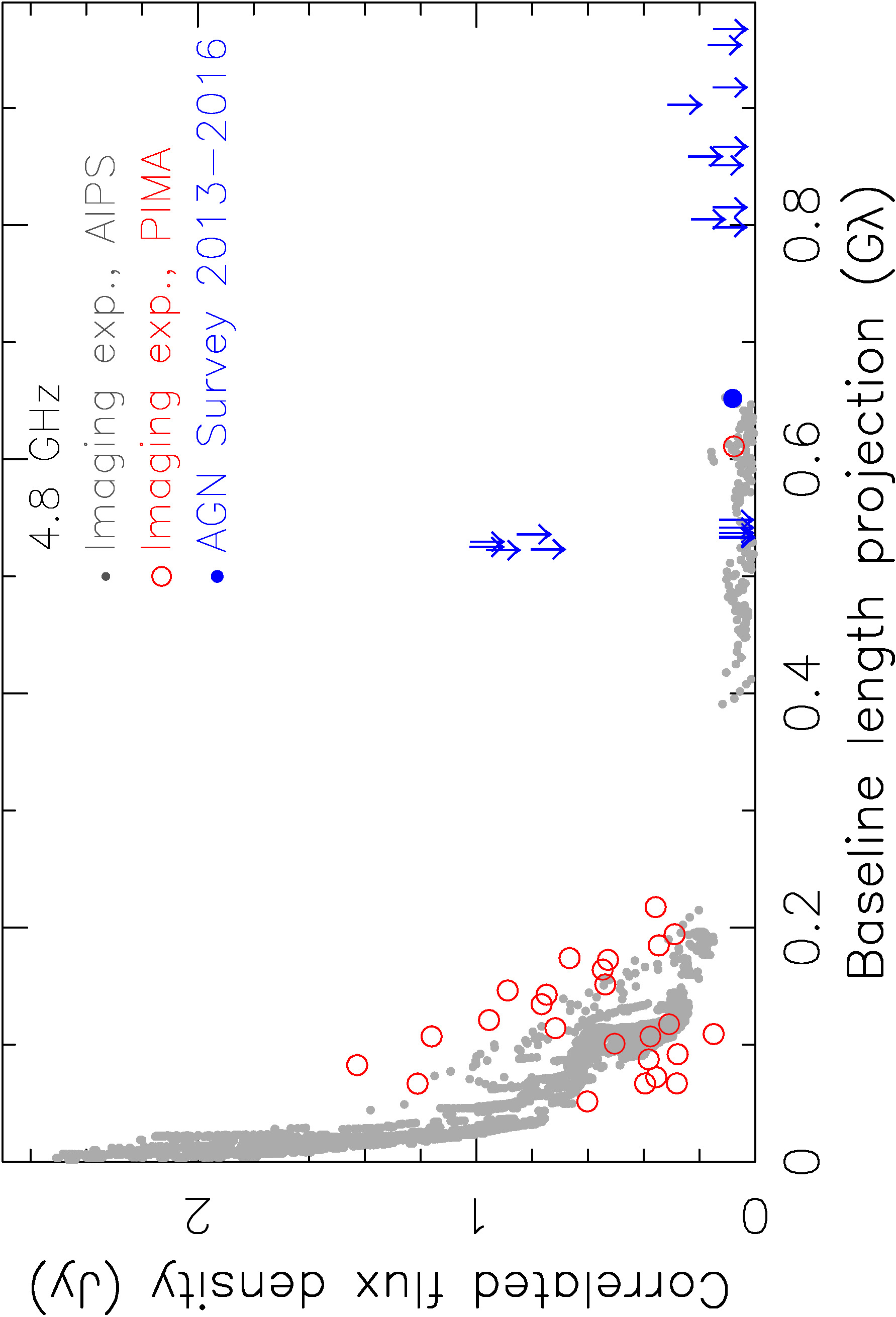}
\caption{(Top) The left part of the empirical distribution (i.e. `no signal') of the fringe signal-to-noise ratio derived using \textsc{PIMA} ($\text{S/N}_\text{PIMA}$) from the imaging observations. The red curve represents the fit by a theoretical distribution \citep{2017isra.book.....T}. The inset presents the correspondence between the probability of false detection and $\text{S/N}_\text{PIMA}$ threshold for the given set of data parameters.
(Bottom) Visibility amplitudes as a function of projected $(u,v)$-distance for all observations of M87 with the \ra{} at 4.8\,GHz. Blue circle indicates AGN survey detection with the \ra{}. Arrows show upper limits for non-detections. }
\label{fig:raplot}
\end{figure}

The first space-VLBI observations of M87 were conducted by the VLBI Space Observatory Programme (VSOP), which involved the HALCA satellite equipped with the 8-m radio telescope in a highly elliptical orbit with an apogee of 21400\,km above the Earth's surface \citep{2000PASJ...52..955H}. M87 was targeted in 2000 at 1.7 and 5.0\,GHz \citep{2006PASJ...58..243D}.
The jet was resolved transversely into three ridgelines, denominated as two outer and one inner morphological component \citep{2016ApJ...833...56A}. Estimated to be narrower than predictions for the parabolic flow, the inner ridgeline was assumed to be associated with the spine produced by the spinning black hole.
However, there is evidence for another inner break in the observed geometry profile that could be associated with the internal MHD jet structure \citep{2024MNRAS.528.6046B}.

The jet of M87 exhibits a filamentary double-helix structure at distances from 300\,pc to 1\,kpc \citep{2003NewAR..47..629L, 2011ApJ...735...61H,2021ApJ...923L...5P}.
It has been suggested that the jet is shaped by Kelvin-Helmholtz (KH) instabilities. The most recent studies are consistent with the presence of KH modes also at parsec-scales, which have been detected in the 1.7\,GHz \ra{} image (Savolainen et al., in prep.) and high-sensitivity 8\,GHz VLBA observations \citep{2023MNRAS.526.5949N}. 
The central filament has been observed previously in several studies within 25\,mas of the core \citep{2017Galax...5....2H, 2018ApJ...855..128W, 2018AA...616A.188K}. 
Simulations \citep{2023MNRAS.523.1247P} show that the central filament observed in edge-brightened models may arise from a \texttt{CLEAN} imaging artefact.
However, there is also a marginal detection of the triple ridge in the 22\,GHz jet image obtained by the East Asian VLBI Network using a regularized maximum likelihood imaging method \citep{2023Galax..11...39T}. 
The nature of this inner ridge and helical jet structure remains an open question.

The \ra{} space VLBI mission \citep{2013ARep...57..153K} involved a 10-m space radio telescope (SRT) on board the Spektr-R satellite in a highly eccentric orbit with the apogee of $\thicksim$370~000\,km.
When the Spektr-R passes perigee, the shortest space-ground baselines become comparable in length to the longest ground-ground baselines, aiding calibration, while space-ground baselines change rapidly, improving the $(u,v)$-coverage. These two factors work together to facilitate efficient imaging of celestial radio sources with high fidelity at an unprecedented resolution of a few tens of microarcseconds \citep[$\mu$as; e.g.][]{2020AdSpR..65..772S,2020ApJ...893...68K}.
One of the \ra{} Key Science Programs was focused on imaging the nearby AGN (see \citealt{2020AdSpR..65..712B} for a description of the \ra{} KSPs). Within this program, three radio galaxies have been targeted over seven near-perigee imaging segments: M87, 3C84 \citep[][Benke et al. in prep.]{2018NatAs...2..472G,2023A&A...676A.114S} and Cen~A. 
M87 was observed in 2014 at 1.7\,GHz (Savolainen et al. in prep.), 4.8\,GHz and 22\,GHz \citep[][hereafter Paper~I]{2023ApJ...952...34K}, as well as in 2018 at 22\,GHz.
In this paper, we present results of the \ra{} observations at 4.8\,GHz made in February 2014. 

\section{Observations and data reduction} \label{sec:obs}
\subsection{\ra{} imaging observations}
M87 was observed by \ra{} simultaneously at 4.8\,GHz ($\lambda=6.2$\,cm, this paper) and at 22\,GHz ($\lambda=1.3$\,cm,   
\citetalias{2023ApJ...952...34K}) on 2014 February 4-5 (\ra{} AO1 observational code: RAKS03B).
While the space telescope and Kalyazin 64-m were obtaining data simultaneously at 4.8\,GHz and 22\,GHz utilizing their co-axial feed horns, the VLBA antennas utilized their frequency agility to switch bands between scans, and the rest of the array was split into telescopes observing exclusively at 4.8\,GHz or 22\,GHz.
The 4.8\,GHz ground segment of the VLBI array consisted of 21 participating telescopes listed in Table~\ref{t:telescp}. 
The observing session lasted from 16\,UT on Feb 4 to 12\,UT on Feb 5 (global experiment code GS032C). 
Over the full time range of 20\,h, the SRT was observing during ten 30\,min segments separated by thermal management breaks necessary for the satellite's high-bandwidth communication antenna drive. 
The orbital segments were scheduled in a range from the perigee (common visibility with VLBA antennas) up to projected baselines of 11.5\,Earth diameters (ED, 2.362 G$\lambda$; common visibility with Australian antennas).
The 128\,Mbps, one-bit-sampled data stream from the SRT was downlinked to the dedicated \ra{} tracking stations in Pushchino, Russia, and Green Bank, USA.
The SRT data were acquired with a total bandwidth of 32\,MHz split into two 16\,MHz wide sub-bands (hereafter IFs), and observed in left circular polarization (LCP) only. 
The ground radio telescopes used two-bit sampling, recording both left and right circular polarization, resulting in a 256\,Mbps recording rate.

Each 30\,min-long observing segment was split into 10\,min-long scans -- the duration determined by the expected coherence time at the K-band due to the Earth's atmosphere.
The recorded data were correlated at the Max Planck Institute for Radio Astronomy using a version of the \textsc{DiFX} software correlator \citep{2011PASP..123..275D} modified for space-VLBI and taking into account parameters of the orbiting antenna and both special and general relativistic effects \citep{2014evn..confE.119B,2016Galax...4...55B}.
An output from the correlator had 256 spectral channels per IF and 0.5\,s integration time.

\begin{table}
\centering
\caption{Radio telescopes which participated in the \ra{} imaging observations of M87 at 5~GHz on 2014 Feb 4--5. }
\label{t:telescp}
\begin{tabular}{lccc}
\hline
Telescope & Code & Diameter & SEFD\\
          &      & (m)      & (Jy)\\
\hline
Spektr-R SRT (RU) & RA & 10 & 4600\\
VLBA-Brewster (USA) &BR & 25 & 210\\
VLBA-Fort Davis (USA) &FD & 25 & 210\\
VLBA-Hancock (USA)& HN & 25 & 210\\
VLBA-Kitt Peak (USA)& KP & 25 & 210\\
VLBA-Los Alamos (USA) & LA & 25 & 210\\
VLBA-Mauna Kea (USA) & MK & 25 & 210\\
VLBA-North Liberty (USA) & NL & 25 & 210\\
VLBA-Owens Valley (USA) & OV & 25 & 210\\
VLBA-Pie Town (USA) & PT & 25 & 210\\
VLBA-St. Croix (USA) & SC & 25 & 210\\
ATCA (AU) & AT & 54$^a$ & 42\\
Ceduna (AU) & CD& 30 & 450\\
Hobart (AU) & HO & 26 & 640\\
Mopra (AU) & MP & 22 & 350\\
Kalyazin (RU) & KL & 64 & 150\\
Onsala (SE) & ON & 25 & 480\\
Hartebeesthoek (ZA) & HH & 26 & 650\\
Medicina (IT) & MC & 32 & 170\\
Noto (IT) & NT & 32 & 260\\
WSRT (NL) & WB & 93$^a$ & 120\\
Effelsberg (DE) & EF& 100 & 20\\
\hline
\end{tabular}

$^{a}$The effective synthesised diameter is given for a phased array.
\end{table}

\subsection{Post-correlation data reduction}
The post-correlation calibration was performed within the \textsc{AIPS} software package \citep{aips} in the \ra{} specific approach \citep[see e.g.][]{2016ApJ...817...96G, 2020ApJ...893...68K,2023A&A...676A.114S}. 
First, \textit{a priori} calibration of the ground array was performed in a standard manner. 
This included corrections for the parallactic angle rotation, removal of the dispersive ionospheric delay using global total electron content maps, and manual phase calibration. The gain amplitudes were calibrated using system temperatures measured throughout the observation and a priori gain curves. Biases due to sampler threshold level variations were removed using autocorrelations.
Global fringe fitting was carried out to remove delay and rate residuals. 
The receiver bandpasses were corrected.
M87 was then imaged in the \textsc{Difmap} package \citep{1997ASPC..125...77S}, and at this stage, the ground array was fully self-calibrated.

\subsection{\textsc{PIMA} fringe fitting}

The software package \textsc{PIMA}\footnote{Available at \url{http://astrogeo.org/pima}.} \citep{2011AJ....142...35P} was primarily used for the AGN Survey data analysis, section~\ref{sec:survey}. 
We also used \textsc{PIMA} for the initial fringe search at ground-space baselines.
\textsc{PIMA} capabilities include phase correction for the residual acceleration: after applying a delay model, it can search for a correlation
coefficient peak in the `residual delay, residual rate, and the rate of
change of the residual rate' space. The search in this third dimension is normally not required for ground stations but can be important for SRT, particularly near perigee, where imperfect knowledge of the spacecraft orbit leads to a non-negligible acceleration term. 

The correspondence between the derived fringe signal-to-noise ratio (S/N) and the probability of false detection (PFD) for every observing scan was derived using the approach described by
\citet{2011AJ....142...35P,2020AdSpR..65..705K}; \citetalias{2023ApJ...952...34K}; \citet{2023A&A...676A.114S}, namely: the theoretical probability density distribution \citep{2017isra.book.....T} of the noise-generated fringe amplitude on the fringe delay-rate plane was fitted to the left part of the full set of estimated S/N values (i.e. `no signal'). 
The $\text{S/N}_\text{PIMA}$ distribution obtained during the acceleration term search and its theoretical fit are shown in Fig.~\ref{fig:raplot} (top).
The following ranges of parameters were used for a fringe search: 
delay of $\pm4~\mu$as, delay rate of $\pm2\times10^{-10}$~s/s, and acceleration of $\pm1\times10^{-14}$~s/s$^2$.
Fringes with the probability of a false detection being less than 0.1~per~cent, which corresponds to $\textrm{S/N}_\textrm{PIMA}>5.8$, were classified as detections.

\begin{figure*}
\centering
\includegraphics[angle=270,width=0.65\columnwidth]{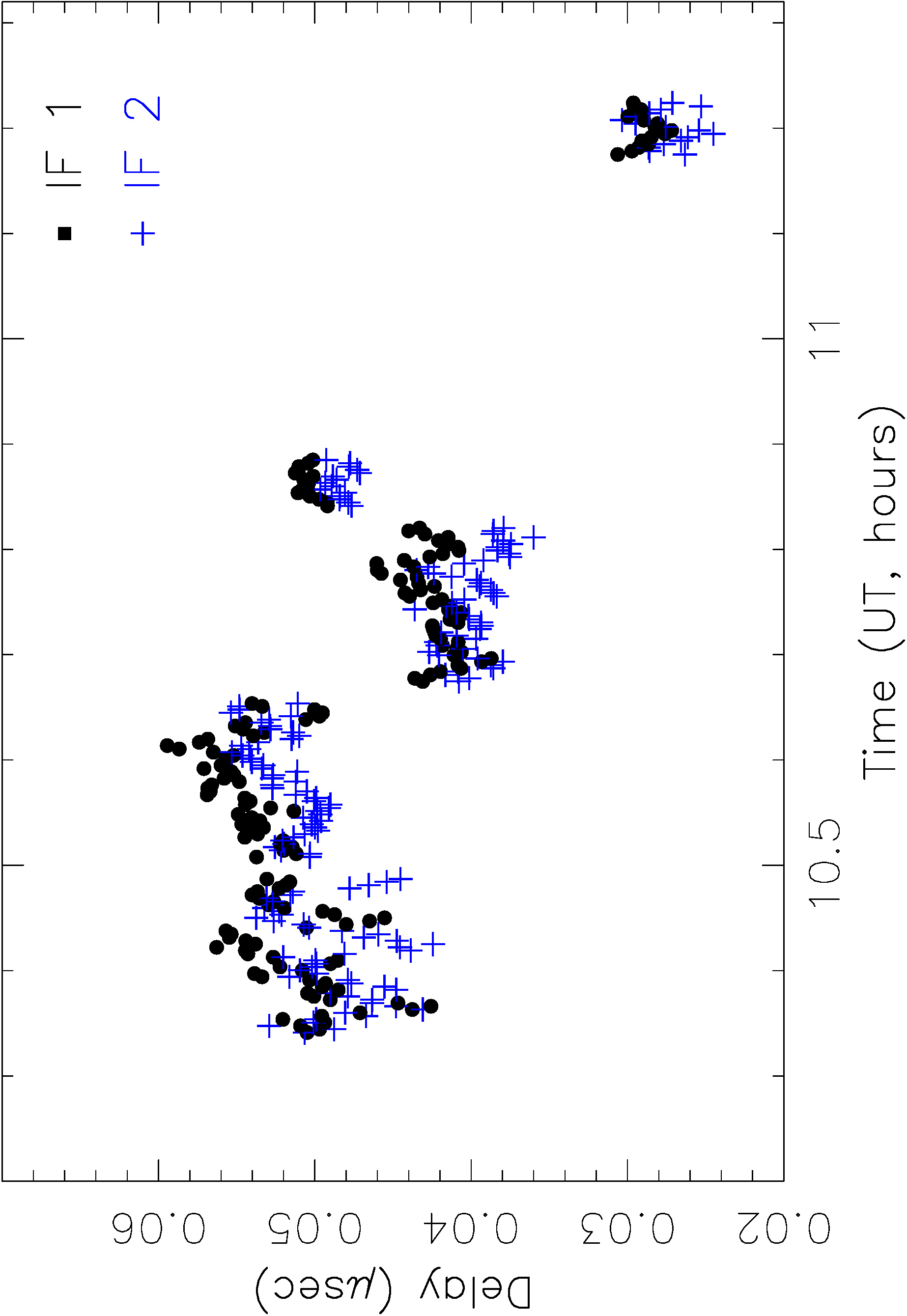}
\includegraphics[angle=270,width=0.65\columnwidth]{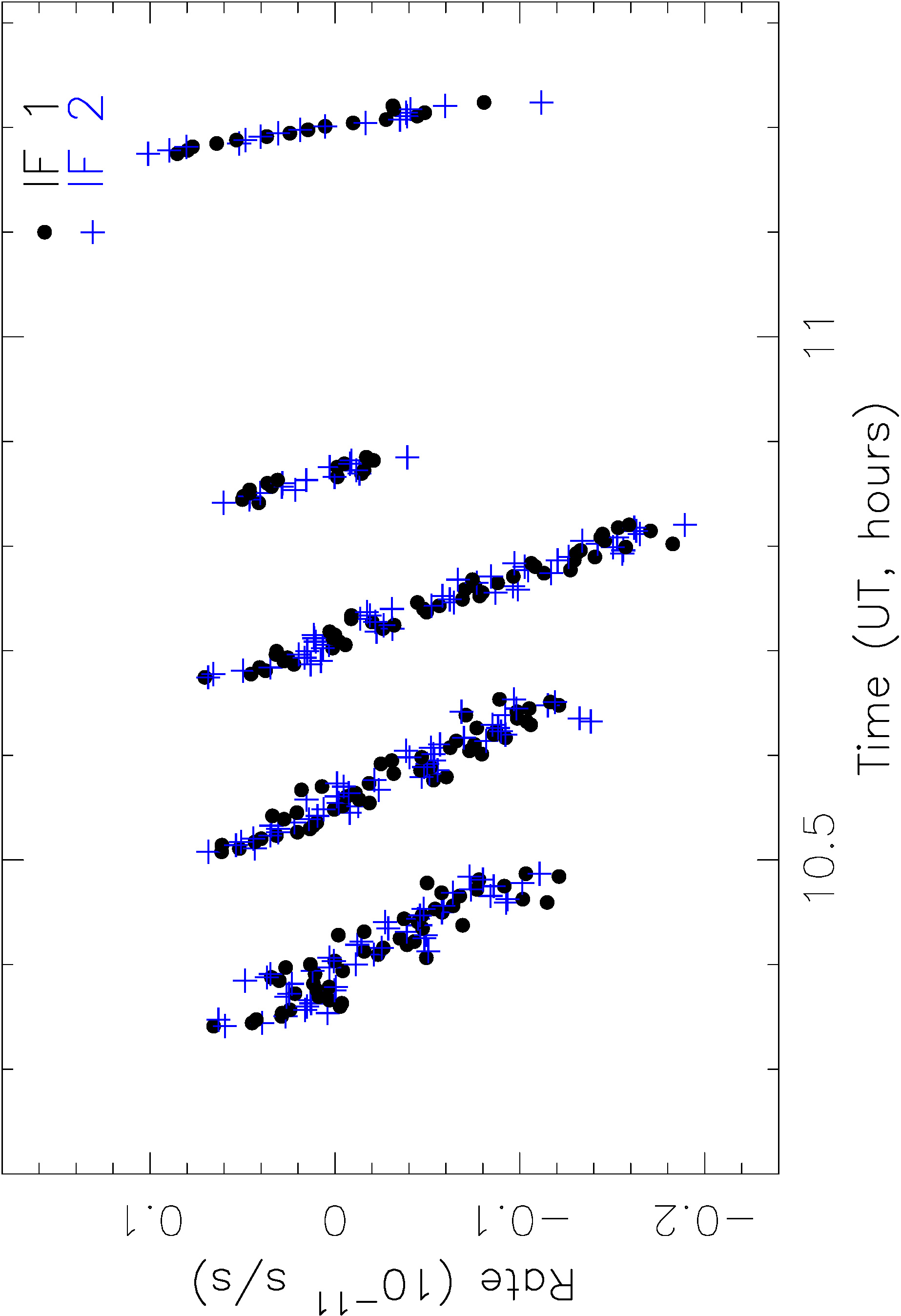}
\includegraphics[angle=270,width=0.65\columnwidth]{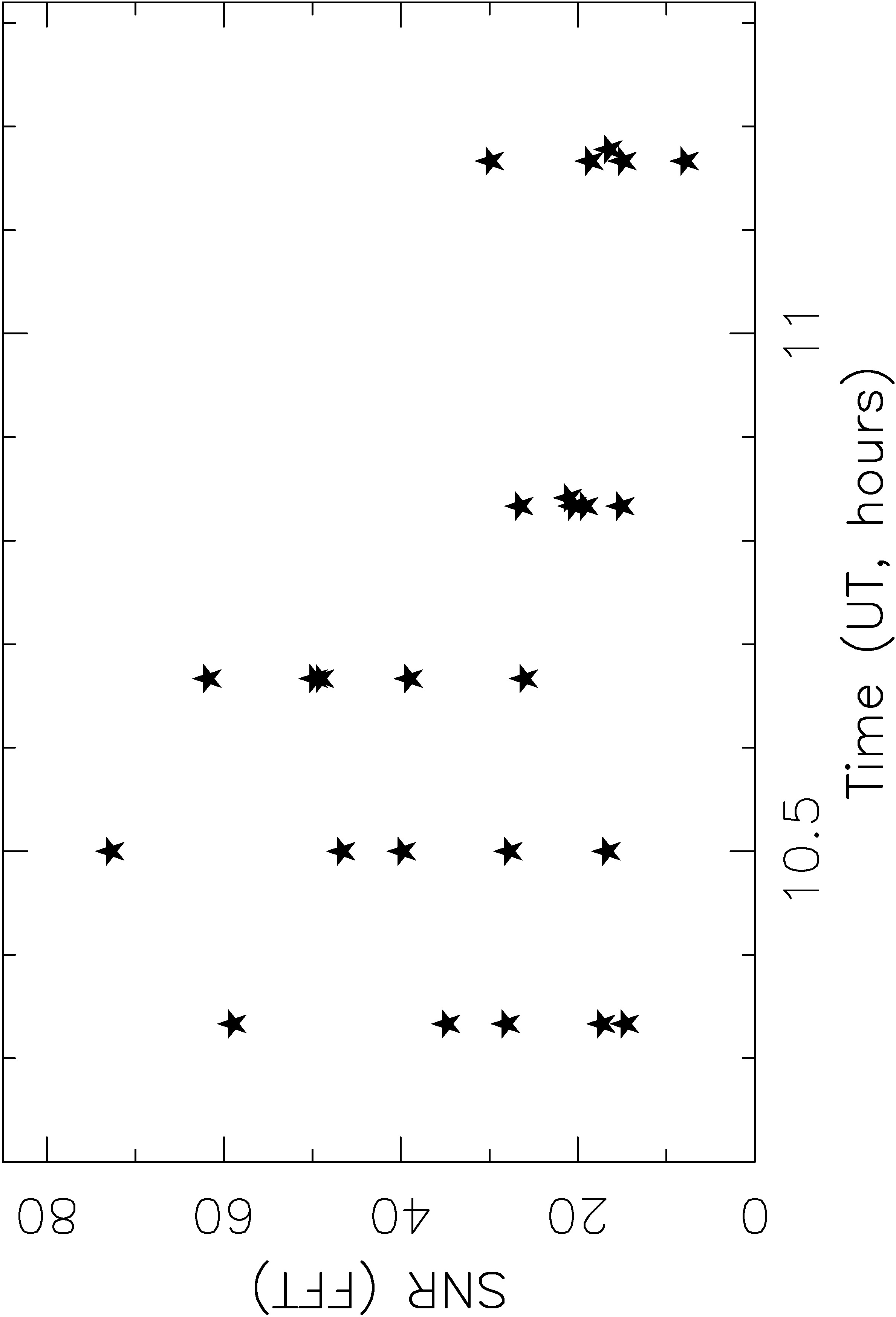}
\caption{Residual delay (left) and rate (middle) \textsc{AIPS} \texttt{FRING} solutions for short space baselines ($\lesssim1$\,ED) during the imaging experiment. The solution interval of 30\,s was used. 
The delay jumps between scans are determined by the accuracy of corrections made to connect the \textit{RadioAstron} data recording and synchronization systems: the recording of observations starts according to the atomic clock at the ground station, and the digitization of the data takes place on the SRT with synchronization according to the on-board frequency standard. The rapid change in the interference frequency inside the scan is caused by an uncompensated residual acceleration of the SRT.
(Right) Fringe S/N derived in \textsc{PIMA}, using the scan length of 10\,min as an integration time. Significant detections have been obtained between the SRT and five VLBA antennas (BR, LA, NL, MK, SC).}
\label{fig:rafrn}
\end{figure*}

Using \textsc{PIMA}, we detected fringes on Earth-space baselines $\lesssim1$\,ED for the five VLBA antennas (BR, LA, MK, NL, SC; $\text{S/N}_\text{PIMA}$ values are given in Fig.~\ref{fig:rafrn}) and on $\thicksim3$\,ED only between the SRT and Effelsberg, the most sensitive ground telescope (with $\text{S/N}_\text{PIMA}=6.0$, which corresponds to a probability of $\approx 3\times10^{-4}$).
Correlated flux densities as a function of projected $(u,v)$ distance are shown in Fig.~\ref{fig:raplot} (bottom).

\subsection{\textsc{AIPS} fringe fitting}
The resultant ground-based image of M87 was loaded back to \textsc{AIPS} and used as a model to improve fringe searching at space baselines.
The global fringe fitting with the \textsc{AIPS} task \texttt{FRING} was performed using the self-calibrated ground array data, model of M87, multiple baseline combinations (\texttt{DPARM}(1)=3) and an exhaustive baseline search with the most sensitive ground antennas. 
It results in significant detections at baselines up to $\sim 1$~ED between the SRT and five VLBA antennas (BR, LA, MK, NL, SC) when adopting the solution interval of 30\,s.
The solutions, shown in Fig.~\ref{fig:rafrn}, are perfectly consistent with the results obtained using \textsc{PIMA}.

Initial attempts to find solutions in the next segment ($\thicksim (2-3)$\,ED) failed when using longer solution intervals (up to 30\,min). 
Then, we made use of \textsc{PIMA} detection between the SRT and EF ($\text{S/N}_\text{PIMA} = 6$), and applied this solution to the data (delay = $1.691~\mu$s, rate = 36.86\,ps/s).
Then the data were averaged coherently over the scan length of 10~min, and the search window was narrowed in the delay-rate space.
This allowed us to detect fringes with $\text{S/N}_\text{AIPS}\thicksim4$, as defined by \textsc{AIPS}, using \texttt{FRING} between the SRT and ten ground-based telescopes (Europe: ON, MC, NT, WB, EF; VLBA: BR, LA, NL, MK, SC).
The described fringe solutions were transferred to the parallel effort of fringe search with the data obtained simultaneously at 22\,GHz \citepalias{2023ApJ...952...34K}.

\begin{figure}
\centering
\includegraphics[angle=0,width=0.82\columnwidth, angle=270]{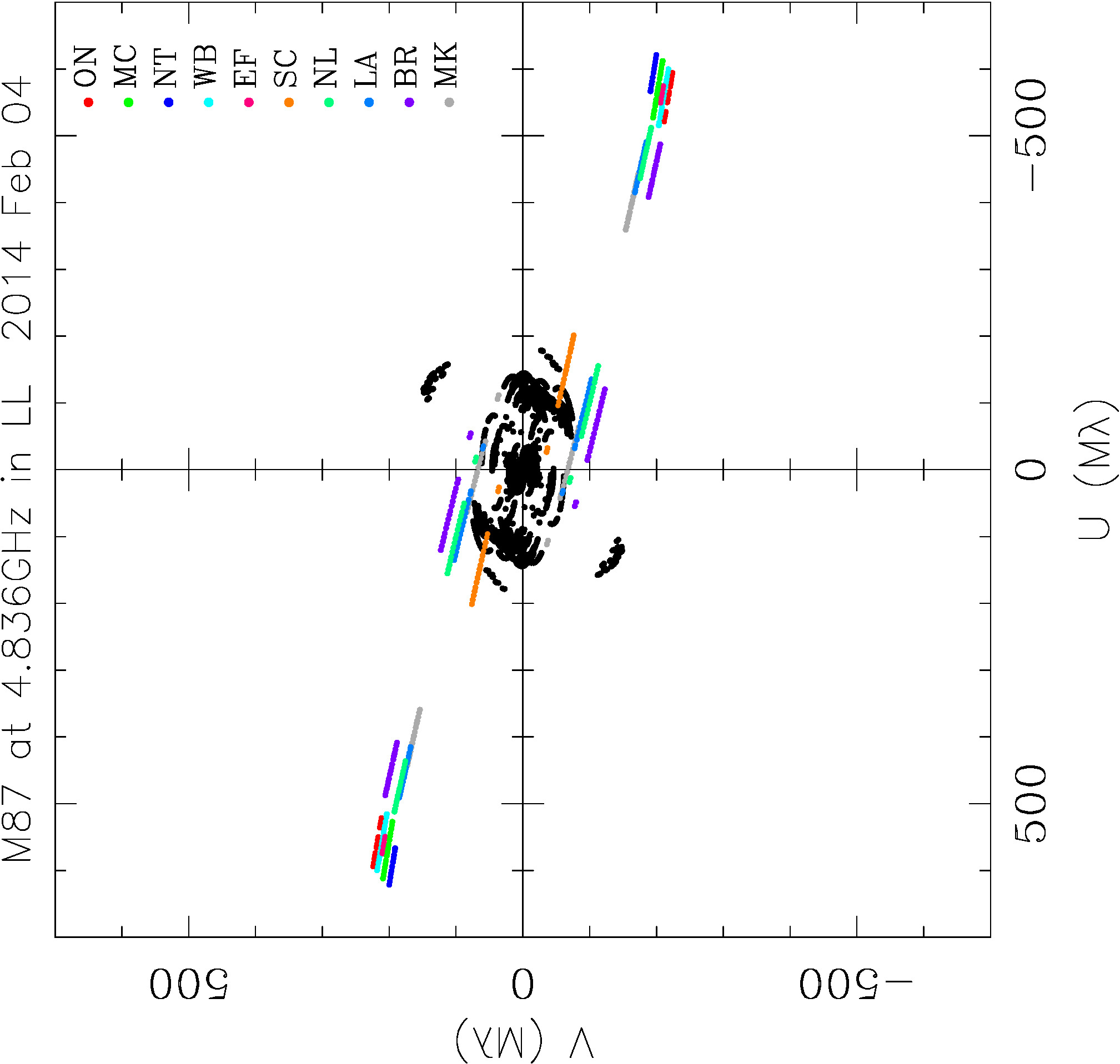}\\
\includegraphics[angle=0,width=0.82\columnwidth, angle=270]{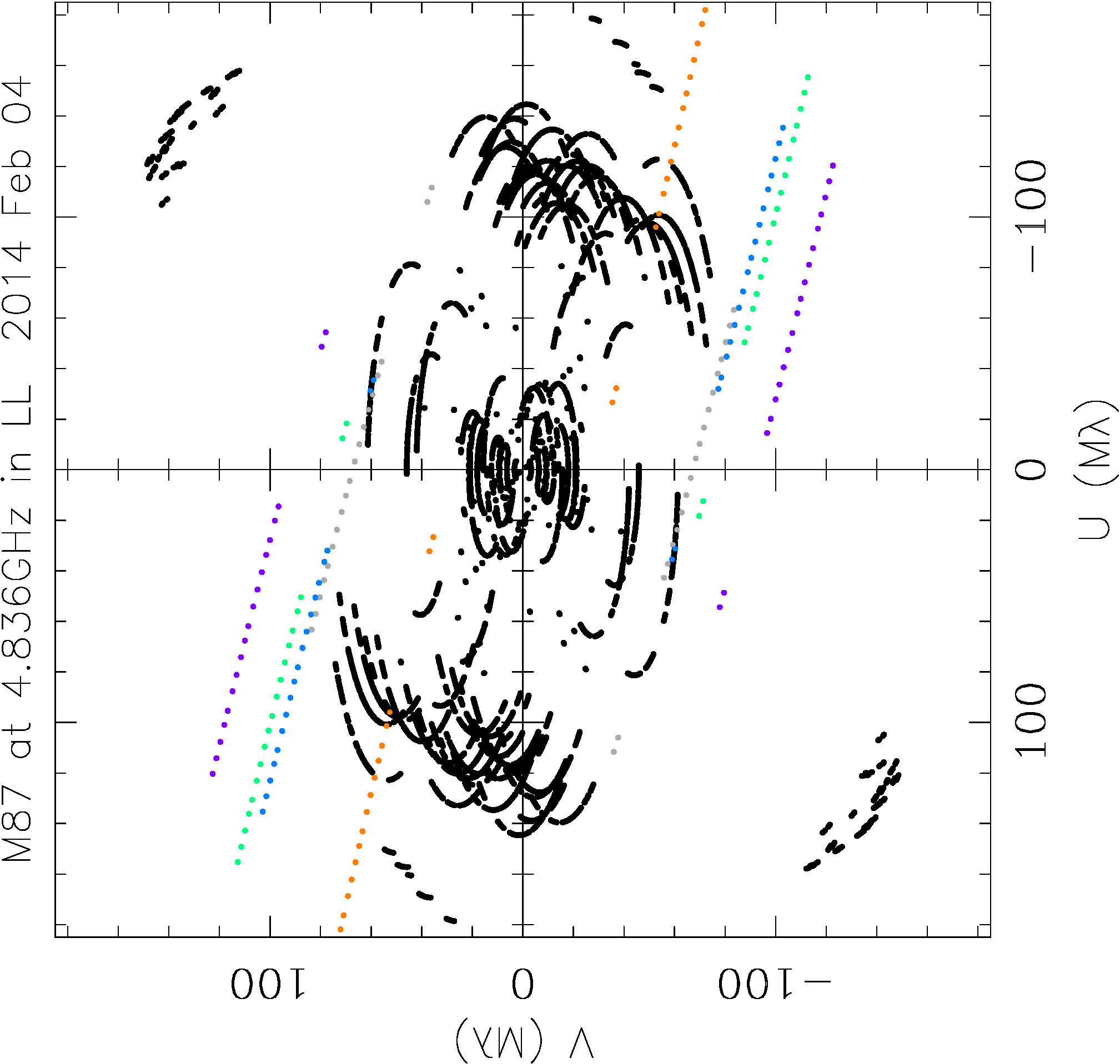}
\caption{The $(u,v)$ coverage of the fringe-fitted interferometric visibilities for the \ra{} observations of M87 at 4.8\,GHz. \textit{Top:} Ground-to-space baselines colour-coded for the different ground-based telescopes. \textit{Bottom:} Zoom into the $(u,v)$ plane area for the baselines between ground-based telescopes.}
\label{fig:uvplt}
\end{figure}

The resulting coverage of the fringe-fitted data in the Fourier domain for the ground and long-baseline detections is shown in Fig.~\ref{fig:uvplt}.
Imaging was performed in \textsc{Difmap} following the \texttt{CLEAN} \citep{CLEAN, 1980A&A....89..377C} imaging and self-calibration procedure. 
Data were averaged over 10~sec, and self-calibration was performed down to 1~min. 

\subsection{Observations within the \ra{} AGN Survey}
\label{sec:survey}

The AGN Survey was one of the \ra{} Early Science Programs and was later continued as a Key Science Program \citep[for a discussion of the survey strategy and preliminary results see][]{2012evn..confE.024, 2020AdSpR..65..705K}. Observations started in 2012.5 and were running for four years. 
The program consisted of one-hour-long snapshot observations of AGNs with the SRT scheduled at 1.7\,GHz ($\lambda=18$\,cm), 4.8\,GHz ($\lambda=6.2$\,cm) and 22\,GHz ($\lambda=1.3$\,cm).
Each survey snapshot was supported by a few ground telescopes, ideally one large telescope (to have the best shot at fringe detection at the long space-ground baseline) and a medium-class telescope (to get visibility amplitude on a ground-ground baseline useful in further analysis and as a fringe-check for the large telescope), depending on the availability of ground stations. The $(u,v)$ coverage of an individual snapshot is too sparse for imaging but allows for a coarse estimate of the source flux density and size \citep[see e.g.][]{2015A&A...574A..84L, 2018MNRAS.474.3523P, 2018MNRAS.475.4994K}. 
The AGN Survey project was completed in 2016 and followed up by an AGN Monitoring program through 2018. 
The detailed discussion of the project will be presented by Kovalev et al., in prep.

M87 was observed within the \ra{} AGN Survey Key Science Program between 2013 and 2016 at 4.8\,GHz in 21 individual snapshots. 
The signal was detected at the only space baseline between Arecibo and the space radio telescope at the projected baseline of 3.2\,ED (experiment code raes03ry, 2013 April 29, $(\text{S/N}_\text{PIMA}=9.9$), in agreement with the imaging experiment described here (Fig.~\ref{fig:raplot}).

\begin{figure}
\centering
\includegraphics[width=0.99\columnwidth]{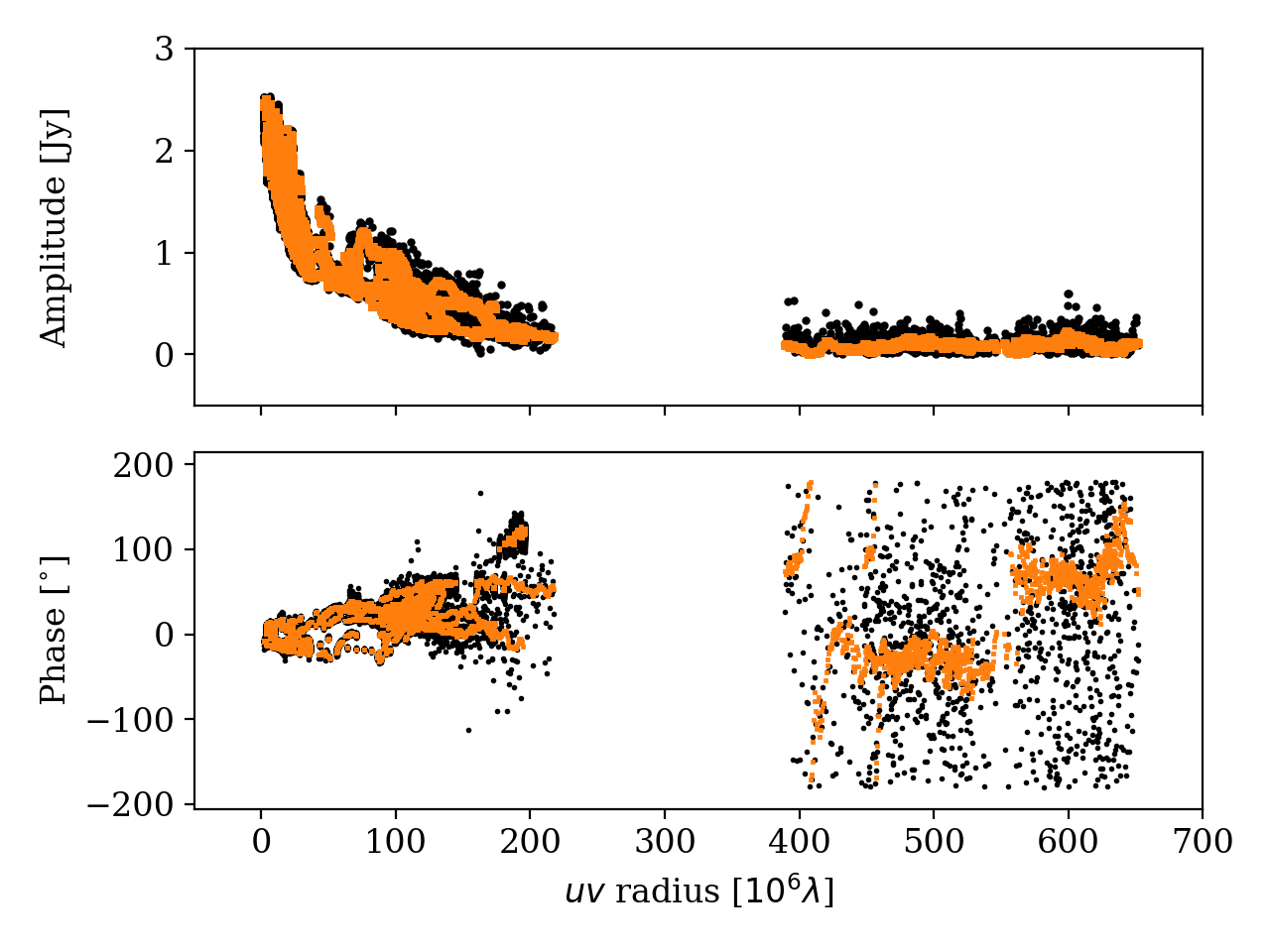}
\caption{Self-calibrated visibility amplitudes and phases as a function of projected $(u,v)$-distance (black) of the \ra{} observations of M87 at 4.8\,GHz. Orange dots indicate the fit of the \texttt{CLEAN} model into the data.}
\label{fig:radpl}
\end{figure}

\section{Results} \label{sec:res}

\subsection{Images and jet structure}

\begin{figure}
\centering
\includegraphics[angle=-90,width=0.9\columnwidth]{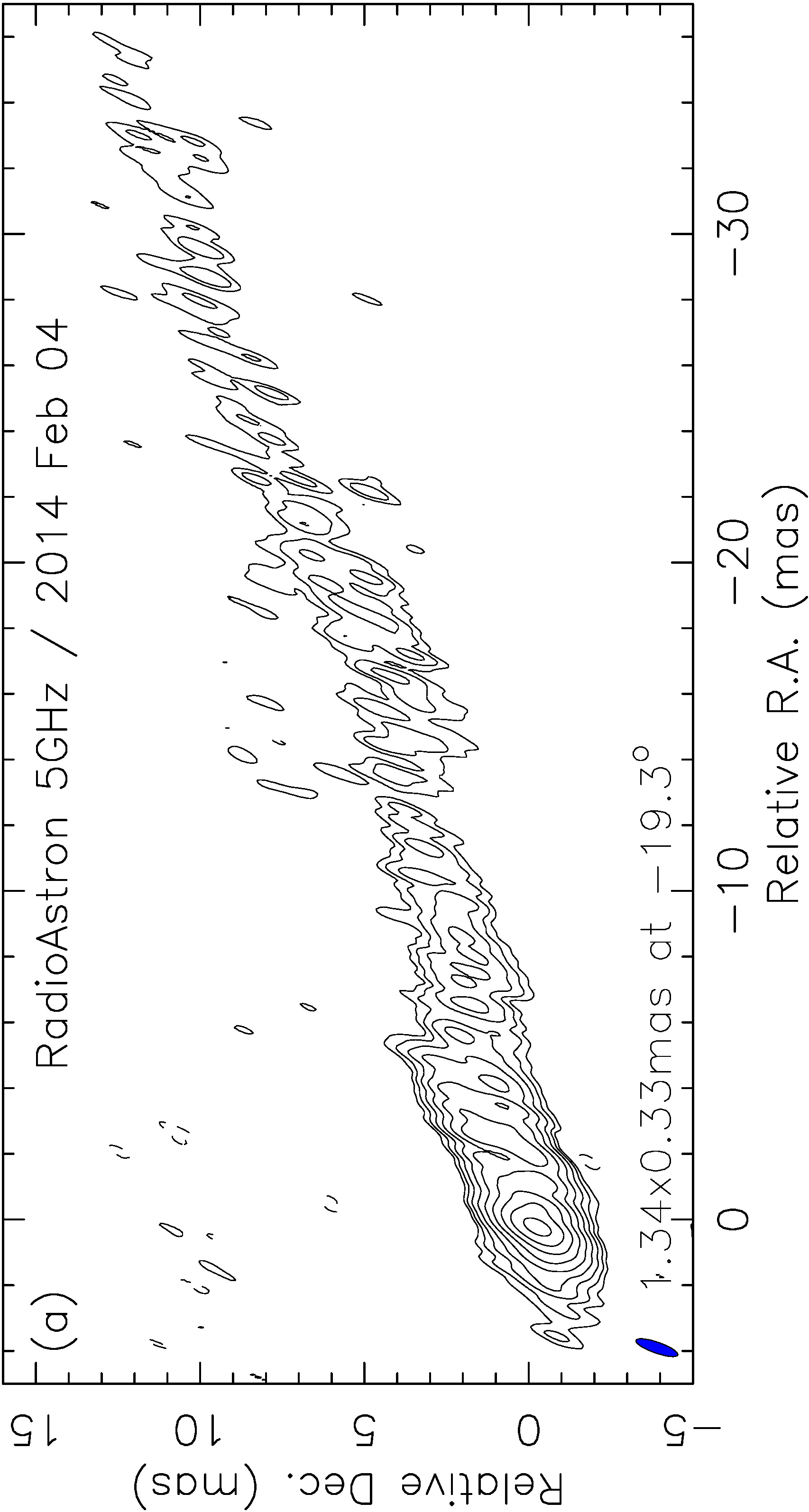}\\
\includegraphics[angle=-90,width=0.9\columnwidth]{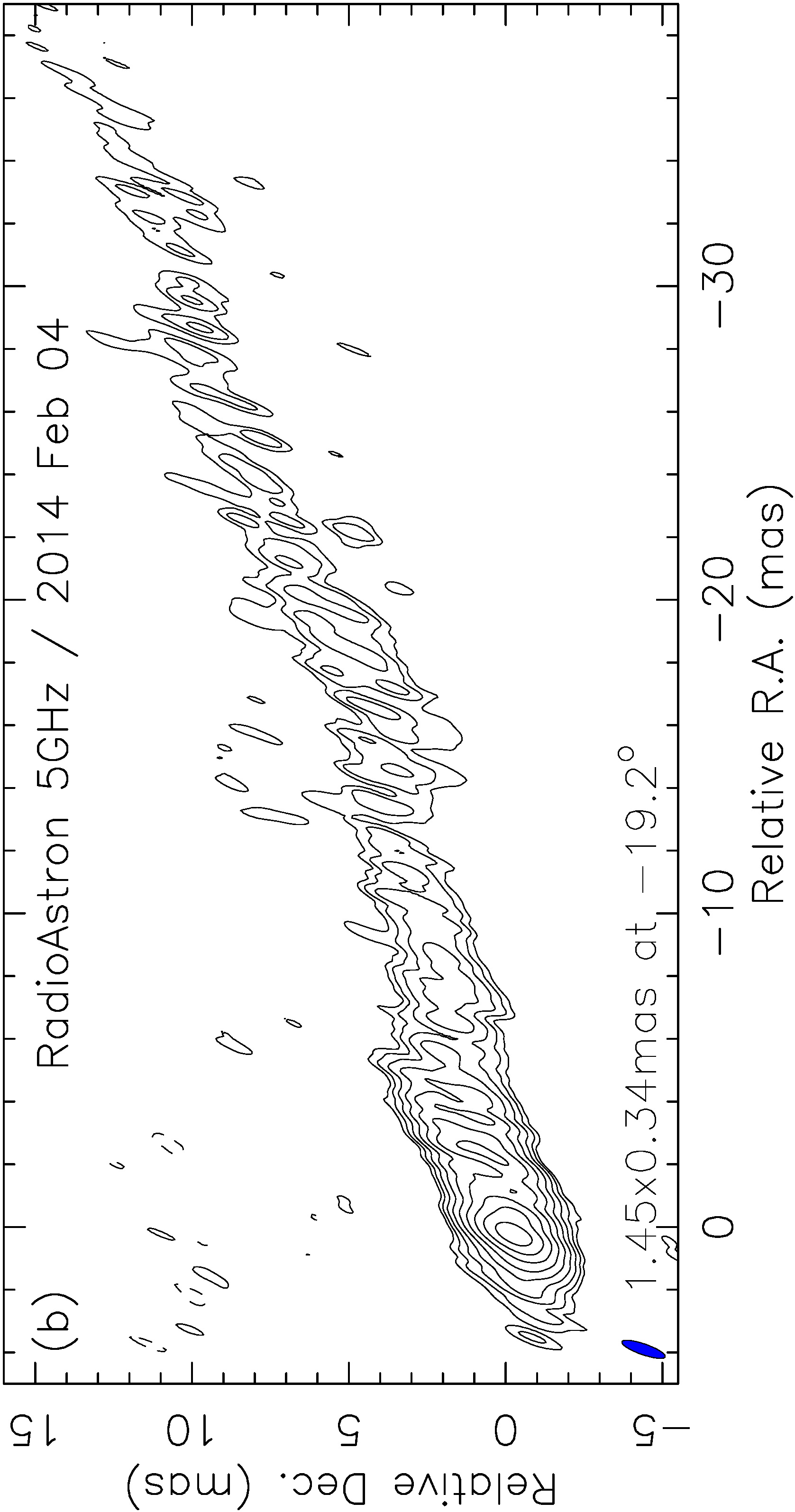}\\
\includegraphics[angle=-90,width=0.9\columnwidth]{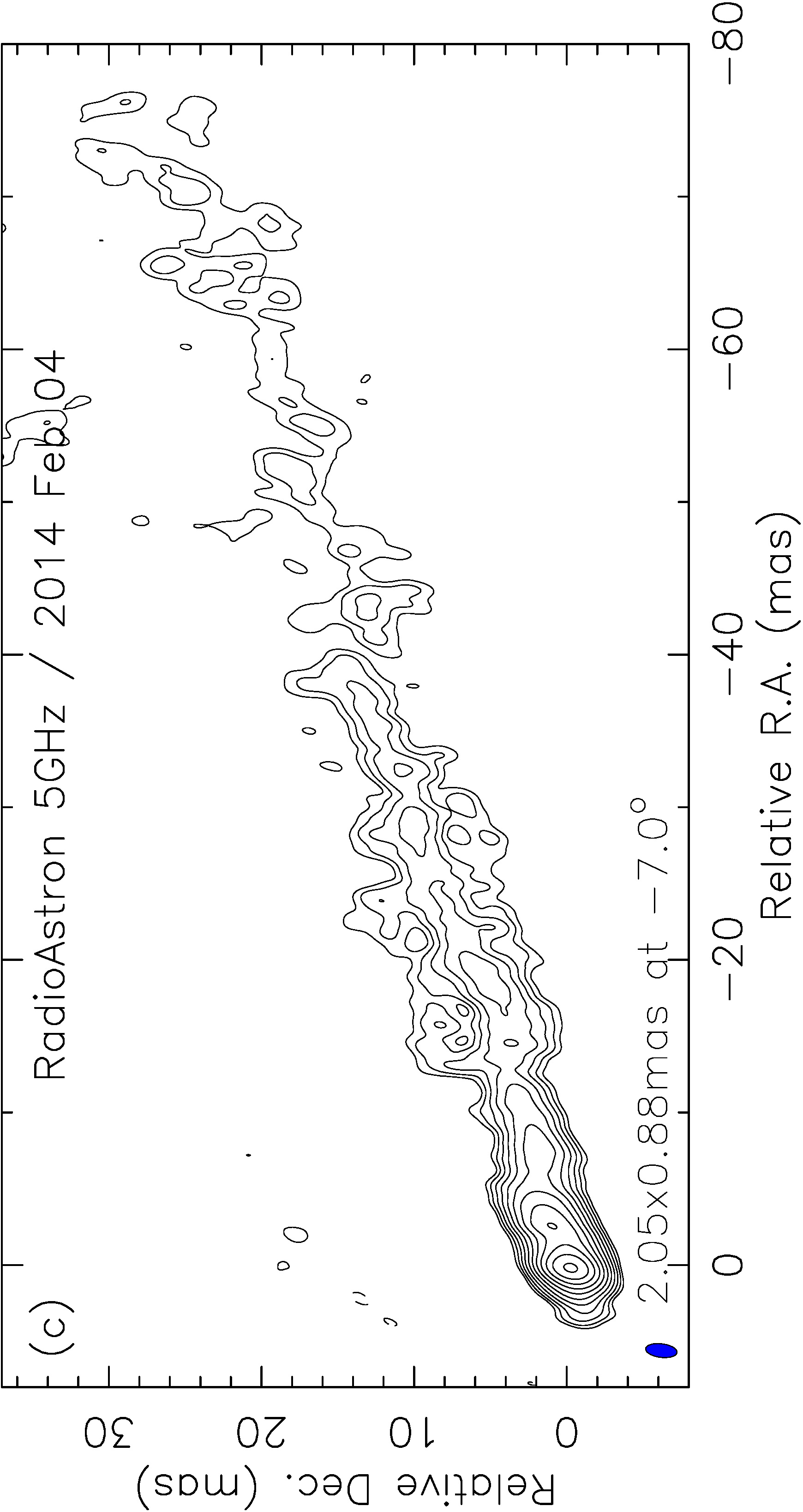}
\caption{ \ra{} 5\,GHz total intensity images obtained on 2014 Feb 04 using different weighting schemes: superuniform (a), uniform (b) and natural (c). The contour levels increase by factors of 2 starting from 1.0 (a), 0.88 (b) and 0.63 (c) mJy~beam$^{-1}$. The noise level $S_{\rm rms}$ is 0.28 (a), 0.29 (b) and 0.18 (c)~mJy~beam$^{-1}$. The peak flux densities are 382 (a), 396 (b) and 725 (c) mJy~beam$^{-1}$. The restoring beam full width at half maximum (FWHM) and position angle (shown by shaded ellipse in the bottom left corner) are: 1.34 $\times$ 0.33\,mas at $-19.3^{\circ}$ (a), 1.45 $\times$ 0.34\,mas at $-19.2^{\circ}$ (b), and 2.05 $\times$ 0.88\,mas at $-7.0^{\circ}$ (c).}
\label{fig:suim}
\end{figure}

Self-calibrated visibility amplitudes and phases as a function of projected Fourier spacing ($(u,v)$ distance) and \texttt{CLEAN} model fit to these data are shown in Fig.~\ref{fig:radpl}.
The resulting \ra{} images using superuniform (in which the bin size in the $(u, v)$ plane is 5 pixels and the weights are scaled by the visibility amplitude error to the power $-1$, i.e. \textsc{UVWEIGHT}$ = (5, -1)$), uniform, and natural weighting schemes are shown in Fig.~\ref{fig:suim}. 
The obtained synthesized beam size corresponding to the superuniform weighting is 1.34$\times$0.3\,mas at a position angle $-$19\degr.

The transverse jet structure in the inner 5~mas from the 4.8\,GHz core appears to be edge-brightened, with the northern edge being brighter closer to the core. Further down the jet, the structure becomes helically-shaped.
The position angle of the jet within the $\thicksim5$~mas off the core is about $-69^{\circ}$. 
There is a weak structure extending away from the core toward the south-east, which can be traced out to $\thicksim4$~mas from the core. The spectral index and the nature of this region are discussed in section~\ref{sec:spi}.

\subsection{Brightness temperature}
\label{sec:tb}

\begin{figure}
\includegraphics[width=0.95\columnwidth]{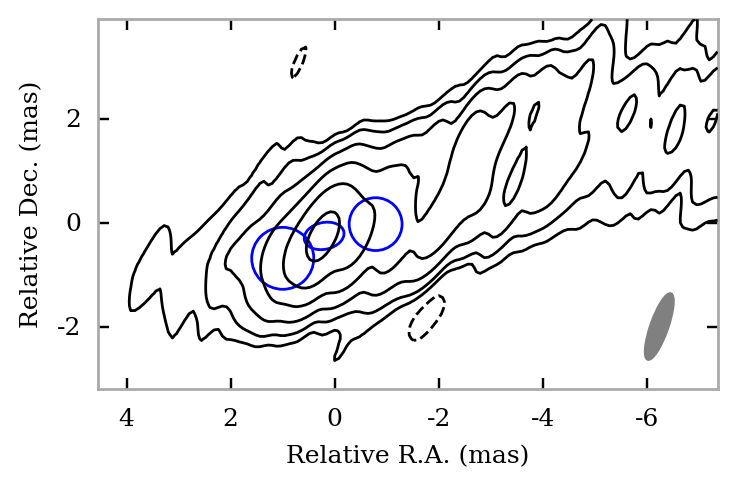}
\caption{Total intensity contours as in Fig.~\ref{fig:suim}a, but increasing by factors of 3 starting from 1.13 mJy~beam$^{-1}$, with overlaid model-fit components in the nuclear region. See section~\ref{sec:tb} for details.} 
\label{fig:nucl_mod}
\end{figure}

\begin{table}
    \centering
    \caption{Parameters of the model-fit core component.}
    \label{tab:nucl_reg}
    \begin{tabular}{lcccc}
    \hline
    $S$ (mJy) & ${\rm FWHM}_{\rm maj}$ (mas) & ${\rm FWHM}_{\rm min}$ (mas) &  PA ($^{\circ}$) \\
    \hline
    $865\pm254$ & $0.775\pm0.15$ & $0.516\pm0.24$ &  $-78\pm18$\\
        \hline
    \end{tabular}
\end{table}

To calculate the apparent brightness temperature $T_{\rm b}$ of the VLBI core, we model-fitted the emission of the nuclear region with a number of Gaussians in the (u,v)-plane, using the \textsc{modelfit} procedure implemented in \textsc{Difmap}.
Before doing so, we subtracted \texttt{CLEAN} components from the visibility data outside 1 mas from the peak along the jet, to avoid influence and fitting the extended jet emission.
The nuclear region is best described by three components: the elliptical VLBI core, the circular jet and counter-jet components, as shown in Fig.~\ref{fig:nucl_mod}. The obtained model-fit parameters are listed in Table~\ref{tab:nucl_reg}.
Using equation~1 of \citetalias{2023ApJ...952...34K}, we obtain $T_{\rm b, core} = (1.15\pm0.64)\times 10^{11}$~K.

\begin{figure}
\includegraphics[width=0.95\columnwidth]{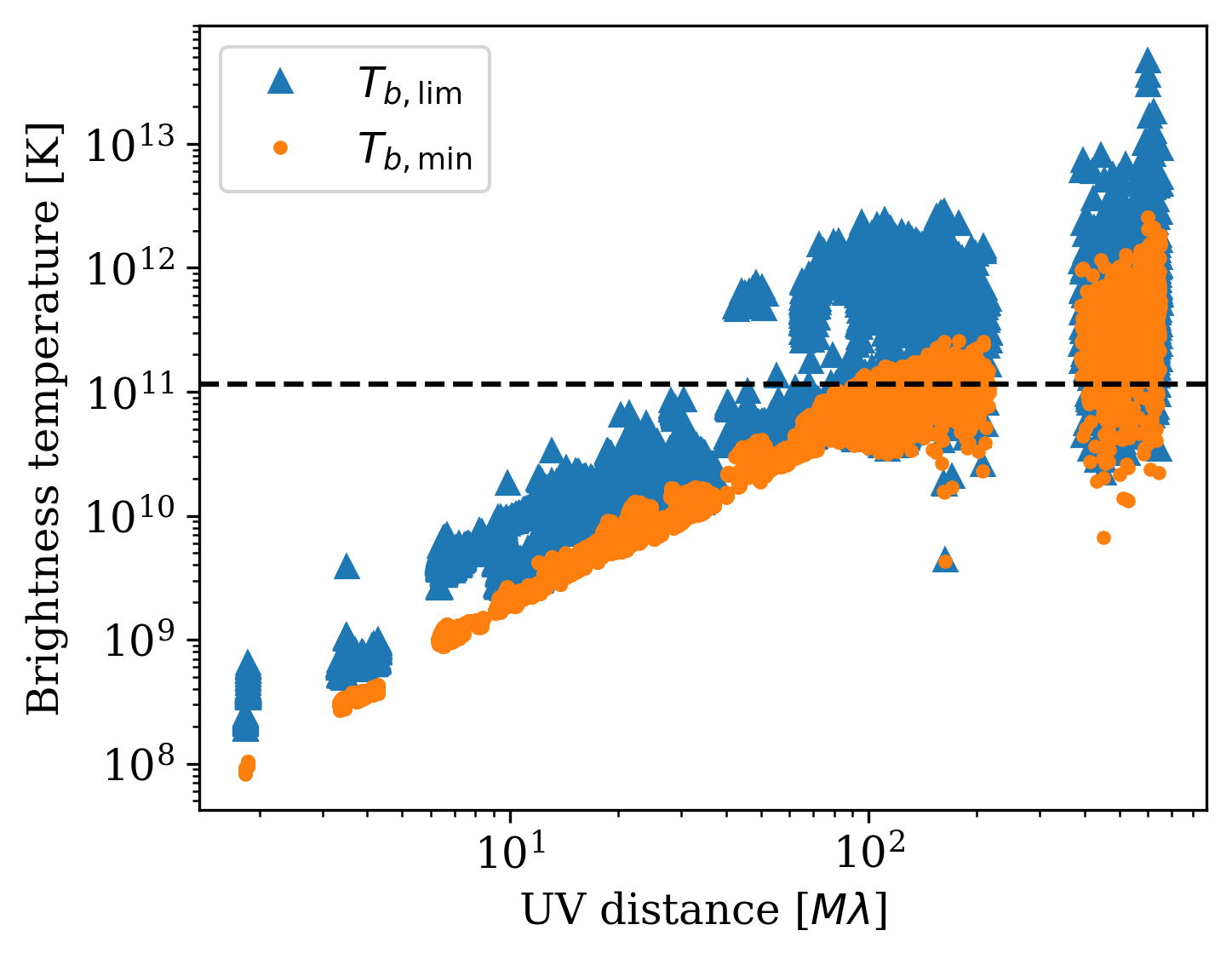}
\caption{Minimum ($T_{\rm b,min}$; orange) and limiting ($T_{\rm b,lim}$; blue) brightness temperature values calculated using visibility amplitudes \citep{2015A&A...574A..84L}. The black dashed horizontal line indicates
$T_{\rm b} = 1.15 \times 10^{11}$~K from the Gaussian model-fitting.} 
\label{fig:tb_uvdist}
\end{figure}

We supplemented our analysis by directly estimating a lower-limit range for the brightness temperature using the method of \citet{2015A&A...574A..84L}. This technique is well-suited for sparse (u,v)-data, as it depends mainly on correlated flux densities without requiring a detailed source model. The results, presented in Fig.~\ref{fig:tb_uvdist} alongside the estimate from the geometric model fit, agree well.

\subsection{Helical jet structure and plasma instabilities} 
\label{sec:instability}

To perform the analysis of the jet helical structure, we followed the procedure described in \citet{2023MNRAS.526.5949N} (see their section~3.1 for details).
The transverse profiles of the total intensity have been taken along the straight ridge-line with the position angle PA$ = -67.7^{\circ}$.
This PA value corresponds to the position angle between the 450\,mas scale structure visible at 8\,GHz and the HST-1 component \citep{2023MNRAS.526.5949N}, and was chosen to compare the results of other studies (\cite{2023MNRAS.526.5949N}; Savolainen et al. in prep.).
Meanwhile, this value of the PA is larger than what is measured using only mas-scale (inner 30 mas) VLBI observations (e.g. PA$ = -70^{\circ}$ in \citealt{2016AA...595A..54M} or PA$ = -72^{\circ}$ in \citealt{2018ApJ...855..128W}) due to a jet bend within about inner 80\,mas.

\begin{figure*}
\centering
\includegraphics[width=0.95\textwidth]{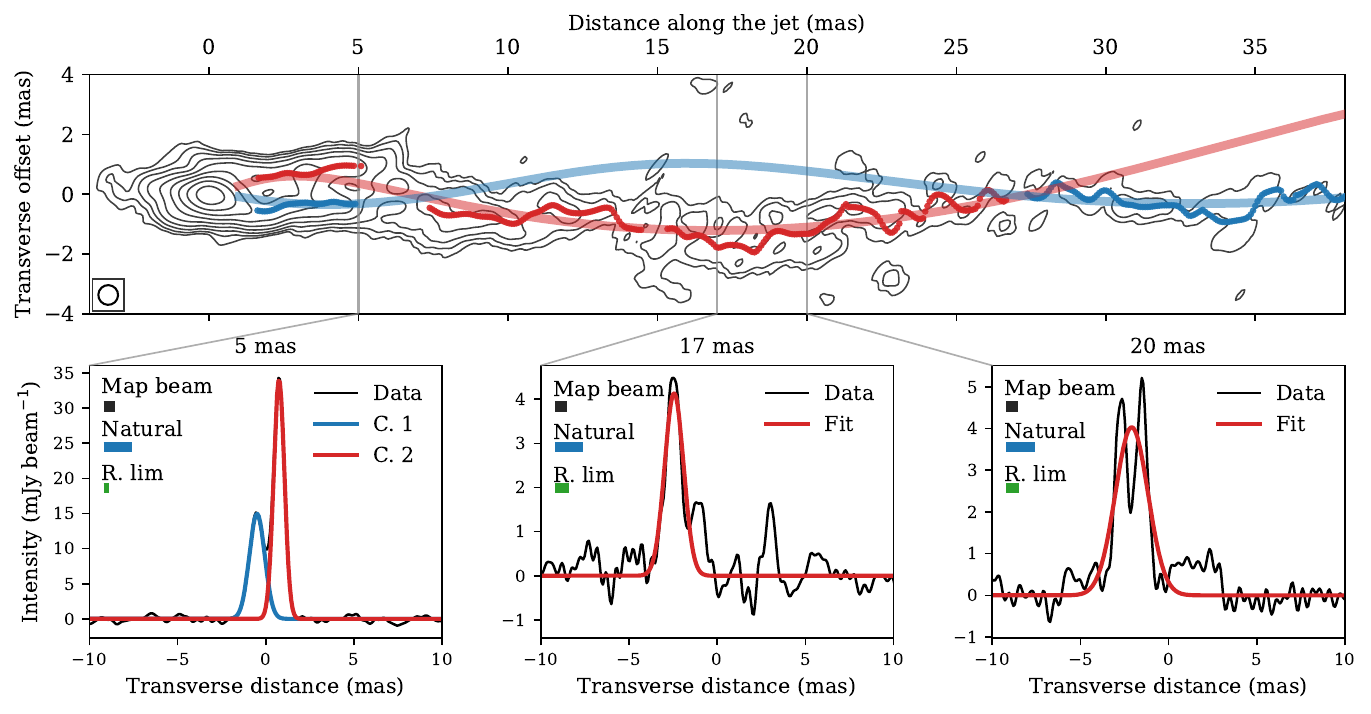}
\caption{\textit{Top panel:} \ra{} Stokes $I$ intensity map at 4.8~GHz, shown by black contours, overlaid with Kelvin--Helmholtz instability threads. The thick red and blue curves show the best-fitting three-dimensional helical model projected on to the plane of the sky, assuming the M87 jet viewing angle of $\theta = 17^{\circ}$. The darker thin curves mark the positions of the fitted Gaussian peaks in total intensity. The Stokes $I$ contours increase by successive factors of $2$, starting from $3\sigma = 1$~mJy~beam$^{-1}$. The circular restoring beam used for the image, with FWHM $0.7$~mas and an area equivalent to the average area of the superuniform synthesized beam, is shown in the lower-left corner. The image has been rotated clockwise by $18^{\circ}$ for presentation, to display the jet and filaments more clearly. The grey vertical lines mark the locations of the transverse slices shown in the lower panels, taken perpendicular to the jet axis at projected distances of 5, 17, and 20~mas from the core.
\textit{Bottom panels:} Transverse intensity profiles at the three slice positions marked in the upper panel (5, 17, and 20~mas). The measured profiles are shown by thin black curves, and the Gaussian model fits by thick coloured curves. In the left-hand panel, two Gaussian components are fitted and are shown individually by the blue and red curves. In the upper-left corner of each profile panel, the horizontal bars indicate the relevant angular scales. The black bar labelled ``Map beam'' corresponds to the 0.7~mas beam adopted for the profile analysis, while the blue bar labelled ``Natural'' shows the projected transverse size of the natural beam ($2\times0.9$~mas), corresponding to 1.6~mas in this direction. The green bar labelled ``R. lim'' indicates the effective resolution limit projected along the transverse-profile direction (see Sect.~\ref{apdx:snr_maps} for details) and amounts to 0.3, 0.8, and 0.7~mas at distances of 5, 17, and 20~mas along the jet, respectively.} 
\label{fig:khmodes}
\end{figure*}

Transverse profiles were fitted with one to three Gaussian functions. The optimal number of components was selected by minimizing the $\chi^2_{\rm red}$ \citep[e.g.][]{2003drea.book.....B} at 95~per~cent confidence level.
The location of the fitted Gaussian peaks and examples of transverse intensity profiles are illustrated on the over-resolved map, Fig.~\ref{fig:khmodes}. A statistically significant double-peaked transverse structure is detected near the VLBI core (within $\thicksim 5$\,mas) and over a short interval around 19--21 mas along the jet.

The fitted Gaussian centroids were used as the input points for the filament analysis. Their intensities, widths, and positions were used to assign the components to possible threads.
To parametrize the filaments, we introduce a geometrical model of a three-dimensional helix: 
\begin{equation}
    [x,y] = A[\sin,\cos]{\left(\frac{2 \pi d}{\lambda_\mathrm{KH}} + \phi \right)},
    \label{eq:helix}
\end{equation}
where $d$ is a distance from the jet origin, $A$ is an amplitude, $\lambda_\mathrm{KH}$ is the wavelength, and $\phi$ is the phase. In the case of the KH instability, the observed instability mode develops within the jet, and the corresponding threads take the form of expanding helices, where their amplitude $A$ and wavelengths $\lambda_\mathrm{KH}$ increase with the jet radius as $[A,\, \lambda_{\textrm{KH}}] \propto R_{\textrm{jet}}$ \citep{2000ApJ...533..176H}, with the jet geometry described by a power law $R_{\textrm{jet}} \propto d^{k}$. In this paper, $k = 0.5$ is fixed to have compatible fitting results with previous works. The instability model was also oriented and projected onto the sky plane using the M87 jet angle to the line of sight $\theta = 17^{\circ}$ \citep{2023MNRAS.526.5949N}.  

\begin{table}
    \centering
    \caption{Identification of Kelvin-Helmholtz plasma instability modes. The parameters are defined for an oscillatory pattern in R$_{\textrm{jet}} = 1$\,mas region.}
    \label{tab:KH_fit}
    \begin{tabular}{lcccc}
        \hline
Thread$^{a}$  & $A_0$ (mas) & $\lambda_\mathrm{KH,0}$ (mas) & $\phi_0$ ($^{\circ}$) &  Mode$^{b}$ \\
\hline
Red & $0.38 \pm 0.03$ & $17.3 \pm 1.4$ & $311 \pm 21$ & E$_{b1}$ \\
Blue & $0.16 \pm 0.06$ & $14.7 \pm 0.3$ & $156 \pm 8$ &E$_{b1}$ \\
        \hline
    \end{tabular}
    
$^{a}$Thread name by colour according to Fig.~\ref{fig:khmodes}.
$^{b}$E$_{b1}$---First order elliptical body mode of KH instability.
\end{table}

Under the hypothesis that the filamentary helical structure is produced by a Kelvin--Helmholtz instability, the detected Gaussian components were assigned to at most two ridge groups. We emphasize that this assignment is model-dependent and not unique.
Due to the gradual jet plasma MHD acceleration up to 3-5 mas projected distance \citep{2016AA...595A..54M}, we expect more or less equal jet sides illumination on smaller scales, and strong boosting effect, highlighting one jet side, on larger scales. This means the suppressed northern (receding) branch of a disturbance of any type in a jet. That is why we attribute the 17-20~mas profiles, being either locally one- or two-peaked, to one branch of a jet instability.

The grouping was based on three criteria. First, we required continuity of the Gaussian-component position, intensity, and width between neighboring transverse slices. Second, we avoided connecting isolated local maxima that were not statistically significant or were not persistent over adjacent slices. Third, we required the resulting pattern to be broadly consistent with the expected behavior of a KH-driven helical or elliptical mode, for which the amplitude and characteristic wavelength increase with the jet radius.

The components assigned to the two groups are shown in Fig.~\ref{fig:khmodes} as red and blue dots. The smooth red and blue curves are the projected best-fitting helical models.
In the inner jet, the two ridge groups are separated most clearly and can be identified with the least ambiguity. Farther downstream, however, the intensity decreases, and one of the possible threads, in particular the northern one, is not significantly detected over part of the jet. 
Therefore, the helical model is not fitted to blank regions of the image, but only to the detected Gaussian components assigned to each thread. 
The continuous curves in Fig.~\ref{fig:khmodes} are model interpolations or extrapolations across regions of non-detection.

We evaluated whether this short, double-peaked region at $15$--$21$\,mas should be considered as two independent ridges. This interpretation does not improve the consistency of the KH model. The FWHM of the two local peaks in this region is at the level of the limiting angular resolution (see Sect.~\ref{apdx:snr_maps} for details and Fig.~\ref{fig:khmodes}), and their separation is smaller than that between the two filaments closer to the core, whereas a KH-driven expanding pattern is expected to show increasing amplitude and transverse separation with distance from the core. 
The transverse structure on the Stokes $I$ image produced using \texttt{eht-imaging} (see Sect.~\ref{apdx:ehtim} for details) is single-peaked. This indicates that low-sensitivity double-peaked region at $15$--$21$\,mas may represent imaging artifact of the over-resolved \texttt{CLEAN} image \citep{2023MNRAS.523.1247P}.
If the same feature were instead interpreted as local edge brightening, the implied jet width would decrease downstream, which is difficult to reconcile with the observed large-scale expansion of the M87 jet.
Alternatively, this double-peaked structure could be associated with a jet bend that can be produced by the precession at the launch \citep{2023Natur.621..711C}. However, this scenario requires more complicated modelling and higher-sensitivity VLBI observations.

Another ambiguity occurs beyond $\approx 27$\,mas from the VLBI core. In this region, there is no clear ridge intersection, but there is a noticeable change in the local position angle of the downstream emission. The outer filament can therefore be interpreted either as a continuation of one of the two ridge groups or as a separate downstream segment affected by the large-scale jet bending. We tested the helical fit both with and without the Gaussian components beyond $\approx 27$\,mas. These points behave as outliers with respect to the smooth helical pattern, and excluding them gives the most stable representation of the inner ridge geometry. Including them does not significantly change the fitted wavelength, amplitude, or phase within the uncertainties.

The grouping of the Gaussian components revealed zones of apparent thread intersections at $\approx 6$\,mas and $\approx 28$\,mas downstream of the jet origin. We interpret these intersections as a projection effect rather than as physical collisions between the threads. Because thread identification is ambiguous in these regions, the corresponding data were excluded from the analysis.

In addition, to account for the jet bend at $\sim 80$\,mas, we subtracted from the component positions the ridgeline predicted by a KH helical surface mode model (Savolainen et al., in prep.), 
thereby obtaining the offset of each component from the ridgeline. Although the helical mode cannot be constrained directly from our data, the observations of Savolainen et al. (in prep.) were obtained in the same year (2014 June 4), only four months apart from ours. 
The measured pattern speed is $v_{p} \sim 0.5c$ \citep{2016AA...595A..54M}, where $c$ is the speed of light. 
For M87 this apparent motion constitutes $\sim 1$\,mas\,yr$^{-1}$. Thus, the expected shift in the pattern position is negligible within the uncertainties. 
Nevertheless, this correction substantially reduces the influence of the large-scale jet bending on the fitted helices.

\begin{figure}
\centering
\includegraphics[angle=-90,width=0.8\columnwidth]{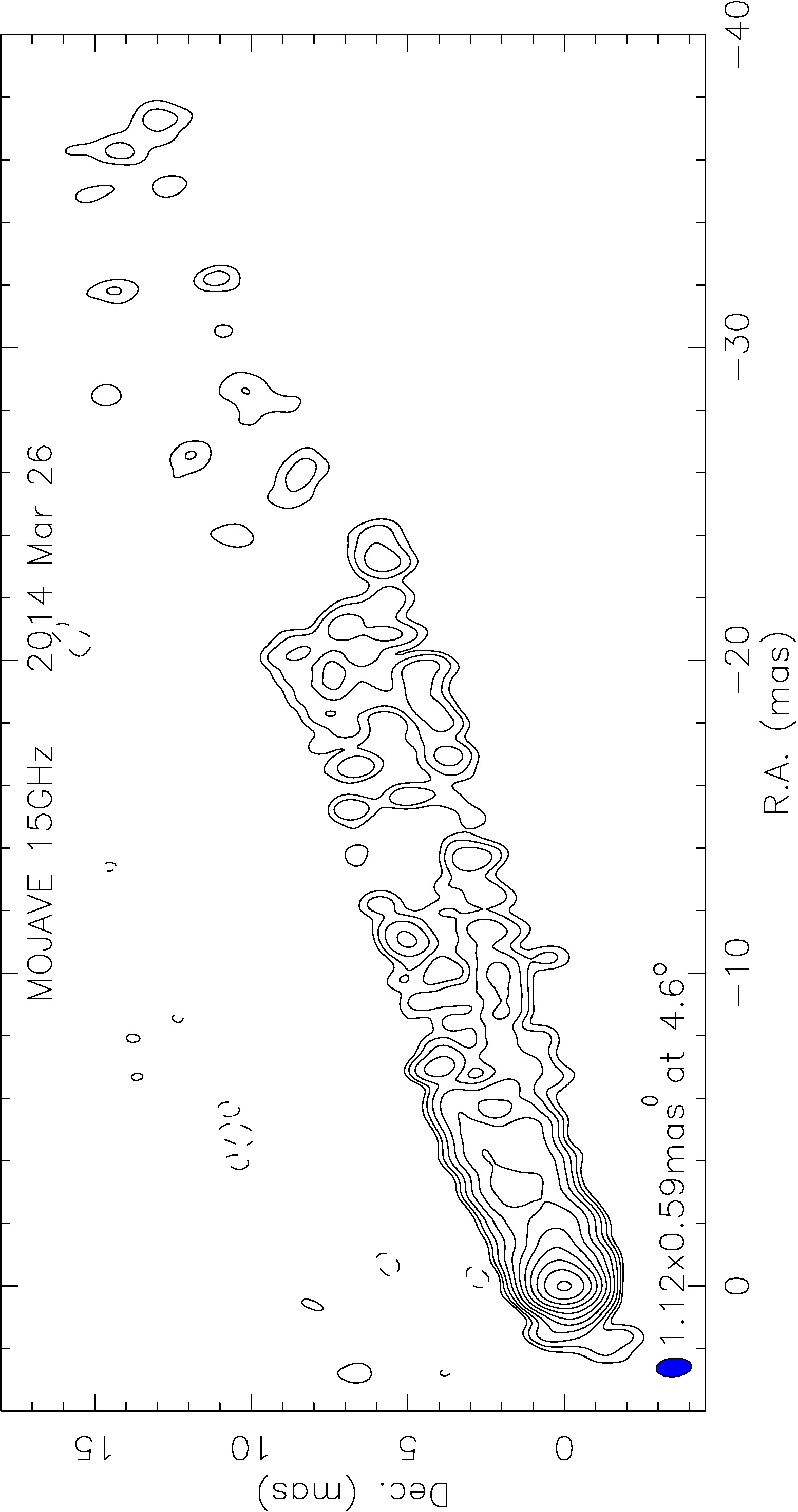}\\
\includegraphics[angle=-90,width=0.8\columnwidth]{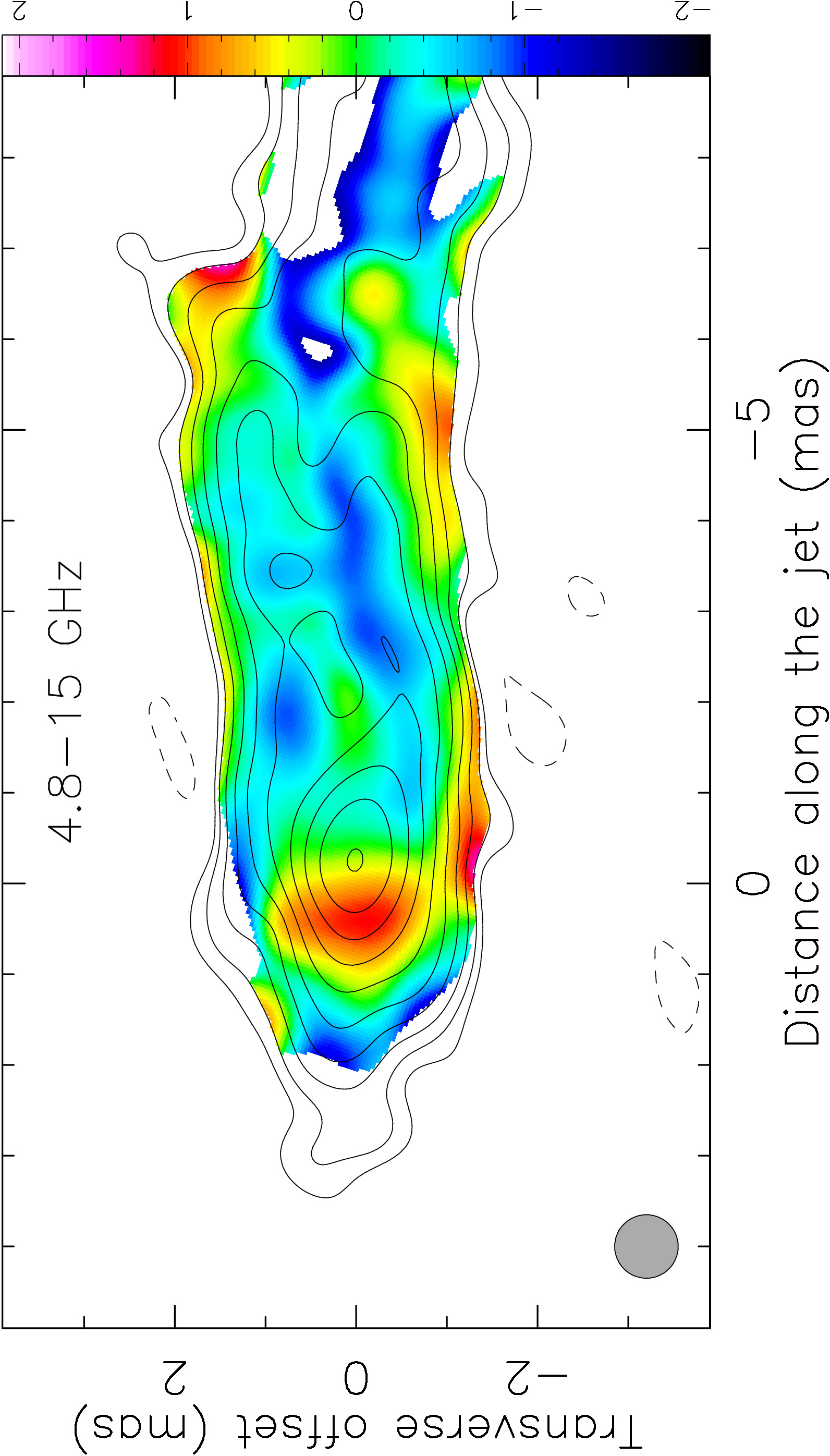}\\
\includegraphics[angle=-90,width=0.8\columnwidth]{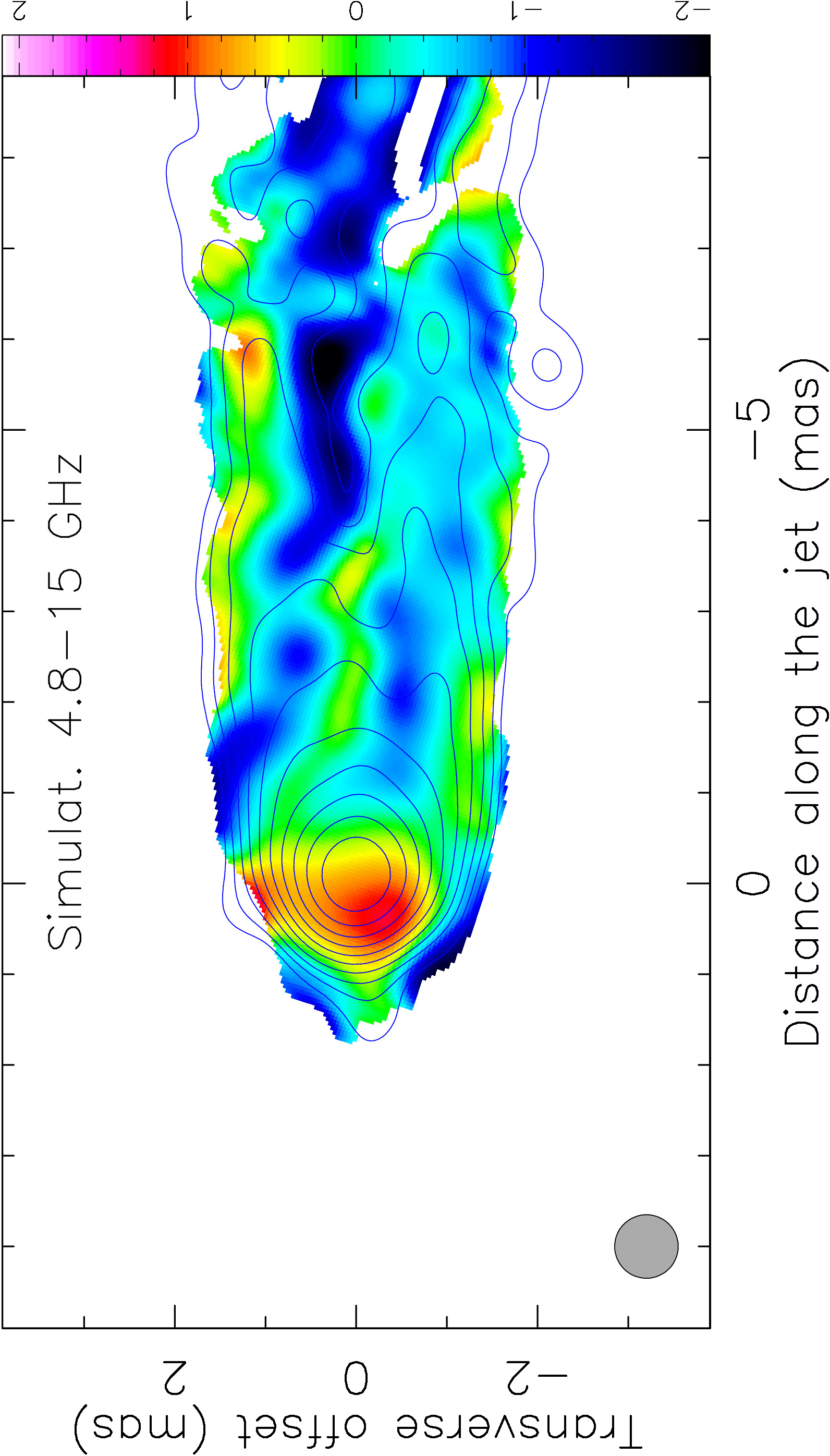}
\caption{(Top) The naturally weighted MOJAVE 15\,GHz image of M87 obtained on 2014 March 26. The lowest contour is drawn at 1.1\,mJy~beam$^{-1}$, $I$ peak 1026\,mJy~beam$^{-1}$. (Middle) 4.8 -- 15\,GHz spectral index map overlaid with 15\,GHz total intensity contours. The $(u,v)$ ranges are matched. A convolving circular beam size of 0.7\,mas is used, which has an average equivalent area to the synthesized 4.8 -- 15\,GHz beams. The image is rotated by $18^{\circ}$ clockwise. Bias compensation is applied, see section~\ref{sec:spi} for details.
(Bottom) Simulated spectral index image for 4.8 -- 15\,GHz $(u,v)$ coverages assuming the double-edge-brightened model (see Sec.~\ref{sec:spi} for details). The image is rotated by $18^{\circ}$ clockwise. Bias compensation is applied, see section~\ref{sec:spi} for details. The $I$ contours increase by factors of 2 in all maps.}
\label{fig:spih}
\end{figure}

Finally, we applied the helical model fit separately to the two groups of detected components, obtaining two intertwined helices. The final results are shown in Fig.~\ref{fig:khmodes} as transparent blue and red curves and are listed in Table~\ref{tab:KH_fit}. We emphasize that these curves show the continuous best-fitting helical model; they are not intended to imply that both threads are directly detected over the full displayed distance range. In particular, the northern part of the blue thread is not detected between $\approx 6$ and $\approx 28$\,mas, so the corresponding fit is constrained mainly by the upstream and downstream detections and may therefore be biased in amplitude and wavelength. However, the 15\,GHz MOJAVE image shown in Fig.~\ref{fig:spih} reveals extended emission with a similar north--south transverse asymmetry in this region, suggesting that fainter emission associated with the northern side of the structure may be present but remains below the sensitivity limit of the RadioAstron image. Nevertheless, the fitted parameters indicate similar wavelengths and a phase difference of $\sim 180^{\circ}$, providing evidence for an elliptical instability mode. These results (Table~\ref{tab:KH_fit}) are in good agreement with the elliptical body mode (E$_{b1}$) observed, measured, and identified by Savolainen et al. (in prep.) and \citet{2023MNRAS.526.5949N}. Our analysis therefore provides evidence for significant emission associated with both phases of the E$_{b1}$ pattern, although the two threads are not continuously detected over the full distance range shown in Fig.~\ref{fig:khmodes}. This extends the comparison with \citet{2023MNRAS.526.5949N}, where primarily the red thread in Fig.~\ref{fig:khmodes} was detected. We cannot exclude an alternative interpretation in which part of the downstream curvature is associated with a precession-driven bend of the jet spine \citep{2023Natur.621..711C}; however, modelling such a scenario would require an additional dynamical component that is not constrained by the present data.

\subsection{Spectra}
\label{sec:spi}
To analyse the spectral properties of the jet, we utilized observations conducted in close time proximity at the nearest frequencies.
These include \ra{} 1.7\,GHz and MOJAVE 15\,GHz data sets.
The 1.7\,GHz observations were obtained with the \ra{} and global VLBI array on 2014 June 4 (four months apart; the superuniformly weighted beam size is 1.2$\times$0.6~mas, PA=$29^{\circ}$; see Savolainen et al., in prep. for details). 
Observations at 15.4\,GHz were made within the MOJAVE monitoring program \citep{2018ApJS..234...12L} on 2014 March 26 (seven weeks apart), the corresponding total intensity image is shown in Fig.~\ref{fig:spih} (naturally weighted beam size is 1.1$\times$0.6~mas, PA=$5^{\circ}$;).

The estimated high brightness temperature values of the M87 VLBI core $T_{\rm b, 5\,GHz~core} \thicksim 10^{11}$~K (section \ref{sec:tb}) and $T_{\rm b,min, 22\,GHz} \thicksim 10^{12}$~K imply some activity in the nuclear region. 
Moreover, from the 24, 43, and 86\,GHz multi-epoch observations in 2014, \citet{2016ApJ...817..131H} obtained the mean speed of the M87 jet components of $\beta_{\rm app}=0.32c$.
Considering the structural variations, as well as the variable nature of the individual components in the extended jet, and active state of the nuclear region, it is difficult to obtain an accurate map of the spectral index distribution.
The non-simultaneous 1.7/4.8 and 4.8/15.4\,GHz images are still useful for examining the spectra.

\begin{figure}
\centering
\includegraphics[angle=-90,width=0.8\columnwidth]{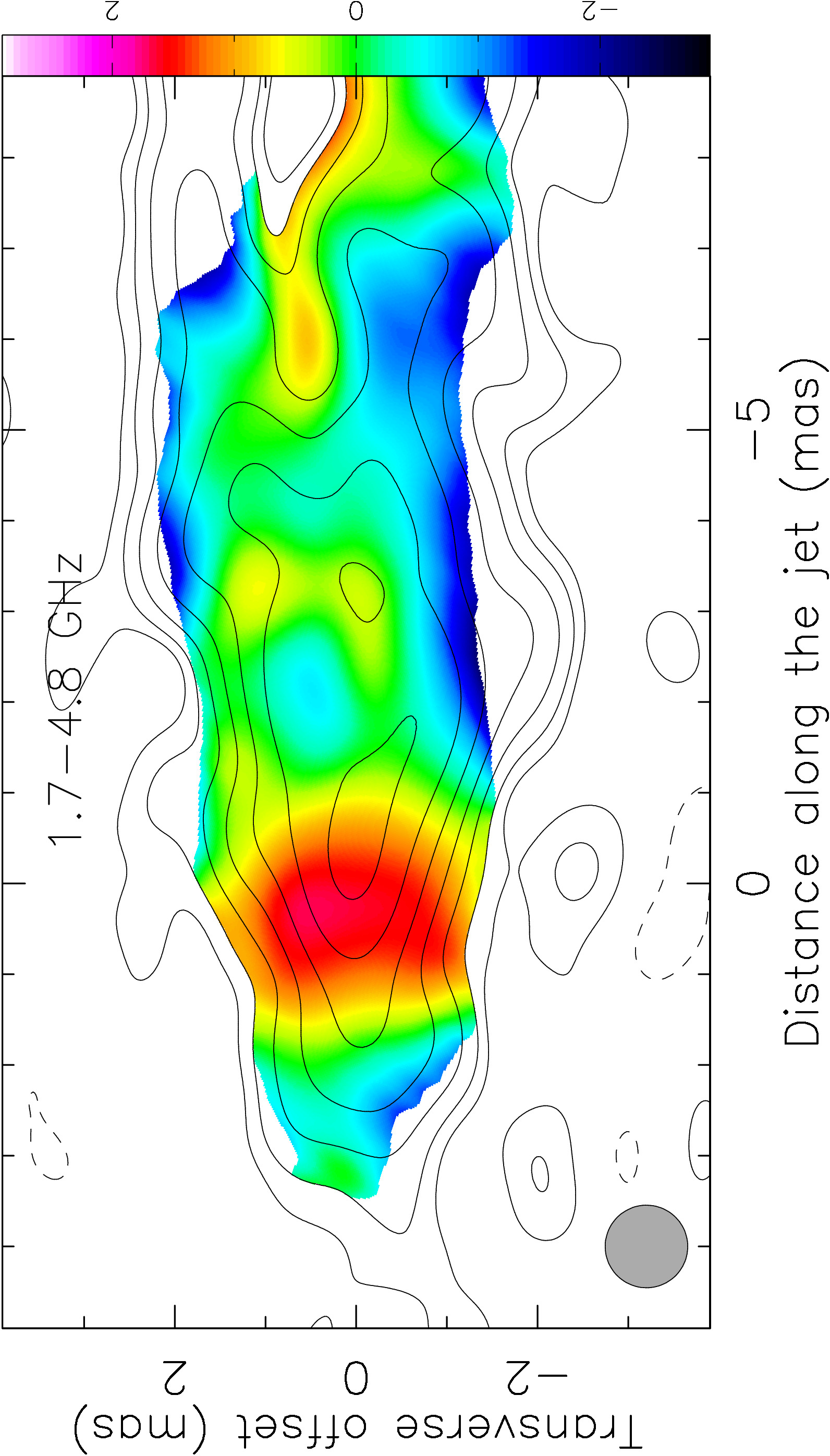}\\
\includegraphics[angle=-90,width=0.8\columnwidth]{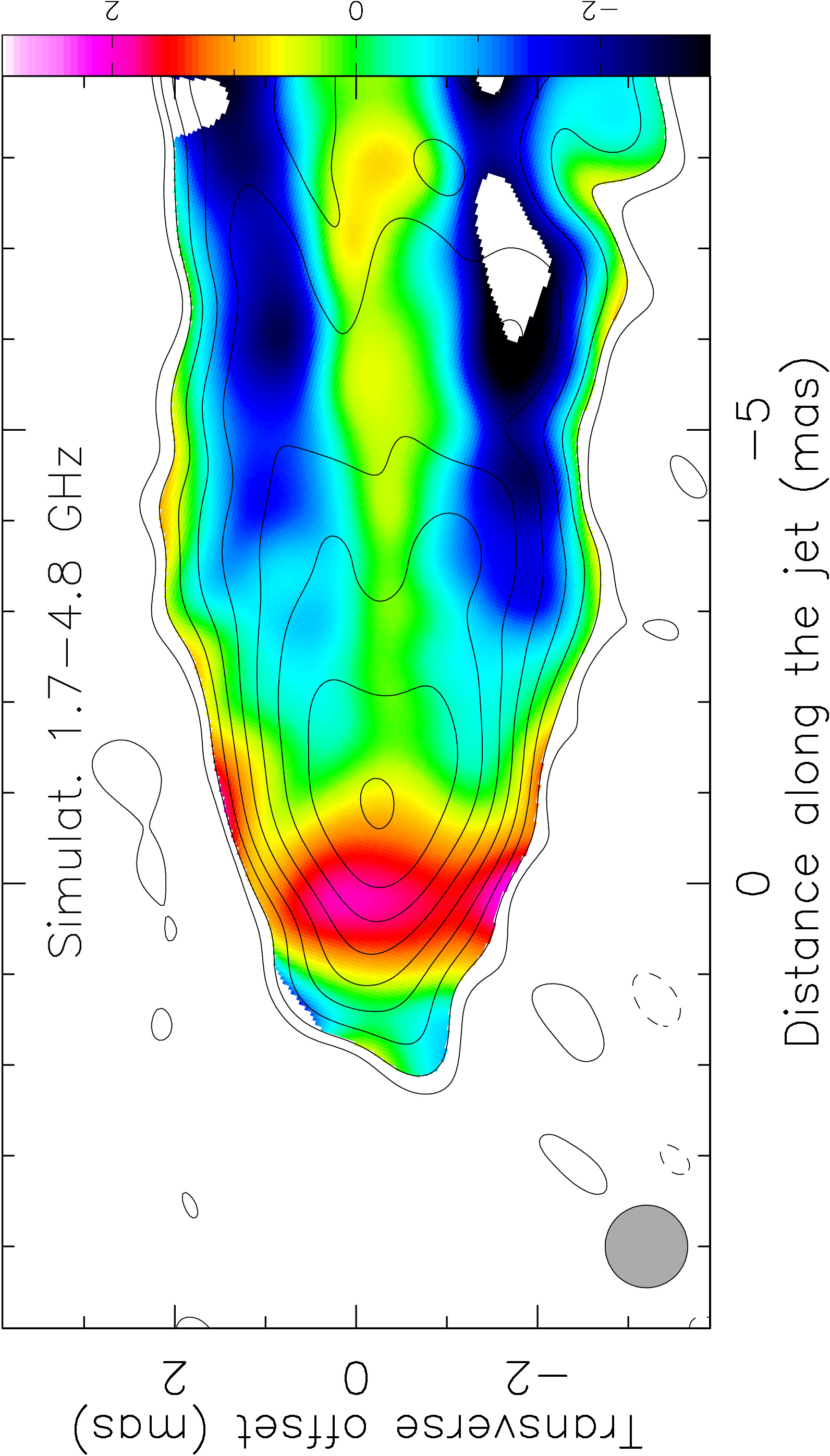}\\
\caption{(Top) 1.7 -- 4.8\,GHz spectral index map overlaid with 1.7\,GHz total intensity contours. The $(u,v)$ ranges are matched. A convolving circular beam size of 0.91\,mas is used. 
(Bottom) Simulated spectral index image for 1.7 -- 4.8\,GHz $(u,v)$-coverages assuming the double-edge-brightened model. The $I$ contours increase by factors of 2 in both maps.
Both images are rotated by $18^{\circ}$ clockwise, and include the bias compensation, see section~\ref{sec:spi} for details.}
\label{fig:spil}
\end{figure}

\begin{figure}
\centering
\includegraphics[width=\columnwidth]{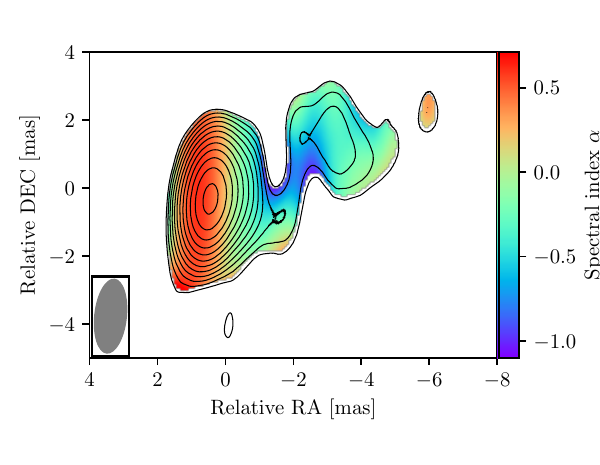}
\caption{Spectral index map between 4.8 and 22\,GHz overlaid by 22\,GHz \citepalias{2023ApJ...952...34K} intensity contours. 
The contours increase by factor of $\sqrt{2}$ starting from 30\,mJy\,beam$^{-1}$. 
A beam FWHM is shown at the lower left corner as an ellipse with size $2 \times 1$\,mas, ${\rm PA} = -6^{\circ}$.
The map was constructed using the bias compensation method, section~\ref{sec:spi}.}
\label{fig:C-K_sp_index}
\end{figure}

Since information about the absolute position is lost when phase self-calibration \citep{1989ASPC....6..185C,1999ASPC..180..187C} is applied during hybrid imaging, the alignment of the Stokes~$I$ images was performed using a cross-correlation technique that assumes optically thin features in the jet have frequency-independent positions \citep{walker_etal2000, 2008MNRAS.386..619C}. 
The use of this method is justified because the M87 jet has a complex double-edge structure and does not have distinct components at each frequency to anchor an alignment. For the pair of images at 4.8 and 15\,GHz, the cross-correlation was carried out over the jet region of 5~mas from the core.
The cross-check was made using core shift estimates by \citet{2011Natur.477..185H} and the model-fitted Gaussian component to the core.
For the pair of images at 4.8 and 22\,GHz, it was hard to align the maps using the cross-correlation technique because there is not much structure present at 22\,GHz. 
The problem was circumvented by implementing a manual shift based on the core shift measurements in M87 by \cite{2011Natur.477..185H}. The total shift between images was calculated as the sum of a core shift between corresponding frequencies and the distance from the phase centre and the core component. 
The applied shifts along R.A. and Dec. for each frequency pair are the following: $-$0.28, 0.03~mas for 1.7--4.8\,GHz, $-$0.55, 0.4~mas for 4.8--15\,GHz, and $-$0.6, 0.4~mas at 4.8 and 22\,GHz.
The 4.8--22\,GHz spectral index map is presented in Fig.~\ref{fig:C-K_sp_index}.

For the spectral analysis, each pair of maps was convolved with the common beam size of the corresponding low-frequency image. We used the pixel size, which equals 1/20 of the beam size of the corresponding high-frequency image.
We also matched the $(u,v)$ ranges for each pair of low and high-frequency images used in the spectral analysis.
The resultant spectral index images for 4.8--15\,GHz, 1.7--4.8\,GHz, and 4.8 and 22\,GHz are shown in Figs.~\ref{fig:spih}, \ref{fig:spil}, and \ref{fig:C-K_sp_index}, respectively.
The images display an inverted spectrum in the upstream region, consistent with a self-absorbed core. 
There is a faint structure seen on the opposite side of the approaching jet. A detailed simulation is needed to address the question of whether this upstream optically thin emission might represent the substructure of the core. We follow the generally accepted view that this feature is the counter-jet.
Within 2\,mas of the core, the jet emission flattens near the jet axis. This is consistent with the spectral behaviour seen in the VSOP 1.6$-$4.8\,GHz \citep{2016ApJ...833...56A} and in the VLBA 8$-$15\,GHz observations \citep{2023MNRAS.526.5949N}.

\section{Discussion}

The earlier high-resolution 5\,GHz images of the M87 jet have been presented by \citet{2016ApJ...833...56A} using the VSOP observations. 
The resolution obtained in the VSOP imaging was 0.76$\times$0.61\,mas at $-$68\degr\ (2000 Mar 20) and 0.85$\times$0.46\,mas at 57\degr\ (2000 Mar 23). 
In terms of an equivalent-area circular beam, our new observations (performed 14 years later) have similar resolution.

Our estimates of the jet position angle within the $\thicksim5$~mas off the core of about $-69^{\circ}$ is close to the jet position angle of $-67^{\circ}$ observed with the VSOP at 5\,GHz 14 years earlier \citep{2016ApJ...833...56A}. Both these values are consistent with the variations of jet position angle tracked over more than 20 years \citep{2018ApJ...855..128W, 2023Natur.621..711C}.

Our estimate of the brightness temperature in the core $T_{\rm b, core} = (1.15\pm0.64)\times 10^{11}$~K is higher than $T_{\rm b, 22\,GHz~core} = (4.4 \pm 0.8) \times 10^{10}$~K measured in the simultaneous 22\,GHz \ra{} observations \citepalias{2023ApJ...952...34K} from the Gaussian model-fit. 
At 22\,GHz, \citetalias{2023ApJ...952...34K} estimated the minimum brightness temperature of the source at the ground-to-space baseline of $T_{\rm b,min} \thicksim 10^{12}$~K, which nearly approaches the inverse-Compton catastrophe limit \citep{1969ApJ...155L..71K}. This is one of the highest values for M87 reported in the literature, and likely linked to the compact substructure in the VLBI core (see discussions in \citealt{Nok17A}; \citetalias{2023ApJ...952...34K}).

\subsection{Jet structure}

The transverse jet structure in the inner 5~mas from the 4.8\,GHz core appears to be edge-brightened, with the northern edge being brighter closer to the core. Further, the jet becomes southern-edge-brightened. 
Such an asymmetric structure can be explained by the clockwise rotation of the jet \citep{2016AA...595A..54M} that boosts emission from the southern limb and de-boosts it at the northern one. 

The jet edge brightening in the first 5~mas from the core points to a stratification of the emitting plasma, with a higher concentration preferably located at the jet edges. This may be due to either the non-thermal electrons accelerating near the jet boundaries or their distribution according to `Ohmic heating' \citep{2005MNRAS.360..869L}.
The latter assumes that the number density of relativistic particles is proportional to the square of the co-moving electric current density, which results in non-uniform distribution of emitting plasma across the jet.

Another possible factor contributing to the observed asymmetry in total intensity is the presence of a helical magnetic field. As \citet{2011MNRAS.415.2081C} show, even for a negligible toroidal velocity, one side of a jet is brighter due to dependence of emission coefficient on an angle between the line of sight and B-field direction. This effect is pronounced if the proper frame toroidal and poloidal magnetic field components are comparable. This is the case in the acceleration and collimation zone \citep{Komissarov_2009, Nokhrina_Pushkarev_2024, N24b}, which is expected to extend in the M87 jet to more than 100\,mas deprojected \citep{2018ApJ...868..146N, Nokhrina+2019MNRAS}. Magnetic field-induced asymmetry works in the same way as the rotation-induced asymmetry. However, as toroidal velocity decreases at large distances, the effects of a helical B-field structure may be dominant.
Due to the limited sensitivity of the array, the northern limb is not detected uninterruptedly. Nevertheless, the curvature in the southern limb is observed to $\approx$ 28~mas down the jet stream.
Edge-brightening itself may be explained by the fast spine --- slow sheath jet structure that de-boosts the emission from the central part of a jet. However, this requires the emissivity and, correspondingly, the physical parameters that define it (emitting plasma number density and a magnetic field) to decrease not too fast towards the jet edge.
Modelling the synchrotron emission of stratified jets by \cite{2023MNRAS.523..887F} demonstrates that the transverse double- and triple-peaked brightness profiles appear when the emitting plasma is distributed mainly at the jet edges. This may be either due to the preferential acceleration of non-thermal electrons near the jet boundary or when the non-thermal electron distribution is proportional to the square of an electric current density in a plasma proper frame. The last scenario is in qualitative agreement with the polarization data \citep{2005MNRAS.360..869L}. 

The position angle of the jet within the $\thicksim5$~mas off the core is about $-69^{\circ}$. 
This value is close to the jet position angle of $-67^{\circ}$ observed with the VSOP at 5\,GHz 14 years earlier \citep{2016ApJ...833...56A}. Both these values are consistent with the variations of the jet position angle tracked over more than 20 years \citep{2018ApJ...855..128W, 2023Natur.621..711C}.

The detected orderly patterns must be classified by their physical nature and by their possible driver (close to the jet base). We discuss below the following possibilities. 

(i) Our analysis of the  bright structures supports the idea that the Kelvin-Helmholtz instability can be behind the formation of the observed helical structure of the jet in M87  \citep{2003NewAR..47..629L, 2011ApJ...735...61H, 2021ApJ...923L...5P}, similarly to the parsec-scale jets of the quasars 3C\,273 \citep{2001Sci...294..128L}, 3C\,279 \citep{2023NatAs...7.1359F}, and 0836+710 \citep{2020A&A...641A..40V}. 
Indeed, the modelled KH-instability wavelength $\lambda_{\mathrm{KH}}$ dependence on the local jet radius $R_{\textrm{jet}}$ is in agreement with the expected instability behaviour \citep[see e.g.][]{2011ApJ...735...61H}. At the same time, the brightness pattern of such instability should reflect the discussed above magnetic field and toroidal velocity components direction that affect the observed transversal asymmetry. 

(ii) The wiggling patterns may be observed due to differential `heating' of plasma moving along the helical field lines \citep{1992A&A...255...59C,2007RMxAC..27...91K,2008Natur.452..966M}. In this case both peculiarities -- the transversal differential brightness and growing with the distance wavelength --- are also naturally explained. Indeed, the toroidal plasma magnetohydrodynamic velocity consists of two components: movement along the magnetic field lines (mainly toroidal magnetic field far from the light cylinder radius $R_\mathrm{L}=c/\Omega_\mathrm{F}$ and for a constant magnetic field angular velocity $\Omega_\mathrm{F}$) and sliding backwards due to the magnetic surface rotation. At the same time, 
as the jet propagates, the plasma bulk motion Lorentz factor, defined mainly by the poloidal motion, grows \citep[see e.g.][]{Beskin06, Komissarov_2009, 2018ApJ...868..146N}. These two effects leads to growing along the jet length between the neighbour steps of a line, following the magnetic flux tube with presumably more emitting plasma (Nokhrina et al., in prep.). 

(iii) Ballistic motion of a pattern started by the jet base precession \citep{2023Natur.621..711C}. In this case it is hard to explain the differential brightness of ejections across the jet, that is usually attributed to the presence of toroidal components of both velocity and magnetic field. Besides, the growing wavelength for such ballistic motion suggests the gradual acceleration. 

(iv) Developing kink instability \citep{Mizuno12} would lead to the jet wiggling in contrast with the spiralling pattern in the jet, and one does not expect the transversal differential brightness between the northern and southern jet limbs.

Thus, we conclude that Kelvin--Helmholtz and loaded by the emitting plasma magnetic tube are the most promising scenarios explaining the observed brightness distribution patterns in M87. The observed phase difference $\sim 180^{\circ}$ of the helical threads, which is expected in the presence of the first order elliptical body mode of K--H instability \citep{2011ApJ...735...61H}, favours the first scenario.

The plasma instabilities can be generated by the SMBH activity, e.g. due to Lense--Thirring (LT) precession \citep{2023Natur.621..711C}, or perturbed mass accretion processes occurring in the magnetically arrested disc \citep{2023Galax..11...33R}.
This is supported by the recent high-cadence monitoring observations of KVN$+$VERA array at 22\,GHz, which show that the parsec-scale jet of M87 exhibits short-term (weekly and monthly) structural variations with an average period of $0.94\pm0.12$ years \citep{2023Galax..11...33R}.

Let us investigate the possible origin of K--H instability. Its wavelength $\lambda_\mathrm{KH}\propto R_\mathrm{jet}$ \citep{2000ApJ...533..176H}, and one can extrapolate the wavelength to the scale where the jet and the location of instability footprint has a radius ${R_\mathrm{base}}$:  $\lambda_\mathrm{KH,\,base}=\lambda_\mathrm{KH,0}R_\mathrm{base}$. Here we introduce the dimensionless value $\lambda_\mathrm{KH,0}$ equal to the K--H wavelength (in mas) at $1$~mas jet radius, its value is equal to $17.3$ for the red fit or $14.7$ for the blue fit from Table~\ref{tab:KH_fit}. The corresponding period $T_\mathrm{KH,\,base}=\lambda_\mathrm{KH,\,base}/\left(\beta_\mathrm{p}c\right)$ depends on a characteristic pattern speed $\beta_\mathrm{p}c$.
If the origin of this instability mode is connected to some process in an accretion disc, the corresponding period of Keplerian motion is equal to $T_\mathrm{K}=2\pi R_\mathrm{base}^{3/2}/\left(\sqrt{r_\mathrm{g}}c\right)$, where the gravitational radius $r_\mathrm{g}=R_\mathrm{s}/2$. Equating the K--H period at the base with the Keplerian period, we obtain the characteristic radius in a disc 
\begin{equation}
R_\mathrm{base}=\left(\frac{\lambda_\mathrm{KH,0}}{2\pi\beta_\mathrm{p}}\right)^2 r_\mathrm{g}.
\end{equation}
The corresponding Keplerian period 
\begin{equation}
T_\mathrm{K}=\frac{r_\mathrm{g}}{4\pi^2 c}\left(\frac{\lambda_\mathrm{KH,0}}{\beta_\mathrm{p}}\right)^3.
\end{equation}
Substituting in this equation $r_\mathrm{g}=9.6\cdot 10^{14}$~cm, dimensionless $\lambda_\mathrm{KH,1}$, and assuming constant pattern speed $\beta_{\mathrm{p}} \sim 0.5$ \citep{2011ApJ...735...61H, 2016AA...595A..54M}, we obtain 
$T_\mathrm{K}\approx 1.1$~yr for the red filament and  $0.65$~yr for the blue one are in excellent agreement with the results by \citet{2023Galax..11...33R, 2026ApJ...999..169R}.
If the angular velocity, defined by this period, is close to the magnetohydrodynamical $c/R_\mathrm{L}$, the expected light cylinder radius $R_\mathrm{L}\approx 150\;r_\mathrm{g}$ is somewhat larger than estimated by \citet{Nokhrina+2019MNRAS} and \citet{2022ApJ...939...83K}. On the other hand, the obtained values for $T_\mathrm{K}$ allow estimating the instability footprint radius as $\displaystyle R_\mathrm{base}=r_\mathrm{g}\left(\frac{cT_\mathrm{K}}{2\pi r_\mathrm{g}}\right)^{2/3}$. The values $R_\mathrm{base}=31.0\;r_\mathrm{g}$ and $R_\mathrm{base}=21.8\;r_\mathrm{g}$, corresponding to $T_\mathrm{K}=1.1$ and $0.65$~yr, agree with $30-80$\,per~cent accuracy with the millimetre photon ring radius $R_\mathrm{ph}=16.8\;r_\mathrm{g}$, measured by \citet{2023Natur.616..686L}. 

Thus, we propose that the first-order elliptical body mode may originate in an accretion disc.
Recent analysis of the EHT observations \citep{2026A&A...706A..27S} shows that the emission of the jet is registered already at distance of 0.325~mas from the ring, which may indicate the disc--jet connection in M87, reconcilable with the scenario of disc origin of KH modes.

On the other hand, one can connect the perturbation not to the disc, but to the precessing jet, as observed by \citep{2023Natur.621..711C} up to scales of the order of 3~mas projected distance along the jet. This length scale corresponds to a local jet radius $R_\mathrm{jet}\sim 0.38$~mas \citep{Nokhrina+2019MNRAS}. Using the estimated $\lambda_\mathrm{KH}$, we assess the mode propagation speed as $0.15$ ($0.13$)~$c$.
Polarized radiative transfer simulations by \cite{2023ARep...67.1275T} demonstrated that a model of a precessing jet with a helical magnetic field reproduces the observed polarization profiles of parsec-scale AGN jets.

\subsection{Spectra}
\cite{2023MNRAS.523.1247P} conducted a series of simulations with various jet brightness models and dual frequency VLBI data sets in a frequency range 1.6$-$43\,GHz and found that the spectral index images of the M87 parsec-scale jet are heavily affected by the systematic effects. For the VSOP 1.6$-$4.8\,GHz data set \citep{2006PASJ...58..243D,2016ApJ...833...56A} the systematics appear as spectral flattening at the spine that is clearly visible up to 5~mas, and at the jet edges.
To check the reliability of the obtained spectral index images, we produced 1.7, 4.8 and 15.4\,GHz artificial data sets using the edge brightened jet model from \cite{2023MNRAS.523.1247P} with intrinsic optically thin spectral index $\alpha = -0.5$. Then, we imaged these data sets in the same way as the original data and obtained the artificial spectral index images (bottom plots in Fig.~\ref{fig:spih} and Fig.~\ref{fig:spil}). The generated maps display the same patterns as the observed spectral index distribution: flattening of the spectra at the spine within 3\,mas from the core and at the jet edges. 
We conclude that these features in the observed spectral index image are the result of systematic errors in the imaging process. The spectral steepening in low surface brightness extended jet regions is also evident in the real and simulated images. The same is also visible in spectral index maps from other studies \citep{2007ApJ...660..200L, 2014AJ....147..143H, 2019Galax...7...86Z, 2020A&A...637L...6K}. 
This is a common systematic effect of the \texttt{CLEAN} procedure that arises in the VLBI spectral index maps, which is studied in detail in \citet{2023MNRAS.523.1247P}. 

\section{Conclusions}

Using the \ra{} 4.8\,GHz observations, we constructed one of the highest angular resolution images of M87 jet at this frequency to date. 
The orientation of the jet within the 5\,mas down the stream of the core shifts northward, consistent with the long-term trend that might be associated with the Lense-Thirring precession of the jet \citep{2023Natur.621..711C}. 
Modelling the transverse intensity profiles, we provide the first evidence for the two elliptical body modes of the Kelvin-Helmholtz instability.
The results are in good agreement with the recent structural analysis of the high-resolution images obtained from the 1.7\,GHz \ra{} (Savolainen et al., in prep.) and 8\,GHz VLBA \citep{2023MNRAS.526.5949N} observations, where helical surface and two elliptical surface modes were registered.
The estimated instability parameters and periodic jet swings support the idea that the Kelvin-Helmholtz instability, which may be generated by perturbations in an accretion disc, can be behind the observed triple-ridge and helical structure of the M87 jet.
Extrapolation of the $E_{b1}$-mode wavelength on~to the scales of a few gravitational radii provides the corresponding period to be on the order of Keplerian velocities in the inner disc. Thus, we propose that the first-order elliptical body mode origin may be connected with some processes in an accretion disc. 
The long-term, high-resolution, high-sensitivity VLBI observations enabled us to investigate the disc-jet connection in M87. 
Hopefully, the accumulated data series will allow us making the same study toward other AGN jets in near future (e.g. Principe et al., subm.).

Precession is a common phenomenon in AGN jets. It's apparent manifestations at different spatial scales in the same source are known (e.g. the case of  M\,81$^*$, \citealt{von_Fellenberg+2023AA}). Such bi- or multi-mode spatial variations corresponding to different temporal modes of the triggering mechanisms, including those controlled by the Lense-Thirring effect, might have several explanations. These include disc tearing or broken discs \citep{Musoke+2023MNRAS}, a combination of LT-driven in an inner disc and tidal torquing in an outer disc \citep{Steinle+2024PhRvD}, and the LT effect triggered by a tidal disruption of a star in the accretion disc \citep{Pasham+2024Nature}. Essentially, the question is whether precession (or `wobbling') develops in the jet due to some instability in the plasma flow, e.g. the KH instability, or the jet is `born' to appear precessing. 
Further conclusive disentanglement of various causes of jet precession and wobbling will require a next-generation ground or space VLBI facility operating at mm/sub-mm wavelengths and providing imaging with sufficiently high dynamic range.

The jet edge brightening in the first 5~mas from the core points to a stratification of the emitting plasma, with a higher concentration preferably located at the jet edges. 
The constructed (1.7 -- 4.8) and (4.8 -- 15.4)\,GHz spectral index images are consistent with the recently presented high-resolution spectral index distribution, with the flattening of the spectra along the jet spine and its steepening toward the jet edges \citep{2023MNRAS.526.5949N}.
Simulations of the VLBI dual frequency spectral index images, assuming various jet brightness models, show that the observed steepening of the spectra is strongly affected by the imaging systematics. 
Meanwhile, its partial steepening can be produced by the propagation of instability modes causing higher pressure regions near the jet boundary and in its interior.
The jet emission is consistent with the underlying constant spectral index along the flow.

\section*{Acknowledgements}

The authors thank Guang-Yao Zhao for his help in producing Fig.~\ref{fig:radpl} as well as the anonymous referee for valuable comments that substantially improved this paper.
ASN, YYK, MML, APL were supported by the M2FINDERS project funded by the European Research Council (ERC) under the European Union's Horizon 2020 Research and Innovation Programme (Grant Agreement No. 101018682). MML acknowledges partial support from the ANID FONDECYT Iniciaci{\'o}n grant no. 11251078. ASN received financial support for this research from the International Max Planck Research School (IMPRS) for Astronomy and Astrophysics at the Universities of Bonn and Cologne.  
J.Y.K. is supported for this research by the National Research Foundation of Korea (NRF) grant funded by the Korean government (Ministry of Science and ICT; grant no. 2022R1C1C1005255, RS-2022-NR071771) and by the Korea Astronomy and Space Science Institute under the R\&D program (Project No. 2025-9-844-00) supervised by the Korea AeroSpace Administration.
The \ra{} project is led by the Astro Space Center of the Lebedev Physical Institute of the Russian Academy of Sciences and the Lavochkin Scientific and Production Association under a contract with the State Space Corporation ROSCOSMOS, in collaboration with partner organizations in Russia and other countries.
This research is based on observations correlated at the Bonn Correlator, jointly operated by the Max Planck Institute for Radio Astronomy (MPIfR) and the Federal Agency for Cartography and Geodesy (BKG).
Partly based on observations with the 100-m telescope of the MPIfR at Effelsberg.
The European VLBI Network is a joint facility of independent European, African, Asian, and North American radio astronomy institutes.
Scientific results from the data presented in this publication are derived from the following EVN project code(s): gs032c.
The National Radio Astronomy Observatory is a facility of the National Science Foundation operated under cooperative agreement by Associated Universities, Inc.
The VLBA is a facility of the National Science Foundation operated under cooperative agreement by Associated Universities, Inc.
The Australia Telescope Compact Array (Parkes radio telescope / Mopra radio telescope / Long Baseline Array) is part of the Australia Telescope National Facility which is funded by the Australian Government for operation as a National Facility managed by CSIRO.
This research has made use of data from the MOJAVE database, which is maintained by the MOJAVE team \citep{2018ApJS..234...12L}.

\section*{Data Availability}

The calibrated uv fits file and RadioAstron images are available online at \url{https://odin.asc.rssi.ru/~ekravchenko/ram87_5ghz_feb2014/}.



\bibliographystyle{mnras_vanHack} 
\bibliography{RA_M87_Cband}

@string{june = {June}}

@article{1918PLicO..13....9C,
 adsurl = {http://adsabs.harvard.edu/abs/1918PLicO..13....9C},
 author = {{Curtis}, H.~D.},
 journal = {Publications of Lick Observatory},
 pages = {9-42},
 title = {{Descriptions of 762 Nebulae and Clusters Photographed with the Crossley Reflector}},
 volume = {13},
 year = {1918}
}

@article{1965R&QE....8..461M,
 adsurl = {https://ui.adsabs.harvard.edu/abs/1965SvRP....8..461M},
 author = {{Matveenko}, L.~I. and {Kardashev}, N.~S. and {Sholomitskii}, G.~B.},
 doi = {10.1007/BF01038318},
 journal = {Soviet Radiophysics},
 month = {July},
 number = {4},
 pages = {461-463},
 title = {{Large base-line radio interferometers}},
 volume = {8},
 year = {1965}
}

@article{1969ApJ...155L..71K,
 adsurl = {http://adsabs.harvard.edu/abs/1969ApJ...155L..71K},
 author = {{Kellermann}, K.~I. and {Pauliny-Toth}, I.~I.~K.},
 doi = {10.1086/180305},
 journal = {\apjl},
 month = {February},
 pages = {L71},
 title = {{The Spectra of Opaque Radio Sources}},
 volume = {155},
 year = {1969}
}

@article{1980A&A....89..377C,
 adsurl = {https://ui.adsabs.harvard.edu/abs/1980A&A....89..377C},
 author = {{Clark}, B.~G.},
 journal = {\aap},
 month = {September},
 number = {3},
 pages = {377},
 title = {{An efficient implementation of the algorithm 'CLEAN'}},
 volume = {89},
 year = {1980}
}

@inproceedings{1989ASPC....6..185C,
 adsurl = {https://ui.adsabs.harvard.edu/abs/1989ASPC....6..185C},
 author = {{Cornwell}, Tim and {Fomalont}, Edward B.},
 booktitle = {Synthesis Imaging in Radio Astronomy},
 editor = {{Perley}, Richard A. and {Schwab}, Frederic R. and {Bridle}, Alan H.},
 month = {January},
 pages = {185},
 series = {Astronomical Society of the Pacific Conference Series},
 title = {{Self-Calibration}},
 volume = {6},
 year = {1989}
}

@article{1992A&A...255...59C,
 adsurl = {https://ui.adsabs.harvard.edu/abs/1992A&A...255...59C},
 author = {{Camenzind}, M. and {Krockenberger}, M.},
 journal = {\aap},
 month = {February},
 pages = {59-62},
 title = {{The lighthouse effect of relativistic jets in blazars. A geometric originof intraday variability.}},
 volume = {255},
 year = {1992}
}

@inproceedings{1997ASPC..125...77S,
 adsurl = {http://adsabs.harvard.edu/abs/1997ASPC..125...77S},
 author = {{Shepherd}, M.~C.},
 booktitle = {Astronomical Data Analysis Software and Systems VI},
 editor = {{Hunt}, G. and {Payne}, H.},
 pages = {77},
 series = {Astronomical Society of the Pacific Conference Series},
 title = {{Difmap: an Interactive Program for Synthesis Imaging}},
 volume = {125},
 year = {1997}
}

@inproceedings{1999ASPC..180..187C,
 adsurl = {https://ui.adsabs.harvard.edu/abs/1999ASPC..180..187C},
 author = {{Cornwell}, Tim and {Fomalont}, Ed B.},
 booktitle = {Synthesis Imaging in Radio Astronomy II},
 editor = {{Taylor}, G.~B. and {Carilli}, C.~L. and {Perley}, R.~A.},
 month = {January},
 pages = {187},
 series = {Astronomical Society of the Pacific Conference Series},
 title = {{Self-Calibration}},
 volume = {180},
 year = {1999}
}

@article{2000ApJ...533..176H,
 adsurl = {https://ui.adsabs.harvard.edu/abs/2000ApJ...533..176H},
 author = {{Hardee}, Philip E.},
 doi = {10.1086/308656},
 journal = {\apj},
 month = {April},
 number = {1},
 pages = {176-193},
 title = {{On Three-dimensional Structures in Relativistic Hydrodynamic Jets}},
 volume = {533},
 year = {2000}
}

@article{2000PASJ...52..955H,
 adsurl = {https://ui.adsabs.harvard.edu/abs/2000PASJ...52..955H},
 author = {{Hirabayashi}, Hisashi and {Hirosawa}, Haruto and {Kobayashi}, Hideyuki and {Murata}, Yasuhiro and {Asaki}, Yoshiharu and {Avruch}, Ian M. and {Edwards}, Philip G. and {Fomalont}, Edward B. and {Ichikawa}, Tsutomu and {Kii}, Tsuneo and {Okayasu}, Rikako and {Wajima}, Kiyoaki and {Inoue}, Makoto and {Kawaguchi}, Noriyuki and {Chikada}, Yoshihiro and {Bushimata}, Takeshi and {Fujisawa}, Kenta and {Horiuchi}, Shinji and {Kameno}, Seiji and {Miyaji}, Takeshi and {Shibata}, Kazunori M. and {Shen}, Zhi-Qiang and {Umemoto}, Tomofumi and {Kasuga}, Takashi and {Nakajima}, Jun'ichi and {Takahashi}, Yukio and {Enome}, Shinzou and {Morimoto}, Masaki and {Ellis}, Jordan and {Meier}, David L. and {Murphy}, David W. and {Preston}, Robert A. and {Smith}, Joel G. and {Wietfeldt}, Rick D. and {Benson}, John M. and {Claussen}, Mark J. and {Flatters}, Chris and {Moellenbrock}, George A. and {Romney}, Jonathan D. and {Ulvestad}, James S. and {Langston}, Glen I. and {Minter}, Anthony H. and {D'Addario}, Larry R. and {Dewdney}, Peter E. and {Dougherty}, Sean M. and {Jauncey}, David L. and {Lovell}, James E.~J. and {Tingay}, Steven J. and {Tzioumis}, Anastasios K. and {Taylor}, A. Russell and {Cannon}, Wayne H. and {Gurvits}, Leonid I. and {Schilizzi}, Richard T. and {Booth}, Roy S. and {Popov}, Misha V.},
 doi = {10.1093/pasj/52.6.955},
 journal = {\pasj},
 month = {December},
 pages = {955-L965},
 title = {{The VLBI Space Observatory Programme and the Radio-Astronomical Satellite HALCA}},
 volume = {52},
 year = {2000}
}

@article{2001Sci...294..128L,
 adsurl = {https://ui.adsabs.harvard.edu/abs/2001Sci...294..128L},
 author = {{Lobanov}, A.~P. and {Zensus}, J.~A.},
 doi = {10.1126/science.1063239},
 journal = {Science},
 month = {October},
 number = {5540},
 pages = {128-131},
 title = {{A Cosmic Double Helix in the Archetypical Quasar 3C273}},
 volume = {294},
 year = {2001}
}

@article{2003NewAR..47..629L,
 adsurl = {https://ui.adsabs.harvard.edu/abs/2003NewAR..47..629L},
 author = {{Lobanov}, Andrei and {Hardee}, Philip and {Eilek}, Jean},
 doi = {10.1016/S1387-6473(03)00109-X},
 journal = {\nar},
 month = {October},
 number = {6-7},
 pages = {629-632},
 title = {{Internal structure and dynamics of the kiloparsec-scale jet in M87}},
 volume = {47},
 year = {2003}
}

@article{2005MNRAS.360..869L,
 adsurl = {http://adsabs.harvard.edu/abs/2005MNRAS.360..869L},
 author = {{Lyutikov}, M. and {Pariev}, V.~I. and {Gabuzda}, D.~C.},
 doi = {10.1111/j.1365-2966.2005.08954.x},
 eprint = {astro-ph/0406144},
 journal = {\mnras},
 month = {July},
 pages = {869-891},
 title = {{Polarization and structure of relativistic parsec-scale AGN jets}},
 volume = {360},
 year = {2005}
}

@article{2006PASJ...58..243D,
 adsurl = {https://ui.adsabs.harvard.edu/abs/2006PASJ...58..243D},
 archiveprefix = {arXiv},
 author = {{Dodson}, Richard and {Edwards}, Philip G. and {Hirabayashi}, Hisashi},
 doi = {10.1093/pasj/58.2.243},
 eprint = {astro-ph/0511383},
 journal = {\pasj},
 month = {April},
 pages = {243-251},
 primaryclass = {astro-ph},
 title = {{Milliarcsecond-Scale Spectral Properties and Jet Motions in M 87}},
 volume = {58},
 year = {2006}
}

@article{2007ApJ...660..200L,
 adsurl = {https://ui.adsabs.harvard.edu/abs/2007ApJ...660..200L},
 archiveprefix = {arXiv},
 author = {{Ly}, Chun and {Walker}, R. Craig and {Junor}, William},
 doi = {10.1086/512846},
 eprint = {astro-ph/0701511},
 journal = {\apj},
 month = {May},
 number = {1},
 pages = {200-205},
 primaryclass = {astro-ph},
 title = {{High-Frequency VLBI Imaging of the Jet Base of M87}},
 volume = {660},
 year = {2007}
}

@article{2007ApJ...668L..27K,
 adsurl = {https://ui.adsabs.harvard.edu/abs/2007ApJ...668L..27K},
 archiveprefix = {arXiv},
 author = {{Kovalev}, Y.~Y. and {Lister}, M.~L. and {Homan}, D.~C. and
{Kellermann}, K.~I.},
 doi = {10.1086/522603},
 eprint = {0708.2695},
 journal = {\apjl},
 month = {Oct},
 number = {1},
 pages = {L27-L30},
 primaryclass = {astro-ph},
 title = {{The Inner Jet of the Radio Galaxy M87}},
 volume = {668},
 year = {2007}
}

@inproceedings{2007RMxAC..27...91K,
 adsurl = {https://ui.adsabs.harvard.edu/abs/2007RMxAC..27...91K},
 author = {{K{\"o}nigl}, A.},
 booktitle = {Revista Mexicana de Astronomia y Astrofisica, Volume 27},
 month = {March},
 pages = {91-101},
 series = {Revista Mexicana de Astronomia y Astrofisica Conference Series},
 title = {{MHD Driving of Relativistic Jets}},
 volume = {27},
 year = {2007}
}

@article{2008MNRAS.386..619C,
 adsurl = {https://ui.adsabs.harvard.edu/abs/2008MNRAS.386..619C},
 archiveprefix = {arXiv},
 author = {{Croke}, S.~M. and {Gabuzda}, D.~C.},
 doi = {10.1111/j.1365-2966.2008.13087.x},
 eprint = {0809.3313},
 journal = {\mnras},
 month = {May},
 number = {2},
 pages = {619-626},
 primaryclass = {astro-ph},
 title = {{Aligning VLBI images of active galactic nuclei at different frequencies}},
 volume = {386},
 year = {2008}
}

@article{2008Natur.452..966M,
 adsurl = {https://ui.adsabs.harvard.edu/abs/2008Natur.452..966M},
 author = {{Marscher}, Alan P. and {Jorstad}, Svetlana G. and
{D'Arcangelo}, Francesca D. and {Smith}, Paul S. and
{Williams}, G. Grant and {Larionov}, Valeri M. and {Oh}, Haruki and
{Olmstead}, Alice R. and {Aller}, Margo F. and {Aller}, Hugh D. and
{McHardy}, Ian M. and {L{\"a}hteenm{\"a}ki}, Anne and
{Tornikoski}, Merja and {Valtaoja}, Esko and
{Hagen-Thorn}, Vladimir A. and {Kopatskaya}, Eugenia N. and
{Gear}, Walter K. and {Tosti}, Gino and {Kurtanidze}, Omar and
{Nikolashvili}, Maria and {Sigua}, Lorand and {Miller}, H. Richard and
{Ryle}, Wesley T.},
 doi = {10.1038/nature06895},
 journal = {\nat},
 month = {Apr},
 number = {7190},
 pages = {966-969},
 title = {{The inner jet of an active galactic nucleus as revealed by a radio-to-{\ensuremath{\gamma}}-ray outburst}},
 volume = {452},
 year = {2008}
}

@article{2010A&A...524A..71B,
 adsurl = {https://ui.adsabs.harvard.edu/abs/2010A&A...524A..71B},
 archiveprefix = {arXiv},
 author = {{Bird}, S. and {Harris}, W.~E. and {Blakeslee}, J.~P. and {Flynn}, C.},
 doi = {10.1051/0004-6361/201014876},
 eid = {A71},
 eprint = {1009.3202},
 journal = {\aap},
 month = {December},
 pages = {A71},
 primaryclass = {astro-ph.GA},
 title = {{The inner halo of M 87: a first direct view of the red-giant population}},
 volume = {524},
 year = {2010}
}

@article{2011AJ....142...35P,
 adsurl = {https://ui.adsabs.harvard.edu/abs/2011AJ....142...35P},
 archiveprefix = {arXiv},
 author = {{Petrov}, L. and {Kovalev}, Y.~Y. and {Fomalont}, E.~B. and {Gordon}, D.},
 doi = {10.1088/0004-6256/142/2/35},
 eid = {35},
 eprint = {1101.1460},
 journal = {\aj},
 month = {August},
 number = {2},
 pages = {35},
 primaryclass = {astro-ph.CO},
 title = {{The Very Long Baseline Array Galactic Plane Survey{\textemdash}VGaPS}},
 volume = {142},
 year = {2011}
}

@article{2011ApJ...735...61H,
 adsurl = {https://ui.adsabs.harvard.edu/abs/2011ApJ...735...61H},
 archiveprefix = {arXiv},
 author = {{Hardee}, P.~E. and {Eilek}, J.~A.},
 doi = {10.1088/0004-637X/735/1/61},
 eid = {61},
 eprint = {1104.4480},
 journal = {\apj},
 month = {July},
 number = {1},
 pages = {61},
 primaryclass = {astro-ph.CO},
 title = {{Using Twisted Filaments to Model the Inner Jet in M 87}},
 volume = {735},
 year = {2011}
}

@article{2011MNRAS.415.2081C,
 adsurl = {https://ui.adsabs.harvard.edu/abs/2011MNRAS.415.2081C},
 archiveprefix = {arXiv},
 author = {{Clausen-Brown}, E. and {Lyutikov}, M. and {Kharb}, P.},
 doi = {10.1111/j.1365-2966.2011.18757.x},
 eprint = {1101.5149},
 journal = {\mnras},
 month = {August},
 number = {3},
 pages = {2081-2092},
 primaryclass = {astro-ph.HE},
 title = {{Signatures of large-scale magnetic fields in active galactic nuclei jets: transverse asymmetries}},
 volume = {415},
 year = {2011}
}

@article{2011Natur.477..185H,
 adsurl = {https://ui.adsabs.harvard.edu/abs/2011Natur.477..185H},
 author = {{Hada}, Kazuhiro and {Doi}, Akihiro and {Kino}, Motoki and
{Nagai}, Hiroshi and {Hagiwara}, Yoshiaki and {Kawaguchi}, Noriyuki},
 doi = {10.1038/nature10387},
 journal = {\nat},
 month = {Sep},
 number = {7363},
 pages = {185-187},
 title = {{An origin of the radio jet in M87 at the location of the central black hole}},
 volume = {477},
 year = {2011}
}

@article{2011PASP..123..275D,
 adsurl = {https://ui.adsabs.harvard.edu/abs/2011PASP..123..275D},
 archiveprefix = {arXiv},
 author = {{Deller}, A.~T. and {Brisken}, W.~F. and {Phillips}, C.~J. and {Morgan}, J. and {Alef}, W. and {Cappallo}, R. and {Middelberg}, E. and {Romney}, J. and {Rottmann}, H. and {Tingay}, S.~J. and {Wayth}, R.},
 doi = {10.1086/658907},
 eprint = {1101.0885},
 journal = {\pasp},
 month = {March},
 number = {901},
 pages = {275},
 primaryclass = {astro-ph.IM},
 title = {{DiFX-2: A More Flexible, Efficient, Robust, and Powerful Software Correlator}},
 volume = {123},
 year = {2011}
}

@article{2012A&A...547A..56D,
 adsurl = {https://ui.adsabs.harvard.edu/abs/2012A&A...547A..56D},
 archiveprefix = {arXiv},
 author = {{de Gasperin}, F. and {Orr{\'u}}, E. and {Murgia}, M. and {Merloni}, A. and {Falcke}, H. and {Beck}, R. and {Beswick}, R. and {B{\^\i}rzan}, L. and {Bonafede}, A. and {Br{\"u}ggen}, M. and {Brunetti}, G. and {Chy{\.z}y}, K. and {Conway}, J. and {Croston}, J.~H. and {En{\ss}lin}, T. and {Ferrari}, C. and {Heald}, G. and {Heidenreich}, S. and {Jackson}, N. and {Macario}, G. and {McKean}, J. and {Miley}, G. and {Morganti}, R. and {Offringa}, A. and {Pizzo}, R. and {Rafferty}, D. and {R{\"o}ttgering}, H. and {Shulevski}, A. and {Steinmetz}, M. and {Tasse}, C. and {van der Tol}, S. and {van Driel}, W. and {van Weeren}, R.~J. and {van Zwieten}, J.~E. and {Alexov}, A. and {Anderson}, J. and {Asgekar}, A. and {Avruch}, M. and {Bell}, M. and {Bell}, M.~R. and {Bentum}, M. and {Bernardi}, G. and {Best}, P. and {Breitling}, F. and {Broderick}, J.~W. and {Butcher}, A. and {Ciardi}, B. and {Dettmar}, R.~J. and {Eisloeffel}, J. and {Frieswijk}, W. and {Gankema}, H. and {Garrett}, M. and {Gerbers}, M. and {Griessmeier}, J.~M. and {Gunst}, A.~W. and {Hassall}, T.~E. and {Hessels}, J. and {Hoeft}, M. and {Horneffer}, A. and {Karastergiou}, A. and {K{\"o}hler}, J. and {Koopman}, Y. and {Kuniyoshi}, M. and {Kuper}, G. and {Maat}, P. and {Mann}, G. and {Mevius}, M. and {Mulcahy}, D.~D. and {Munk}, H. and {Nijboer}, R. and {Noordam}, J. and {Paas}, H. and {Pandey}, M. and {Pandey}, V.~N. and {Polatidis}, A. and {Reich}, W. and {Schoenmakers}, A.~P. and {Sluman}, J. and {Smirnov}, O. and {Sobey}, C. and {Stappers}, B. and {Swinbank}, J. and {Tagger}, M. and {Tang}, Y. and {van Bemmel}, I. and {van Cappellen}, W. and {van Duin}, A.~P. and {van Haarlem}, M. and {van Leeuwen}, J. and {Vermeulen}, R. and {Vocks}, C. and {White}, S. and {Wise}, M. and {Wucknitz}, O. and {Zarka}, P.},
 doi = {10.1051/0004-6361/201220209},
 eid = {A56},
 eprint = {1210.1346},
 journal = {\aap},
 month = {November},
 pages = {A56},
 primaryclass = {astro-ph.GA},
 title = {{M 87 at metre wavelengths: the LOFAR picture}},
 volume = {547},
 year = {2012}
}

@article{2012ApJ...745L..28A,
 adsurl = {https://ui.adsabs.harvard.edu/abs/2012ApJ...745L..28A},
 archiveprefix = {arXiv},
 author = {{Asada}, Keiichi and {Nakamura}, Masanori},
 doi = {10.1088/2041-8205/745/2/L28},
 eid = {L28},
 eprint = {1110.1793},
 journal = {ApJ},
 month = {Feb},
 number = {2},
 pages = {L28},
 primaryclass = {astro-ph.HE},
 title = {{The Structure of the M87 Jet: A Transition from Parabolic to Conical Streamlines}},
 volume = {745},
 year = {2012}
}

@inproceedings{2012evn..confE.024,
 author = {{Sokolovsky}, Kirill V.},
 booktitle = {Proceedings of the 11th European VLBI Network Symposium and Users Meeting (EVN 2012). 9-12 October 2012. Bordeaux.},
 eid = {24},
 month = {January},
 pages = {113},
 title = {{RadioAstron Early Science Program Space-VLBI AGN survey: strategy and first results}},
 year = {2016}
}

@article{2013AJ....146...86T,
 adsurl = {https://ui.adsabs.harvard.edu/abs/2013AJ....146...86T},
 archiveprefix = {arXiv},
 author = {{Tully}, R. Brent and {Courtois}, H{\'e}l{\`e}ne M. and {Dolphin}, Andrew E. and {Fisher}, J. Richard and {H{\'e}raudeau}, Philippe and {Jacobs}, Bradley A. and {Karachentsev}, Igor D. and {Makarov}, Dmitry and {Makarova}, Lidia and {Mitronova}, Sofia and {Rizzi}, Luca and {Shaya}, Edward J. and {Sorce}, Jenny G. and {Wu}, Po-Feng},
 doi = {10.1088/0004-6256/146/4/86},
 eid = {86},
 eprint = {1307.7213},
 journal = {\aj},
 month = {October},
 number = {4},
 pages = {86},
 primaryclass = {astro-ph.CO},
 title = {{Cosmicflows-2: The Data}},
 volume = {146},
 year = {2013}
}

@article{2013ARep...57..153K,
 adsurl = {https://ui.adsabs.harvard.edu/abs/2013ARep...57..153K},
 archiveprefix = {arXiv},
 author = {{Kardashev}, N.~S. and {Khartov}, V.~V. and {Abramov}, V.~V. and {Avdeev}, V. Yu. and {Alakoz}, A.~V. and {Aleksandrov}, Yu. A. and {Ananthakrishnan}, S. and {Andreyanov}, V.~V. and {Andrianov}, A.~S. and {Antonov}, N.~M. and {Artyukhov}, M.~I. and {Arkhipov}, M. Yu. and {Baan}, W. and {Babakin}, N.~G. and {Babyshkin}, V.~E. and {Bartel'}, N. and {Belousov}, K.~G. and {Belyaev}, A.~A. and {Berulis}, J.~J. and {Burke}, B.~F. and {Biryukov}, A.~V. and {Bubnov}, A.~E. and {Burgin}, M.~S. and {Busca}, G. and {Bykadorov}, A.~A. and {Bychkova}, V.~S. and {Vasil'kov}, V.~I. and {Wellington}, K.~J. and {Vinogradov}, I.~S. and {Wietfeldt}, R. and {Voitsik}, P.~A. and {Gvamichava}, A.~S. and {Girin}, I.~A. and {Gurvits}, L.~I. and {Dagkesamanskii}, R.~D. and {D'Addario}, L. and {Giovannini}, G. and {Jauncey}, D.~L. and {Dewdney}, P.~E. and {D'yakov}, A.~A. and {Zharov}, V.~E. and {Zhuravlev}, V.~I. and {Zaslavskii}, G.~S. and {Zakhvatkin}, M.~V. and {Zinov'ev}, A.~N. and {Ilinen}, Yu. and {Ipatov}, A.~V. and {Kanevskii}, B.~Z. and {Knorin}, I.~A. and {Casse}, J.~L. and {Kellermann}, K.~I. and {Kovalev}, Yu. A. and {Kovalev}, Yu. Yu. and {Kovalenko}, A.~V. and {Kogan}, B.~L. and {Komaev}, R.~V. and {Konovalenko}, A.~A. and {Kopelyanskii}, G.~D. and {Korneev}, Yu. A. and {Kostenko}, V.~I. and {Kotik}, A.~N. and {Kreisman}, B.~B. and {Kukushkin}, A. Yu. and {Kulishenko}, V.~F. and {Cooper}, D.~N. and {Kut'kin}, A.~M. and {Cannon}, W.~H. and {Larionov}, M.~G. and {Lisakov}, M.~M. and {Litvinenko}, L.~N. and {Likhachev}, S.~F. and {Likhacheva}, L.~N. and {Lobanov}, A.~P. and {Logvinenko}, S.~V. and {Langston}, G. and {McCracken}, K. and {Medvedev}, S. Yu. and {Melekhin}, M.~V. and {Menderov}, A.~V. and {Murphy}, D.~W. and {Mizyakina}, T.~A. and {Mozgovoi}, Yu. V. and {Nikolaev}, N. Ya. and {Novikov}, B.~S. and {Novikov}, I.~D. and {Oreshko}, V.~V. and {Pavlenko}, Yu. K. and {Pashchenko}, I.~N. and {Ponomarev}, Yu. N. and {Popov}, M.~V. and {Pravin-Kumar}, A. and {Preston}, R.~A. and {Pyshnov}, V.~N. and {Rakhimov}, I.~A. and {Rozhkov}, V.~M. and {Romney}, J.~D. and {Rocha}, P. and {Rudakov}, V.~A. and {R{\"a}is{\"a}nen}, A. and {Sazankov}, S.~V. and {Sakharov}, B.~A. and {Semenov}, S.~K. and {Serebrennikov}, V.~A. and {Schilizzi}, R.~T. and {Skulachev}, D.~P. and {Slysh}, V.~I. and {Smirnov}, A.~I. and {Smith}, J.~G. and {Soglasnov}, V.~A. and {Sokolovskii}, K.~V. and {Sondaar}, L.~H. and {Stepan'yants}, V.~A. and {Turygin}, M.~S. and {Turygin}, S. Yu. and {Tuchin}, A.~G. and {Urpo}, S. and {Fedorchuk}, S.~D. and {Finkel'shtein}, A.~M. and {Fomalont}, E.~B. and {Fejes}, I. and {Fomina}, A.~N. and {Khapin}, Yu. B. and {Tsarevskii}, G.~S. and {Zensus}, J.~A. and {Chuprikov}, A.~A. and {Shatskaya}, M.~V. and {Shapirovskaya}, N. Ya. and {Sheikhet}, A.~I. and {Shirshakov}, A.~E. and {Schmidt}, A. and {Shnyreva}, L.~A. and {Shpilevskii}, V.~V. and {Ekers}, R.~D. and {Yakimov}, V.~E.},
 doi = {10.1134/S1063772913030025},
 eprint = {1303.5013},
 journal = {Astronomy Reports},
 month = {March},
 number = {3},
 pages = {153-194},
 primaryclass = {astro-ph.IM},
 title = {{``RadioAstron''-A telescope with a size of 300 000 km: Main parameters and first observational results}},
 volume = {57},
 year = {2013}
}

@article{2014AJ....147..143H,
 adsurl = {https://ui.adsabs.harvard.edu/abs/2014AJ....147..143H},
 archiveprefix = {arXiv},
 author = {{Hovatta}, Talvikki and {Aller}, Margo F. and {Aller}, Hugh D. and {Clausen-Brown}, Eric and {Homan}, Daniel C. and {Kovalev}, Yuri Y. and {Lister}, Matthew L. and {Pushkarev}, Alexander B. and {Savolainen}, Tuomas},
 doi = {10.1088/0004-6256/147/6/143},
 eid = {143},
 eprint = {1404.0014},
 journal = {\aj},
 month = {June},
 number = {6},
 pages = {143},
 primaryclass = {astro-ph.GA},
 title = {{MOJAVE: Monitoring of Jets in Active Galactic Nuclei with VLBA Experiments. XI. Spectral Distributions}},
 volume = {147},
 year = {2014}
}

@inproceedings{2014evn..confE.119B,
 adsurl = {https://ui.adsabs.harvard.edu/abs/2014evn..confE.119B},
 author = {{Bruni}, G. and {Anderson}, J. and {Alef}, W. and {Lobanov}, A. and {Zensus}, J.~A.},
 booktitle = {Proceedings of the 12th European VLBI Network Symposium and Users Meeting (EVN 2014). 7-10 October 2014. Cagliari},
 eid = {119},
 month = {January},
 pages = {119},
 title = {{Space-VLBI with RadioAstron: new correlator capabilities at MPIfR}},
 year = {2014}
}

@article{2015A&A...574A..84L,
 adsurl = {https://ui.adsabs.harvard.edu/abs/2015A&A...574A..84L},
 archiveprefix = {arXiv},
 author = {{Lobanov}, Andrei},
 doi = {10.1051/0004-6361/201425084},
 eid = {A84},
 eprint = {1412.2121},
 journal = {\aap},
 month = {February},
 pages = {A84},
 primaryclass = {astro-ph.IM},
 title = {{Brightness temperature constraints from interferometric visibilities}},
 volume = {574},
 year = {2015}
}

@article{2016AA...595A..54M,
 adsurl = {https://ui.adsabs.harvard.edu/abs/2016A&A...595A..54M},
 archiveprefix = {arXiv},
 author = {{Mertens}, F. and {Lobanov}, A.~P. and {Walker}, R.~C. and {Hardee}, P.~E.},
 doi = {10.1051/0004-6361/201628829},
 eid = {A54},
 eprint = {1608.05063},
 journal = {\aap},
 month = {Oct},
 pages = {A54},
 primaryclass = {astro-ph.HE},
 title = {{Kinematics of the jet in M 87 on scales of 100-1000 Schwarzschild radii}},
 volume = {595},
 year = {2016}
}

@article{2016ApJ...817...96G,
 adsurl = {https://ui.adsabs.harvard.edu/abs/2016ApJ...817...96G},
 archiveprefix = {arXiv},
 author = {{G{\'o}mez}, Jos{\'e} L. and {Lobanov}, Andrei P. and {Bruni}, Gabriele and {Kovalev}, Yuri Y. and {Marscher}, Alan P. and {Jorstad}, Svetlana G. and {Mizuno}, Yosuke and {Bach}, Uwe and {Sokolovsky}, Kirill V. and {Anderson}, James M. and {Galindo}, Pablo and {Kardashev}, Nikolay S. and {Lisakov}, Mikhail M.},
 doi = {10.3847/0004-637X/817/2/96},
 eid = {96},
 eprint = {1512.04690},
 journal = {\apj},
 month = {February},
 number = {2},
 pages = {96},
 primaryclass = {astro-ph.HE},
 title = {{Probing the Innermost Regions of AGN Jets and Their Magnetic Fields with RadioAstron. I. Imaging BL Lacertae at 21 Microarcsecond Resolution}},
 volume = {817},
 year = {2016}
}

@article{2016ApJ...817..131H,
 adsurl = {https://ui.adsabs.harvard.edu/abs/2016ApJ...817..131H},
 archiveprefix = {arXiv},
 author = {{Hada}, Kazuhiro and {Kino}, Motoki and {Doi}, Akihiro and
{Nagai}, Hiroshi and {Honma}, Mareki and {Akiyama}, Kazunori and
{Tazaki}, Fumie and {Lico}, Rocco and {Giroletti}, Marcello and
{Giovannini}, Gabriele and {Orienti}, Monica and {Hagiwara}, Yoshiaki},
 doi = {10.3847/0004-637X/817/2/131},
 eid = {131},
 eprint = {1512.03783},
 journal = {\apj},
 month = {Feb},
 number = {2},
 pages = {131},
 primaryclass = {astro-ph.HE},
 title = {{High-sensitivity 86 GHz (3.5 mm) VLBI Observations of M87: Deep Imaging of the Jet Base at a Resolution of 10 Schwarzschild Radii}},
 volume = {817},
 year = {2016}
}

@article{2016ApJ...833...56A,
 adsurl = {https://ui.adsabs.harvard.edu/abs/2016ApJ...833...56A},
 author = {{Asada}, Keiichi and {Nakamura}, Masanori and {Pu}, Hung-Yi},
 doi = {10.3847/1538-4357/833/1/56},
 eid = {56},
 journal = {\apj},
 month = {December},
 number = {1},
 pages = {56},
 title = {{Indication of the Black Hole Powered Jet in M87 by VSOP Observations}},
 volume = {833},
 year = {2016}
}

@article{2016Galax...4...55B,
 adsurl = {https://ui.adsabs.harvard.edu/abs/2016Galax...4...55B},
 author = {{Bruni}, Gabriele and {Anderson}, James M. and {Alef}, Walter and {Rottmann}, Helge and {Lobanov}, Andrei P. and {Zensus}, J. Anton},
 doi = {10.3390/galaxies4040055},
 eid = {55},
 journal = {Galaxies},
 month = {October},
 number = {4},
 pages = {55},
 title = {{The RadioAstron Dedicated DiFX Distribution}},
 volume = {4},
 year = {2016}
}

@article{2017Galax...5....2H,
 adsurl = {https://ui.adsabs.harvard.edu/abs/2017Galax...5....2H},
 author = {{Hada}, Kazuhiro},
 doi = {10.3390/galaxies5010002},
 journal = {Galaxies},
 month = {January},
 number = {1},
 pages = {2},
 title = {{The Structure and Propagation of the Misaligned Jet M87}},
 volume = {5},
 year = {2017}
}

@book{2017isra.book.....T,
 adsurl = {https://ui.adsabs.harvard.edu/abs/2017isra.book.....T},
 author = {{Thompson}, A. Richard and {Moran}, James M. and {Swenson}, George W., Jr.},
 doi = {10.1007/978-3-319-44431-4},
 publisher = {Springer Cham},
 title = {{Interferometry and Synthesis in Radio Astronomy, 3rd Edition}},
 year = {2017}
}

@article{2018AA...616A.188K,
 adsurl = {https://ui.adsabs.harvard.edu/abs/2018A&A...616A.188K},
 archiveprefix = {arXiv},
 author = {{Kim}, J. -Y. and {Krichbaum}, T.~P. and {Lu}, R. -S. and {Ros}, E. and {Bach}, U. and {Bremer}, M. and {de Vicente}, P. and {Lindqvist}, M. and {Zensus}, J.~A.},
 doi = {10.1051/0004-6361/201832921},
 eid = {A188},
 eprint = {1805.02478},
 journal = {\aap},
 month = {Sep},
 pages = {A188},
 primaryclass = {astro-ph.GA},
 title = {{The limb-brightened jet of M87 down to the 7 Schwarzschild radii scale}},
 volume = {616},
 year = {2018}
}

@article{2018ApJ...855..128W,
 adsurl = {http://adsabs.harvard.edu/abs/2018ApJ...855..128W},
 archiveprefix = {arXiv},
 author = {{Walker}, R.~C. and {Hardee}, P.~E. and {Davies}, F.~B. and 
{Ly}, C. and {Junor}, W.},
 doi = {10.3847/1538-4357/aaafcc},
 eid = {128},
 eprint = {1802.06166},
 journal = {\apj},
 month = {March},
 pages = {128},
 primaryclass = {astro-ph.HE},
 title = {{The Structure and Dynamics of the Subparsec Jet in M87 Based on 50 VLBA Observations over 17 Years at 43 GHz}},
 volume = {855},
 year = {2018}
}

@article{2018ApJ...868..146N,
 adsurl = {https://ui.adsabs.harvard.edu/abs/2018ApJ...868..146N},
 archiveprefix = {arXiv},
 author = {{Nakamura}, Masanori and {Asada}, Keiichi and {Hada}, Kazuhiro and         {Pu}, Hung-Yi and {Noble}, Scott and {Tseng}, Chihyin and         {Toma}, Kenji and {Kino}, Motoki and {Nagai}, Hiroshi and         {Takahashi}, Kazuya and {Algaba}, Juan-Carlos and {Orienti}, Monica and         {Akiyama}, Kazunori and {Doi}, Akihiro and {Giovannini}, Gabriele and         {Giroletti}, Marcello and {Honma}, Mareki and {Koyama}, Shoko and
{Lico}, Rocco and {Niinuma}, Kotaro and {Tazaki}, Fumie},
 doi = {10.3847/1538-4357/aaeb2d},
 eid = {146},
 eprint = {1810.09963},
 journal = {ApJ},
 month = {Dec},
 number = {2},
 pages = {146},
 primaryclass = {astro-ph.HE},
 title = {{Parabolic Jets from the Spinning Black Hole in M87}},
 volume = {868},
 year = {2018}
}

@article{2018ApJS..234...12L,
 adsurl = {https://ui.adsabs.harvard.edu/abs/2018ApJS..234...12L},
 archiveprefix = {arXiv},
 author = {{Lister}, M.~L. and {Aller}, M.~F. and {Aller}, H.~D. and {Hodge}, M.~A. and {Homan}, D.~C. and {Kovalev}, Y.~Y. and {Pushkarev}, A.~B. and {Savolainen}, T.},
 doi = {10.3847/1538-4365/aa9c44},
 eid = {12},
 eprint = {1711.07802},
 journal = {\apjs},
 month = {January},
 number = {1},
 pages = {12},
 primaryclass = {astro-ph.GA},
 title = {{MOJAVE. XV. VLBA 15 GHz Total Intensity and Polarization Maps of 437 Parsec-scale AGN Jets from 1996 to 2017}},
 volume = {234},
 year = {2018}
}

@article{2018MNRAS.474.3523P,
 adsurl = {https://ui.adsabs.harvard.edu/abs/2018MNRAS.474.3523P},
 archiveprefix = {arXiv},
 author = {{Pilipenko}, S.~V. and {Kovalev}, Y.~Y. and {Andrianov}, A.~S. and {Bach}, U. and {Buttaccio}, S. and {Cassaro}, P. and {Cim{\`o}}, G. and {Edwards}, P.~G. and {Gawro{\'n}ski}, M.~P. and {Gurvits}, L.~I. and {Hovatta}, T. and {Jauncey}, D.~L. and {Johnson}, M.~D. and {Kovalev}, Yu A. and {Kutkin}, A.~M. and {Lisakov}, M.~M. and {Melnikov}, A.~E. and {Orlati}, A. and {Rudnitskiy}, A.~G. and {Sokolovsky}, K.~V. and {Stanghellini}, C. and {de Vicente}, P. and {Voitsik}, P.~A. and {Wolak}, P. and {Zhekanis}, G.~V.},
 doi = {10.1093/mnras/stx2991},
 eprint = {1711.06713},
 journal = {\mnras},
 month = {March},
 number = {3},
 pages = {3523-3534},
 primaryclass = {astro-ph.HE},
 title = {{The high brightness temperature of B0529+483 revealed by RadioAstron and implications for interstellar scattering}},
 volume = {474},
 year = {2018}
}

@article{2018MNRAS.475.4994K,
 adsurl = {https://ui.adsabs.harvard.edu/abs/2018MNRAS.475.4994K},
 archiveprefix = {arXiv},
 author = {{Kutkin}, A.~M. and {Pashchenko}, I.~N. and {Lisakov}, M.~M. and {Voytsik}, P.~A. and {Sokolovsky}, K.~V. and {Kovalev}, Y.~Y. and {Lobanov}, A.~P. and {Ipatov}, A.~V. and {Aller}, M.~F. and {Aller}, H.~D. and {Lahteenmaki}, A. and {Tornikoski}, M. and {Gurvits}, L.~I.},
 doi = {10.1093/mnras/sty144},
 eprint = {1801.04892},
 journal = {\mnras},
 month = {April},
 number = {4},
 pages = {4994-5009},
 primaryclass = {astro-ph.GA},
 title = {{The extreme blazar AO 0235+164 as seen by extensive ground and space radio observations}},
 volume = {475},
 year = {2018}
}

@article{2018NatAs...2..472G,
 adsurl = {https://ui.adsabs.harvard.edu/abs/2018NatAs...2..472G},
 archiveprefix = {arXiv},
 author = {{Giovannini}, G. and {Savolainen}, T. and {Orienti}, M. and
{Nakamura}, M. and {Nagai}, H. and {Kino}, M. and {Giroletti}, M. and
{Hada}, K. and {Bruni}, G. and {Kovalev}, Y.~Y. and {Anderson}, J.~M. and
{D'Ammando}, F. and {Hodgson}, J. and {Honma}, M. and
{Krichbaum}, T.~P. and {Lee}, S. -S. and {Lico}, R. and
{Lisakov}, M.~M. and {Lobanov}, A.~P. and {Petrov}, L. and
{Sohn}, B.~W. and {Sokolovsky}, K.~V. and {Voitsik}, P.~A. and
{Zensus}, J.~A. and {Tingay}, S.},
 doi = {10.1038/s41550-018-0431-2},
 eprint = {1804.02198},
 journal = {Nature Astronomy},
 month = {Apr},
 pages = {472-477},
 primaryclass = {astro-ph.GA},
 title = {{A wide and collimated radio jet in 3C84 on the scale of a few hundred gravitational radii}},
 volume = {2},
 year = {2018}
}

@article{2019ApJ...875L...1E,
 adsurl = {https://ui.adsabs.harvard.edu/abs/2019ApJ...875L...1E},
 archiveprefix = {arXiv},
 author = {{Event Horizon Telescope Collaboration} and {Akiyama}, Kazunori and {Alberdi}, Antxon and {Alef}, Walter and {Asada}, Keiichi and {Azulay}, Rebecca and {Baczko}, Anne-Kathrin and {Ball}, David and {Balokovi{\'c}}, Mislav and {Barrett}, John and {Bintley}, Dan and {Blackburn}, Lindy and {Boland}, Wilfred and {Bouman}, Katherine L. and {Bower}, Geoffrey C. and {Bremer}, Michael and {Brinkerink}, Christiaan D. and {Brissenden}, Roger and {Britzen}, Silke and {Broderick}, Avery E. and {Broguiere}, Dominique and {Bronzwaer}, Thomas and {Byun}, Do-Young and {Carlstrom}, John E. and {Chael}, Andrew and {Chan}, Chi-kwan and {Chatterjee}, Shami and {Chatterjee}, Koushik and {Chen}, Ming-Tang and {Chen}, Yongjun and {Cho}, Ilje and {Christian}, Pierre and {Conway}, John E. and {Cordes}, James M. and {Crew}, Geoffrey B. and {Cui}, Yuzhu and {Davelaar}, Jordy and {De Laurentis}, Mariafelicia and {Deane}, Roger and {Dempsey}, Jessica and {Desvignes}, Gregory and {Dexter}, Jason and {Doeleman}, Sheperd S. and {Eatough}, Ralph P. and {Falcke}, Heino and {Fish}, Vincent L. and {Fomalont}, Ed and {Fraga-Encinas}, Raquel and {Freeman}, William T. and {Friberg}, Per and {Fromm}, Christian M. and {G{\'o}mez}, Jos{\'e} L. and {Galison}, Peter and {Gammie}, Charles F. and {Garc{\'\i}a}, Roberto and {Gentaz}, Olivier and {Georgiev}, Boris and {Goddi}, Ciriaco and {Gold}, Roman and {Gu}, Minfeng and {Gurwell}, Mark and {Hada}, Kazuhiro and {Hecht}, Michael H. and {Hesper}, Ronald and {Ho}, Luis C. and {Ho}, Paul and {Honma}, Mareki and {Huang}, Chih-Wei L. and {Huang}, Lei and {Hughes}, David H. and {Ikeda}, Shiro and {Inoue}, Makoto and {Issaoun}, Sara and {James}, David J. and {Jannuzi}, Buell T. and {Janssen}, Michael and {Jeter}, Britton and {Jiang}, Wu and {Johnson}, Michael D. and {Jorstad}, Svetlana and {Jung}, Taehyun and {Karami}, Mansour and {Karuppusamy}, Ramesh and {Kawashima}, Tomohisa and {Keating}, Garrett K. and {Kettenis}, Mark and {Kim}, Jae-Young and {Kim}, Junhan and {Kim}, Jongsoo and {Kino}, Motoki and {Koay}, Jun Yi and {Koch}, Patrick M. and {Koyama}, Shoko and {Kramer}, Michael and {Kramer}, Carsten and {Krichbaum}, Thomas P. and {Kuo}, Cheng-Yu and {Lauer}, Tod R. and {Lee}, Sang-Sung and {Li}, Yan-Rong and {Li}, Zhiyuan and {Lindqvist}, Michael and {Liu}, Kuo and {Liuzzo}, Elisabetta and {Lo}, Wen-Ping and {Lobanov}, Andrei P. and {Loinard}, Laurent and {Lonsdale}, Colin and {Lu}, Ru-Sen and {MacDonald}, Nicholas R. and {Mao}, Jirong and {Markoff}, Sera and {Marrone}, Daniel P. and {Marscher}, Alan P. and {Mart{\'\i}-Vidal}, Iv{\'a}n and {Matsushita}, Satoki and {Matthews}, Lynn D. and {Medeiros}, Lia and {Menten}, Karl M. and {Mizuno}, Yosuke and {Mizuno}, Izumi and {Moran}, James M. and {Moriyama}, Kotaro and {Moscibrodzka}, Monika and {M{\"u}ller}, Cornelia and {Nagai}, Hiroshi and {Nagar}, Neil M. and {Nakamura}, Masanori and {Narayan}, Ramesh and {Narayanan}, Gopal and {Natarajan}, Iniyan and {Neri}, Roberto and {Ni}, Chunchong and {Noutsos}, Aristeidis and {Okino}, Hiroki and {Olivares}, H{\'e}ctor and {Ortiz-Le{\'o}n}, Gisela N. and {Oyama}, Tomoaki and {{\"O}zel}, Feryal and {Palumbo}, Daniel C.~M. and {Patel}, Nimesh and {Pen}, Ue-Li and {Pesce}, Dominic W. and {Pi{\'e}tu}, Vincent and {Plambeck}, Richard and {PopStefanija}, Aleksandar and {Porth}, Oliver and {Prather}, Ben and {Preciado-L{\'o}pez}, Jorge A. and {Psaltis}, Dimitrios and {Pu}, Hung-Yi and {Ramakrishnan}, Venkatessh and {Rao}, Ramprasad and {Rawlings}, Mark G. and {Raymond}, Alexander W. and {Rezzolla}, Luciano and {Ripperda}, Bart and {Roelofs}, Freek and {Rogers}, Alan and {Ros}, Eduardo and {Rose}, Mel and {Roshanineshat}, Arash and {Rottmann}, Helge and {Roy}, Alan L. and {Ruszczyk}, Chet and {Ryan}, Benjamin R. and {Rygl}, Kazi L.~J. and {S{\'a}nchez}, Salvador and {S{\'a}nchez-Arguelles}, David and {Sasada}, Mahito and {Savolainen}, Tuomas and {Schloerb}, F. Peter and {Schuster}, Karl-Friedrich and {Shao}, Lijing and {Shen}, Zhiqiang and {Small}, Des and {Sohn}, Bong Won and {SooHoo}, Jason and {Tazaki}, Fumie and {Tiede}, Paul and {Tilanus}, Remo P.~J. and {Titus}, Michael and {Toma}, Kenji and {Torne}, Pablo and {Trent}, Tyler and {Trippe}, Sascha and {Tsuda}, Shuichiro and {van Bemmel}, Ilse and {van Langevelde}, Huib Jan and {van Rossum}, Daniel R. and {Wagner}, Jan and {Wardle}, John and {Weintroub}, Jonathan and {Wex}, Norbert and {Wharton}, Robert and {Wielgus}, Maciek and {Wong}, George N. and {Wu}, Qingwen and {Young}, Ken and {Young}, Andr{\'e} and {Younsi}, Ziri and {Yuan}, Feng and {Yuan}, Ye-Fei and {Zensus}, J. Anton and {Zhao}, Guangyao and {Zhao}, Shan-Shan and {Zhu}, Ziyan and {Algaba}, Juan-Carlos and {Allardi}, Alexander and {Amestica}, Rodrigo and {Anczarski}, Jadyn and {Bach}, Uwe and {Baganoff}, Frederick K. and {Beaudoin}, Christopher and {Benson}, Bradford A. and {Berthold}, Ryan and {Blanchard}, Jay M. and {Blundell}, Ray and {Bustamente}, Sandra and {Cappallo}, Roger and {Castillo-Dom{\'\i}nguez}, Edgar and {Chang}, Chih-Cheng and {Chang}, Shu-Hao and {Chang}, Song-Chu and {Chen}, Chung-Chen and {Chilson}, Ryan and {Chuter}, Tim C. and {C{\'o}rdova Rosado}, Rodrigo and {Coulson}, Iain M. and {Crawford}, Thomas M. and {Crowley}, Joseph and {David}, John and {Derome}, Mark and {Dexter}, Matthew and {Dornbusch}, Sven and {Dudevoir}, Kevin A. and {Dzib}, Sergio A. and {Eckart}, Andreas and {Eckert}, Chris and {Erickson}, Neal R. and {Everett}, Wendeline B. and {Faber}, Aaron and {Farah}, Joseph R. and {Fath}, Vernon and {Folkers}, Thomas W. and {Forbes}, David C. and {Freund}, Robert and {G{\'o}mez-Ruiz}, Arturo I. and {Gale}, David M. and {Gao}, Feng and {Geertsema}, Gertie and {Graham}, David A. and {Greer}, Christopher H. and {Grosslein}, Ronald and {Gueth}, Fr{\'e}d{\'e}ric and {Haggard}, Daryl and {Halverson}, Nils W. and {Han}, Chih-Chiang and {Han}, Kuo-Chang and {Hao}, Jinchi and {Hasegawa}, Yutaka and {Henning}, Jason W. and {Hern{\'a}ndez-G{\'o}mez}, Antonio and {Herrero-Illana}, Rub{\'e}n and {Heyminck}, Stefan and {Hirota}, Akihiko and {Hoge}, James and {Huang}, Yau-De and {Impellizzeri}, C.~M. Violette and {Jiang}, Homin and {Kamble}, Atish and {Keisler}, Ryan and {Kimura}, Kimihiro and {Kono}, Yusuke and {Kubo}, Derek and {Kuroda}, John and {Lacasse}, Richard and {Laing}, Robert A. and {Leitch}, Erik M. and {Li}, Chao-Te and {Lin}, Lupin C. -C. and {Liu}, Ching-Tang and {Liu}, Kuan-Yu and {Lu}, Li-Ming and {Marson}, Ralph G. and {Martin-Cocher}, Pierre L. and {Massingill}, Kyle D. and {Matulonis}, Callie and {McColl}, Martin P. and {McWhirter}, Stephen R. and {Messias}, Hugo and {Meyer-Zhao}, Zheng and {Michalik}, Daniel and {Monta{\~n}a}, Alfredo and {Montgomerie}, William and {Mora-Klein}, Matias and {Muders}, Dirk and {Nadolski}, Andrew and {Navarro}, Santiago and {Neilsen}, Joseph and {Nguyen}, Chi H. and {Nishioka}, Hiroaki and {Norton}, Timothy and {Nowak}, Michael A. and {Nystrom}, George and {Ogawa}, Hideo and {Oshiro}, Peter and {Oyama}, Tomoaki and {Parsons}, Harriet and {Paine}, Scott N. and {Pe{\~n}alver}, Juan and {Phillips}, Neil M. and {Poirier}, Michael and {Pradel}, Nicolas and {Primiani}, Rurik A. and {Raffin}, Philippe A. and {Rahlin}, Alexandra S. and {Reiland}, George and {Risacher}, Christopher and {Ruiz}, Ignacio and {S{\'a}ez-Mada{\'\i}n}, Alejandro F. and {Sassella}, Remi and {Schellart}, Pim and {Shaw}, Paul and {Silva}, Kevin M. and {Shiokawa}, Hotaka and {Smith}, David R. and {Snow}, William and {Souccar}, Kamal and {Sousa}, Don and {Sridharan}, T.~K. and {Srinivasan}, Ranjani and {Stahm}, William and {Stark}, Anthony A. and {Story}, Kyle and {Timmer}, Sjoerd T. and {Vertatschitsch}, Laura and {Walther}, Craig and {Wei}, Ta-Shun and {Whitehorn}, Nathan and {Whitney}, Alan R. and {Woody}, David P. and {Wouterloot}, Jan G.~A. and {Wright}, Melvin and {Yamaguchi}, Paul and {Yu}, Chen-Yu and {Zeballos}, Milagros and {Zhang}, Shuo and {Ziurys}, Lucy},
 doi = {10.3847/2041-8213/ab0ec7},
 eid = {L1},
 eprint = {1906.11238},
 journal = {\apjl},
 month = {April},
 number = {1},
 pages = {L1},
 primaryclass = {astro-ph.GA},
 title = {{First M87 Event Horizon Telescope Results. I. The Shadow of the Supermassive Black Hole}},
 volume = {875},
 year = {2019}
}

@article{2019ApJ...875L...4E,
 adsurl = {https://ui.adsabs.harvard.edu/abs/2019ApJ...875L...4E},
 archiveprefix = {arXiv},
 author = {{Event Horizon Telescope Collaboration} and {Akiyama}, Kazunori and {Alberdi}, Antxon and {Alef}, Walter and {Asada}, Keiichi and {Azulay}, Rebecca and {Baczko}, Anne-Kathrin and {Ball}, David and {Balokovi{\'c}}, Mislav and {Barrett}, John and {Bintley}, Dan and {Blackburn}, Lindy and {Boland}, Wilfred and {Bouman}, Katherine L. and {Bower}, Geoffrey C. and {Bremer}, Michael and {Brinkerink}, Christiaan D. and {Brissenden}, Roger and {Britzen}, Silke and {Broderick}, Avery E. and {Broguiere}, Dominique and {Bronzwaer}, Thomas and {Byun}, Do-Young and {Carlstrom}, John E. and {Chael}, Andrew and {Chan}, Chi-kwan and {Chatterjee}, Shami and {Chatterjee}, Koushik and {Chen}, Ming-Tang and {Chen}, Yongjun and {Cho}, Ilje and {Christian}, Pierre and {Conway}, John E. and {Cordes}, James M. and {Crew}, Geoffrey B. and {Cui}, Yuzhu and {Davelaar}, Jordy and {De Laurentis}, Mariafelicia and {Deane}, Roger and {Dempsey}, Jessica and {Desvignes}, Gregory and {Dexter}, Jason and {Doeleman}, Sheperd S. and {Eatough}, Ralph P. and {Falcke}, Heino and {Fish}, Vincent L. and {Fomalont}, Ed and {Fraga-Encinas}, Raquel and {Freeman}, William T. and {Friberg}, Per and {Fromm}, Christian M. and {G{\'o}mez}, Jos{\'e} L. and {Galison}, Peter and {Gammie}, Charles F. and {Garc{\'\i}a}, Roberto and {Gentaz}, Olivier and {Georgiev}, Boris and {Goddi}, Ciriaco and {Gold}, Roman and {Gu}, Minfeng and {Gurwell}, Mark and {Hada}, Kazuhiro and {Hecht}, Michael H. and {Hesper}, Ronald and {Ho}, Luis C. and {Ho}, Paul and {Honma}, Mareki and {Huang}, Chih-Wei L. and {Huang}, Lei and {Hughes}, David H. and {Ikeda}, Shiro and {Inoue}, Makoto and {Issaoun}, Sara and {James}, David J. and {Jannuzi}, Buell T. and {Janssen}, Michael and {Jeter}, Britton and {Jiang}, Wu and {Johnson}, Michael D. and {Jorstad}, Svetlana and {Jung}, Taehyun and {Karami}, Mansour and {Karuppusamy}, Ramesh and {Kawashima}, Tomohisa and {Keating}, Garrett K. and {Kettenis}, Mark and {Kim}, Jae-Young and {Kim}, Junhan and {Kim}, Jongsoo and {Kino}, Motoki and {Koay}, Jun Yi and {Koch}, Patrick M. and {Koyama}, Shoko and {Kramer}, Michael and {Kramer}, Carsten and {Krichbaum}, Thomas P. and {Kuo}, Cheng-Yu and {Lauer}, Tod R. and {Lee}, Sang-Sung and {Li}, Yan-Rong and {Li}, Zhiyuan and {Lindqvist}, Michael and {Liu}, Kuo and {Liuzzo}, Elisabetta and {Lo}, Wen-Ping and {Lobanov}, Andrei P. and {Loinard}, Laurent and {Lonsdale}, Colin and {Lu}, Ru-Sen and {MacDonald}, Nicholas R. and {Mao}, Jirong and {Markoff}, Sera and {Marrone}, Daniel P. and {Marscher}, Alan P. and {Mart{\'\i}-Vidal}, Iv{\'a}n and {Matsushita}, Satoki and {Matthews}, Lynn D. and {Medeiros}, Lia and {Menten}, Karl M. and {Mizuno}, Yosuke and {Mizuno}, Izumi and {Moran}, James M. and {Moriyama}, Kotaro and {Moscibrodzka}, Monika and {M{\"u}ller}, Cornelia and {Nagai}, Hiroshi and {Nagar}, Neil M. and {Nakamura}, Masanori and {Narayan}, Ramesh and {Narayanan}, Gopal and {Natarajan}, Iniyan and {Neri}, Roberto and {Ni}, Chunchong and {Noutsos}, Aristeidis and {Okino}, Hiroki and {Olivares}, H{\'e}ctor and {Oyama}, Tomoaki and {{\"O}zel}, Feryal and {Palumbo}, Daniel C.~M. and {Patel}, Nimesh and {Pen}, Ue-Li and {Pesce}, Dominic W. and {Pi{\'e}tu}, Vincent and {Plambeck}, Richard and {PopStefanija}, Aleksandar and {Porth}, Oliver and {Prather}, Ben and {Preciado-L{\'o}pez}, Jorge A. and {Psaltis}, Dimitrios and {Pu}, Hung-Yi and {Ramakrishnan}, Venkatessh and {Rao}, Ramprasad and {Rawlings}, Mark G. and {Raymond}, Alexander W. and {Rezzolla}, Luciano and {Ripperda}, Bart and {Roelofs}, Freek and {Rogers}, Alan and {Ros}, Eduardo and {Rose}, Mel and {Roshanineshat}, Arash and {Rottmann}, Helge and {Roy}, Alan L. and {Ruszczyk}, Chet and {Ryan}, Benjamin R. and {Rygl}, Kazi L.~J. and {S{\'a}nchez}, Salvador and {S{\'a}nchez-Arguelles}, David and {Sasada}, Mahito and {Savolainen}, Tuomas and {Schloerb}, F. Peter and {Schuster}, Karl-Friedrich and {Shao}, Lijing and {Shen}, Zhiqiang and {Small}, Des and {Sohn}, Bong Won and {SooHoo}, Jason and {Tazaki}, Fumie and {Tiede}, Paul and {Tilanus}, Remo P.~J. and {Titus}, Michael and {Toma}, Kenji and {Torne}, Pablo and {Trent}, Tyler and {Trippe}, Sascha and {Tsuda}, Shuichiro and {van Bemmel}, Ilse and {van Langevelde}, Huib Jan and {van Rossum}, Daniel R. and {Wagner}, Jan and {Wardle}, John and {Weintroub}, Jonathan and {Wex}, Norbert and {Wharton}, Robert and {Wielgus}, Maciek and {Wong}, George N. and {Wu}, Qingwen and {Young}, Andr{\'e} and {Young}, Ken and {Younsi}, Ziri and {Yuan}, Feng and {Yuan}, Ye-Fei and {Zensus}, J. Anton and {Zhao}, Guangyao and {Zhao}, Shan-Shan and {Zhu}, Ziyan and {Farah}, Joseph R. and {Meyer-Zhao}, Zheng and {Michalik}, Daniel and {Nadolski}, Andrew and {Nishioka}, Hiroaki and {Pradel}, Nicolas and {Primiani}, Rurik A. and {Souccar}, Kamal and {Vertatschitsch}, Laura and {Yamaguchi}, Paul},
 doi = {10.3847/2041-8213/ab0e85},
 eid = {L4},
 eprint = {1906.11241},
 journal = {\apjl},
 month = {April},
 number = {1},
 pages = {L4},
 primaryclass = {astro-ph.GA},
 title = {{First M87 Event Horizon Telescope Results. IV. Imaging the Central Supermassive Black Hole}},
 volume = {875},
 year = {2019}
}

@article{2019ApJ...887..147P,
 adsurl = {https://ui.adsabs.harvard.edu/abs/2019ApJ...887..147P},
 archiveprefix = {arXiv},
 author = {{Park}, Jongho and {Hada}, Kazuhiro and {Kino}, Motoki and {Nakamura}, Masanori and {Hodgson}, Jeffrey and {Ro}, Hyunwook and {Cui}, Yuzhu and {Asada}, Keiichi and {Algaba}, Juan-Carlos and {Sawada-Satoh}, Satoko and {Lee}, Sang-Sung and {Cho}, Ilje and {Shen}, Zhiqiang and {Jiang}, Wu and {Trippe}, Sascha and {Niinuma}, Kotaro and {Sohn}, Bong Won and {Jung}, Taehyun and {Zhao}, Guang-Yao and {Wajima}, Kiyoaki and {Tazaki}, Fumie and {Honma}, Mareki and {An}, Tao and {Akiyama}, Kazunori and {Byun}, Do-Young and {Kim}, Jongsoo and {Zhang}, Yingkang and {Cheng}, Xiaopeng and {Kobayashi}, Hideyuki and {Shibata}, Katsunori M. and {Lee}, Jee Won and {Roh}, Duk-Gyoo and {Oh}, Se-Jin and {Yeom}, Jae-Hwan and {Jung}, Dong-Kyu and {Oh}, Chungsik and {Kim}, Hyo-Ryoung and {Hwang}, Ju-Yeon and {Hagiwara}, Yoshiaki},
 doi = {10.3847/1538-4357/ab5584},
 eid = {147},
 eprint = {1911.02279},
 journal = {\apj},
 month = {December},
 number = {2},
 pages = {147},
 primaryclass = {astro-ph.HE},
 title = {{Kinematics of the M87 Jet in the Collimation Zone: Gradual Acceleration and Velocity Stratification}},
 volume = {887},
 year = {2019}
}

@article{2019Galax...7...86Z,
 adsurl = {https://ui.adsabs.harvard.edu/abs/2019Galax...7...86Z},
 archiveprefix = {arXiv},
 author = {{Zhao}, Wei and {Hong}, Xiaoyu and {An}, Tao and {Li}, Xiaofeng and {Cheng}, Xiaopeng and {Wu}, Fang},
 doi = {10.3390/galaxies7040086},
 eprint = {1910.14299},
 journal = {Galaxies},
 month = {October},
 number = {4},
 pages = {86},
 primaryclass = {astro-ph.GA},
 title = {{Features of Structure and Absorption in the Jet-Launching Region of M87}},
 volume = {7},
 year = {2019}
}

@article{2020A&A...637L...6K,
 adsurl = {https://ui.adsabs.harvard.edu/abs/2020A&A...637L...6K},
 archiveprefix = {arXiv},
 author = {{Kravchenko}, E. and {Giroletti}, M. and {Hada}, K. and {Meier}, D.~L. and {Nakamura}, M. and {Park}, J. and {Walker}, R.~C.},
 doi = {10.1051/0004-6361/201937315},
 eid = {L6},
 eprint = {2006.07059},
 journal = {\aap},
 month = {May},
 pages = {L6},
 primaryclass = {astro-ph.GA},
 title = {{Linear polarization in the nucleus of M87 at 7 mm and 1.3 cm}},
 volume = {637},
 year = {2020}
}

@article{2020A&A...641A..40V,
 adsurl = {https://ui.adsabs.harvard.edu/abs/2020A&A...641A..40V},
 archiveprefix = {arXiv},
 author = {{Vega-Garc{\'\i}a}, L. and {Lobanov}, A.~P. and {Perucho}, M. and {Bruni}, G. and {Ros}, E. and {Anderson}, J.~M. and {Agudo}, I. and {Davis}, R. and {G{\'o}mez}, J.~L. and {Kovalev}, Y.~Y. and {Krichbaum}, T.~P. and {Lisakov}, M. and {Savolainen}, T. and {Schinzel}, F.~K. and {Zensus}, J.~A.},
 doi = {10.1051/0004-6361/201935168},
 eid = {A40},
 eprint = {1912.00925},
 journal = {\aap},
 month = {September},
 pages = {A40},
 primaryclass = {astro-ph.GA},
 title = {{Multiband RadioAstron space VLBI imaging of the jet in quasar S5 0836+710}},
 volume = {641},
 year = {2020}
}

@article{2020AdSpR..65..705K,
 adsurl = {https://ui.adsabs.harvard.edu/abs/2020AdSpR..65..705K},
 archiveprefix = {arXiv},
 author = {{Kovalev}, Y.~Y. and {Kardashev}, N.~S. and {Sokolovsky}, K.~V. and {Voitsik}, P.~A. and {An}, T. and {Anderson}, J.~M. and {Andrianov}, A.~S. and {Avdeev}, V. Yu. and {Bartel}, N. and {Bignall}, H.~E. and {Burgin}, M.~S. and {Edwards}, P.~G. and {Ellingsen}, S.~P. and {Frey}, S. and {Garc{\'\i}a-Mir{\'o}}, C. and {Gawro{\'n}ski}, M.~P. and {Ghigo}, F.~D. and {Ghosh}, T. and {Giovannini}, G. and {Girin}, I.~A. and {Giroletti}, M. and {Gurvits}, L.~I. and {Jauncey}, D.~L. and {Horiuchi}, S. and {Ivanov}, D.~V. and {Kharinov}, M.~A. and {Koay}, J.~Y. and {Kostenko}, V.~I. and {Kovalenko}, A.~V. and {Kovalev}, Yu. A. and {Kravchenko}, E.~V. and {Kunert-Bajraszewska}, M. and {Kutkin}, A.~M. and {Likhachev}, S.~F. and {Lisakov}, M.~M. and {Litovchenko}, I.~D. and {McCallum}, J.~N. and {Melis}, A. and {Melnikov}, A.~E. and {Migoni}, C. and {Nair}, D.~G. and {Pashchenko}, I.~N. and {Phillips}, C.~J. and {Polatidis}, A. and {Pushkarev}, A.~B. and {Quick}, J.~F.~H. and {Rakhimov}, I.~A. and {Reynolds}, C. and {Rizzo}, J.~R. and {Rudnitskiy}, A.~G. and {Savolainen}, T. and {Shakhvorostova}, N.~N. and {Shatskaya}, M.~V. and {Shen}, Z. -Q. and {Shchurov}, M.~A. and {Vermeulen}, R.~C. and {de Vicente}, P. and {Wolak}, P. and {Zensus}, J.~A. and {Zuga}, V.~A.},
 doi = {10.1016/j.asr.2019.08.035},
 eprint = {1909.00785},
 journal = {Advances in Space Research},
 month = {January},
 number = {2},
 pages = {705-711},
 primaryclass = {astro-ph.GA},
 title = {{Detection statistics of the RadioAstron AGN survey}},
 volume = {65},
 year = {2020}
}

@article{2020AdSpR..65..712B,
 adsurl = {https://ui.adsabs.harvard.edu/abs/2020AdSpR..65..712B},
 archiveprefix = {arXiv},
 author = {{Bruni}, Gabriele and {Savolainen}, Tuomas and {G{\'o}mez}, Jose Luis and {Lobanov}, Andrei P. and {Kovalev}, Yuri Y. and {RadioAstron AGN Imaging Team} and {KSP Team}},
 doi = {10.1016/j.asr.2019.03.044},
 eprint = {1904.00814},
 journal = {Advances in Space Research},
 month = {January},
 number = {2},
 pages = {712-719},
 primaryclass = {astro-ph.GA},
 title = {{Active galactic nuclei imaging programs of the RadioAstron mission}},
 volume = {65},
 year = {2020}
}

@article{2020AdSpR..65..772S,
 adsurl = {https://ui.adsabs.harvard.edu/abs/2020AdSpR..65..772S},
 archiveprefix = {arXiv},
 author = {{Shakhvorostova}, N.~N. and {Sobolev}, A.~M. and {Moran}, J.~M. and {Alakoz}, A.~V. and {Imai}, H. and {Avdeev}, V.~Y.},
 doi = {10.1016/j.asr.2019.05.011},
 eprint = {1905.02440},
 journal = {Advances in Space Research},
 month = {January},
 number = {2},
 pages = {772-779},
 primaryclass = {astro-ph.GA},
 title = {{RadioAstron probes the ultra-fine spatial structure in the H$_{2}$O maser emission in the star forming region W49N}},
 volume = {65},
 year = {2020}
}

@article{2020ApJ...893...68K,
 adsurl = {https://ui.adsabs.harvard.edu/abs/2020ApJ...893...68K},
 archiveprefix = {arXiv},
 author = {{Kravchenko}, E.~V. and {G{\'o}mez}, J.~L. and {Kovalev}, Y.~Y. and {Lobanov}, A.~P. and {Savolainen}, T. and {Bruni}, G. and {Fuentes}, A. and {Anderson}, J.~M. and {Jorstad}, S.~G. and {Marscher}, A.~P. and {Tornikoski}, M. and {L{\"a}hteenm{\"a}ki}, A. and {Lisakov}, M.~M.},
 doi = {10.3847/1538-4357/ab7dae},
 eid = {68},
 eprint = {2003.08776},
 journal = {\apj},
 month = {April},
 number = {1},
 pages = {68},
 primaryclass = {astro-ph.HE},
 title = {{Probing the Innermost Regions of AGN Jets and Their Magnetic Fields with RadioAstron. III. Blazar S5 0716+71 at Microarcsecond Resolution}},
 volume = {893},
 year = {2020}
}

@article{2021ApJ...910L..12E,
 adsurl = {https://ui.adsabs.harvard.edu/abs/2021ApJ...910L..12E},
 archiveprefix = {arXiv},
 author = {{Event Horizon Telescope Collaboration} and {Akiyama}, Kazunori and {Algaba}, Juan Carlos and {Alberdi}, Antxon and {Alef}, Walter and {Anantua}, Richard and {Asada}, Keiichi and {Azulay}, Rebecca and {Baczko}, Anne-Kathrin and {Ball}, David and {Balokovi{\'c}}, Mislav and {Barrett}, John and {Benson}, Bradford A. and {Bintley}, Dan and {Blackburn}, Lindy and {Blundell}, Raymond and {Boland}, Wilfred and {Bouman}, Katherine L. and {Bower}, Geoffrey C. and {Boyce}, Hope and {Bremer}, Michael and {Brinkerink}, Christiaan D. and {Brissenden}, Roger and {Britzen}, Silke and {Broderick}, Avery E. and {Broguiere}, Dominique and {Bronzwaer}, Thomas and {Byun}, Do-Young and {Carlstrom}, John E. and {Chael}, Andrew and {Chan}, Chi-kwan and {Chatterjee}, Shami and {Chatterjee}, Koushik and {Chen}, Ming-Tang and {Chen}, Yongjun and {Chesler}, Paul M. and {Cho}, Ilje and {Christian}, Pierre and {Conway}, John E. and {Cordes}, James M. and {Crawford}, Thomas M. and {Crew}, Geoffrey B. and {Cruz-Osorio}, Alejandro and {Cui}, Yuzhu and {Davelaar}, Jordy and {De Laurentis}, Mariafelicia and {Deane}, Roger and {Dempsey}, Jessica and {Desvignes}, Gregory and {Dexter}, Jason and {Doeleman}, Sheperd S. and {Eatough}, Ralph P. and {Falcke}, Heino and {Farah}, Joseph and {Fish}, Vincent L. and {Fomalont}, Ed and {Ford}, H. Alyson and {Fraga-Encinas}, Raquel and {Freeman}, William T. and {Friberg}, Per and {Fromm}, Christian M. and {Fuentes}, Antonio and {Galison}, Peter and {Gammie}, Charles F. and {Garc{\'\i}a}, Roberto and {Gentaz}, Olivier and {Georgiev}, Boris and {Goddi}, Ciriaco and {Gold}, Roman and {G{\'o}mez}, Jos{\'e} L. and {G{\'o}mez-Ruiz}, Arturo I. and {Gu}, Minfeng and {Gurwell}, Mark and {Hada}, Kazuhiro and {Haggard}, Daryl and {Hecht}, Michael H. and {Hesper}, Ronald and {Ho}, Luis C. and {Ho}, Paul and {Honma}, Mareki and {Huang}, Chih-Wei L. and {Huang}, Lei and {Hughes}, David H. and {Ikeda}, Shiro and {Inoue}, Makoto and {Issaoun}, Sara and {James}, David J. and {Jannuzi}, Buell T. and {Janssen}, Michael and {Jeter}, Britton and {Jiang}, Wu and {Jimenez-Rosales}, Alejandra and {Johnson}, Michael D. and {Jorstad}, Svetlana and {Jung}, Taehyun and {Karami}, Mansour and {Karuppusamy}, Ramesh and {Kawashima}, Tomohisa and {Keating}, Garrett K. and {Kettenis}, Mark and {Kim}, Dong-Jin and {Kim}, Jae-Young and {Kim}, Jongsoo and {Kim}, Junhan and {Kino}, Motoki and {Koay}, Jun Yi and {Kofuji}, Yutaro and {Koch}, Patrick M. and {Koyama}, Shoko and {Kramer}, Michael and {Kramer}, Carsten and {Krichbaum}, Thomas P. and {Kuo}, Cheng-Yu and {Lauer}, Tod R. and {Lee}, Sang-Sung and {Levis}, Aviad and {Li}, Yan-Rong and {Li}, Zhiyuan and {Lindqvist}, Michael and {Lico}, Rocco and {Lindahl}, Greg and {Liu}, Jun and {Liu}, Kuo and {Liuzzo}, Elisabetta and {Lo}, Wen-Ping and {Lobanov}, Andrei P. and {Loinard}, Laurent and {Lonsdale}, Colin and {Lu}, Ru-Sen and {MacDonald}, Nicholas R. and {Mao}, Jirong and {Marchili}, Nicola and {Markoff}, Sera and {Marrone}, Daniel P. and {Marscher}, Alan P. and {Mart{\'\i}-Vidal}, Iv{\'a}n and {Matsushita}, Satoki and {Matthews}, Lynn D. and {Medeiros}, Lia and {Menten}, Karl M. and {Mizuno}, Izumi and {Mizuno}, Yosuke and {Moran}, James M. and {Moriyama}, Kotaro and {Moscibrodzka}, Monika and {M{\"u}ller}, Cornelia and {Musoke}, Gibwa and {Mej{\'\i}as}, Alejandro Mus and {Michalik}, Daniel and {Nadolski}, Andrew and {Nagai}, Hiroshi and {Nagar}, Neil M. and {Nakamura}, Masanori and {Narayan}, Ramesh and {Narayanan}, Gopal and {Natarajan}, Iniyan and {Nathanail}, Antonios and {Neilsen}, Joey and {Neri}, Roberto and {Ni}, Chunchong and {Noutsos}, Aristeidis and {Nowak}, Michael A. and {Okino}, Hiroki and {Olivares}, H{\'e}ctor and {Ortiz-Le{\'o}n}, Gisela N. and {Oyama}, Tomoaki and {{\"O}zel}, Feryal and {Palumbo}, Daniel C.~M. and {Park}, Jongho and {Patel}, Nimesh and {Pen}, Ue-Li and {Pesce}, Dominic W. and {Pi{\'e}tu}, Vincent and {Plambeck}, Richard and {PopStefanija}, Aleksandar and {Porth}, Oliver and {P{\"o}tzl}, Felix M. and {Prather}, Ben and {Preciado-L{\'o}pez}, Jorge A. and {Psaltis}, Dimitrios and {Pu}, Hung-Yi and {Ramakrishnan}, Venkatessh and {Rao}, Ramprasad and {Rawlings}, Mark G. and {Raymond}, Alexander W. and {Rezzolla}, Luciano and {Ricarte}, Angelo and {Ripperda}, Bart and {Roelofs}, Freek and {Rogers}, Alan and {Ros}, Eduardo and {Rose}, Mel and {Roshanineshat}, Arash and {Rottmann}, Helge and {Roy}, Alan L. and {Ruszczyk}, Chet and {Rygl}, Kazi L.~J. and {S{\'a}nchez}, Salvador and {S{\'a}nchez-Arguelles}, David and {Sasada}, Mahito and {Savolainen}, Tuomas and {Schloerb}, F. Peter and {Schuster}, Karl-Friedrich and {Shao}, Lijing and {Shen}, Zhiqiang and {Small}, Des and {Sohn}, Bong Won and {SooHoo}, Jason and {Sun}, He and {Tazaki}, Fumie and {Tetarenko}, Alexandra J. and {Tiede}, Paul and {Tilanus}, Remo P.~J. and {Titus}, Michael and {Toma}, Kenji and {Torne}, Pablo and {Trent}, Tyler and {Traianou}, Efthalia and {Trippe}, Sascha and {van Bemmel}, Ilse and {van Langevelde}, Huib Jan and {van Rossum}, Daniel R. and {Wagner}, Jan and {Ward-Thompson}, Derek and {Wardle}, John and {Weintroub}, Jonathan and {Wex}, Norbert and {Wharton}, Robert and {Wielgus}, Maciek and {Wong}, George N. and {Wu}, Qingwen and {Yoon}, Doosoo and {Young}, Andr{\'e} and {Young}, Ken and {Younsi}, Ziri and {Yuan}, Feng and {Yuan}, Ye-Fei and {Zensus}, J. Anton and {Zhao}, Guang-Yao and {Zhao}, Shan-Shan},
 doi = {10.3847/2041-8213/abe71d},
 eid = {L12},
 eprint = {2105.01169},
 journal = {\apjl},
 month = {March},
 number = {1},
 pages = {L12},
 primaryclass = {astro-ph.HE},
 title = {{First M87 Event Horizon Telescope Results. VII. Polarization of the Ring}},
 volume = {910},
 year = {2021}
}

@article{2021ApJ...923L...5P,
 adsurl = {https://ui.adsabs.harvard.edu/abs/2021ApJ...923L...5P},
 archiveprefix = {arXiv},
 author = {{Pasetto}, Alice and {Carrasco-Gonz{\'a}lez}, Carlos and {G{\'o}mez}, Jos{\'e} L. and {Mart{\'\i}}, Jos{\'e}-Maria and {Perucho}, Manel and {O'Sullivan}, Shane P. and {Anderson}, Craig and {D{\'\i}az-Gonz{\'a}lez}, Daniel Jacobo and {Fuentes}, Antonio and {Wardle}, John},
 doi = {10.3847/2041-8213/ac3a88},
 eid = {L5},
 eprint = {2112.06971},
 journal = {\apjl},
 month = {December},
 number = {1},
 pages = {L5},
 primaryclass = {astro-ph.GA},
 title = {{Reading M87's DNA: A Double Helix Revealing a Large-scale Helical Magnetic Field}},
 volume = {923},
 year = {2021}
}

@article{2022ApJ...939...83K,
 adsurl = {https://ui.adsabs.harvard.edu/abs/2022ApJ...939...83K},
 archiveprefix = {arXiv},
 author = {{Kino}, Motoki and {Takahashi}, Masaaki and {Kawashima}, Tomohisa and {Park}, Jongho and {Hada}, Kazuhiro and {Ro}, Hyunwook and {Cui}, Yuzhu},
 doi = {10.3847/1538-4357/ac8c2f},
 eid = {83},
 eprint = {2209.07264},
 journal = {\apj},
 month = {November},
 number = {2},
 pages = {83},
 primaryclass = {astro-ph.HE},
 title = {{Implications from the Velocity Profile of the M87 Jet: A Possibility of a Slowly Rotating Black Hole Magnetosphere}},
 volume = {939},
 year = {2022}
}

@article{2022NatAs.tmp..261A,
 adsurl = {https://ui.adsabs.harvard.edu/abs/2022NatAs...6..259A},
 archiveprefix = {arXiv},
 author = {{Arras}, Philipp and {Frank}, Philipp and {Haim}, Philipp and {Knollm{\"u}ller}, Jakob and {Leike}, Reimar and {Reinecke}, Martin and {En{\ss}lin}, Torsten},
 doi = {10.1038/s41550-021-01548-0},
 eprint = {2002.05218},
 journal = {Nature Astronomy},
 month = {January},
 pages = {259-269},
 primaryclass = {astro-ph.IM},
 title = {{Variable structures in M87* from space, time and frequency resolved interferometry}},
 volume = {6},
 year = {2022}
}

@article{2023A&A...673A.159R,
 adsurl = {https://ui.adsabs.harvard.edu/abs/2023A&A...673A.159R},
 archiveprefix = {arXiv},
 author = {{Ro}, Hyunwook and {Kino}, Motoki and {Sohn}, Bong Won and {Hada}, Kazuhiro and {Park}, Jongho and {Nakamura}, Masanori and {Cui}, Yuzhu and {Yi}, Kunwoo and {Chung}, Aeree and {Hodgson}, Jeffrey and {Kawashima}, Tomohisa and {An}, Tao and {Trippe}, Sascha and {Algaba}, Juan-Carlos and {Kim}, Jae-Young and {Sawada-Satoh}, Satoko and {Wajima}, Kiyoaki and {Shen}, Zhiqiang and {Cheng}, Xiaopeng and {Cho}, Ilje and {Jiang}, Wu and {Jung}, Taehyun and {Lee}, Jee-Won and {Niinuma}, Kotaro and {Oh}, Junghwan and {Tazaki}, Fumie and {Zhao}, Guang-Yao and {Akiyama}, Kazunori and {Honma}, Mareki and {Lee}, Jeong Ae and {Lu}, Rusen and {Zhang}, Yingkang and {Asada}, Keiichi and {Cui}, Lang and {Hagiwara}, Yoshiaki and {Hirota}, Tomoya and {Kawaguchi}, Noriyuki and {Koyama}, Shoko and {Lee}, Sang-Sung and {Oh}, Se-Jin and {Sugiyama}, Koichiro and {Takamura}, Mieko and {Wang}, Xuezheng and {Hwang}, Ju-Yeon and {Jung}, Dong-Kyu and {Kim}, Hyo-Ryoung and {Kim}, Jeong-Sook and {Kobayashi}, Hideyuki and {Oh}, Chung-Sik and {Oyama}, Tomoaki and {Roh}, Duk-Gyoo and {Yeom}, Jae-Hwan},
 doi = {10.1051/0004-6361/202142988},
 eid = {A159},
 eprint = {2303.01014},
 journal = {\aap},
 month = {May},
 pages = {A159},
 primaryclass = {astro-ph.HE},
 title = {{Spectral analysis of a parsec-scale jet in M 87: Observational constraint on the magnetic field strengths in the jet}},
 volume = {673},
 year = {2023}
}

@article{2023A&A...676A.114S,
 adsurl = {https://ui.adsabs.harvard.edu/abs/2023A&A...676A.114S},
 archiveprefix = {arXiv},
 author = {{Savolainen}, T. and {Giovannini}, G. and {Kovalev}, Y.~Y. and {Perucho}, M. and {Anderson}, J.~M. and {Bruni}, G. and {Edwards}, P.~G. and {Fuentes}, A. and {Giroletti}, M. and {G{\'o}mez}, J.~L. and {Hada}, K. and {Lee}, S. -S. and {Lisakov}, M.~M. and {Lobanov}, A.~P. and {L{\'o}pez-Miralles}, J. and {Orienti}, M. and {Petrov}, L. and {Plavin}, A.~V. and {Sohn}, B.~W. and {Sokolovsky}, K.~V. and {Voitsik}, P.~A. and {Zensus}, J.~A.},
 doi = {10.1051/0004-6361/202142594},
 eid = {A114},
 eprint = {2111.04481},
 journal = {\aap},
 month = {August},
 pages = {A114},
 primaryclass = {astro-ph.HE},
 title = {{RadioAstron discovery of a mini-cocoon around the restarted parsec-scale jet in 3C 84}},
 volume = {676},
 year = {2023}
}

@article{2023ApJ...952...34K,
 adsurl = {https://ui.adsabs.harvard.edu/abs/2023ApJ...952...34K},
 archiveprefix = {arXiv},
 author = {{Kim}, Jae-Young and {Savolainen}, Tuomas and {Voitsik}, Petr and {Kravchenko}, Evgeniya V. and {Lisakov}, Mikhail M. and {Kovalev}, Yuri Y. and {M{\"u}ller}, Hendrik and {Lobanov}, Andrei P. and {Sokolovsky}, Kirill V. and {Bruni}, Gabriele and {Edwards}, Philip G. and {Reynolds}, Cormac and {Bach}, Uwe and {Gurvits}, Leonid I. and {Krichbaum}, Thomas P. and {Hada}, Kazuhiro and {Giroletti}, Marcello and {Orienti}, Monica and {Anderson}, James M. and {Lee}, Sang-Sung and {Sohn}, Bong Won and {Zensus}, J. Anton},
 doi = {10.3847/1538-4357/accf17},
 eid = {34},
 eprint = {2304.09816},
 journal = {\apj},
 month = {July},
 note = {Paper~I},
 number = {1},
 pages = {34},
 primaryclass = {astro-ph.GA},
 shorthand = {Paper~I},
 title = {{RadioAstron Space VLBI Imaging of the Jet in M87. I. Detection of High Brightness Temperature at 22 GHz} - {Paper I}},
 volume = {952},
 year = {Paper~I, 2023}
}

@article{2023ARep...67.1275T,
 adsurl = {https://ui.adsabs.harvard.edu/abs/2023ARep...67.1275T},
 archiveprefix = {arXiv},
 author = {{Todorov}, R.~V. and {Kravchenko}, E.~V. and {Pashchenko}, I.~N. and {Pushkarev}, A.~B.},
 doi = {10.1134/S1063772923120119},
 eprint = {2402.07500},
 journal = {Astronomy Reports},
 month = {December},
 number = {12},
 pages = {1275-1285},
 primaryclass = {astro-ph.HE},
 title = {{Simulations of Linear Polarization of Precessing AGN Jets at Parsec Scales}},
 volume = {67},
 year = {2023}
}

@article{2023Galax..11...33R,
 adsurl = {https://ui.adsabs.harvard.edu/abs/2023Galax..11...33R},
 archiveprefix = {arXiv},
 author = {{Ro}, Hyunwook and {Yi}, Kunwoo and {Cui}, Yuzhu and {Kino}, Motoki and {Hada}, Kazuhiro and {Kawashima}, Tomohisa and {Mizuno}, Yosuke and {Sohn}, Bong Won and {Tazaki}, Fumie},
 doi = {10.3390/galaxies11010033},
 eid = {33},
 eprint = {2303.01106},
 journal = {Galaxies},
 month = {February},
 number = {1},
 pages = {33},
 primaryclass = {astro-ph.HE},
 title = {{Transverse Oscillations of the M87 Jet Revealed by KaVA Observations}},
 volume = {11},
 year = {2023}
}

@article{2023Galax..11...39T,
 adsurl = {https://ui.adsabs.harvard.edu/abs/2023Galax..11...39T},
 archiveprefix = {arXiv},
 author = {{Tazaki}, Fumie and {Cui}, Yuzhu and {Hada}, Kazuhiro and {Kino}, Motoki and {Cho}, Ilje and {Zhao}, Guang-Yao and {Akiyama}, Kazunori and {Mizuno}, Yosuke and {Ro}, Hyunwook and {Honma}, Mareki and {Lu}, Ru-Sen and {Shen}, Zhi-Qiang and {Cui}, Lang and {Yonekura}, Yoshinori},
 doi = {10.3390/galaxies11020039},
 eid = {39},
 eprint = {2303.01048},
 journal = {Galaxies},
 month = {February},
 number = {2},
 pages = {39},
 primaryclass = {astro-ph.HE},
 title = {{Super-Resolved Image of M87 Observed with East Asian VLBI Network}},
 volume = {11},
 year = {2023}
}

@article{2023MNRAS.523..887F,
 adsurl = {https://ui.adsabs.harvard.edu/abs/2023MNRAS.523..887F},
 archiveprefix = {arXiv},
 author = {{Frolova}, V.~A. and {Nokhrina}, E.~E. and {Pashchenko}, I.~N.},
 doi = {10.1093/mnras/stad1381},
 eprint = {2305.02929},
 journal = {\mnras},
 month = {July},
 number = {1},
 pages = {887-906},
 primaryclass = {astro-ph.HE},
 title = {{Synchrotron intensity plots from a relativistic stratified jet}},
 volume = {523},
 year = {2023}
}

@article{2023MNRAS.523.1247P,
 adsurl = {https://ui.adsabs.harvard.edu/abs/2023MNRAS.523.1247P},
 archiveprefix = {arXiv},
 author = {{Pashchenko}, I.~N. and {Kravchenko}, E.~V. and {Nokhrina}, E.~E. and {Nikonov}, A.~S.},
 doi = {10.1093/mnras/stad1527},
 eprint = {2301.12861},
 journal = {\mnras},
 month = {July},
 number = {1},
 pages = {1247-1267},
 primaryclass = {astro-ph.GA},
 title = {{CLEAN imaging systematics of M87 radio jet}},
 volume = {523},
 year = {2023}
}

@article{2023MNRAS.526.5949N,
 adsurl = {https://ui.adsabs.harvard.edu/abs/2023MNRAS.526.5949N},
 archiveprefix = {arXiv},
 author = {{Nikonov}, A.~S. and {Kovalev}, Y.~Y. and {Kravchenko}, E.~V. and {Pashchenko}, I.~N. and {Lobanov}, A.~P.},
 doi = {10.1093/mnras/stad3061},
 eprint = {2307.11660},
 journal = {\mnras},
 month = {December},
 number = {4},
 pages = {5949-5963},
 primaryclass = {astro-ph.GA},
 title = {{Properties of the jet in M87 revealed by its helical structure imaged with the VLBA at 8 and 15 GHz}},
 volume = {526},
 year = {2023}
}

@article{2023NatAs...7.1359F,
 adsurl = {https://ui.adsabs.harvard.edu/abs/2023NatAs...7.1359F},
 archiveprefix = {arXiv},
 author = {{Fuentes}, Antonio and {G{\'o}mez}, Jos{\'e} L. and {Mart{\'\i}}, Jos{\'e} M. and {Perucho}, Manel and {Zhao}, Guang-Yao and {Lico}, Rocco and {Lobanov}, Andrei P. and {Bruni}, Gabriele and {Kovalev}, Yuri Y. and {Chael}, Andrew and {Akiyama}, Kazunori and {Bouman}, Katherine L. and {Sun}, He and {Cho}, Ilje and {Traianou}, Efthalia and {Toscano}, Teresa and {Dahale}, Rohan and {Foschi}, Marianna and {Gurvits}, Leonid I. and {Jorstad}, Svetlana and {Kim}, Jae-Young and {Marscher}, Alan P. and {Mizuno}, Yosuke and {Ros}, Eduardo and {Savolainen}, Tuomas},
 doi = {10.1038/s41550-023-02105-7},
 eprint = {2311.01861},
 journal = {Nature Astronomy},
 month = {November},
 pages = {1359-1367},
 primaryclass = {astro-ph.HE},
 title = {{Filamentary structures as the origin of blazar jet radio variability}},
 volume = {7},
 year = {2023}
}

@article{2023Natur.616..686L,
 adsurl = {https://ui.adsabs.harvard.edu/abs/2023Natur.616..686L},
 archiveprefix = {arXiv},
 author = {{Lu}, Ru-Sen and {Asada}, Keiichi and {Krichbaum}, Thomas P. and {Park}, Jongho and {Tazaki}, Fumie and {Pu}, Hung-Yi and {Nakamura}, Masanori and {Lobanov}, Andrei and {Hada}, Kazuhiro and {Akiyama}, Kazunori and {Kim}, Jae-Young and {Marti-Vidal}, Ivan and {G{\'o}mez}, Jos{\'e} L. and {Kawashima}, Tomohisa and {Yuan}, Feng and {Ros}, Eduardo and {Alef}, Walter and {Britzen}, Silke and {Bremer}, Michael and {Broderick}, Avery E. and {Doi}, Akihiro and {Giovannini}, Gabriele and {Giroletti}, Marcello and {Ho}, Paul T.~P. and {Honma}, Mareki and {Hughes}, David H. and {Inoue}, Makoto and {Jiang}, Wu and {Kino}, Motoki and {Koyama}, Shoko and {Lindqvist}, Michael and {Liu}, Jun and {Marscher}, Alan P. and {Matsushita}, Satoki and {Nagai}, Hiroshi and {Rottmann}, Helge and {Savolainen}, Tuomas and {Schuster}, Karl-Friedrich and {Shen}, Zhi-Qiang and {de Vicente}, Pablo and {Walker}, R. Craig and {Yang}, Hai and {Zensus}, J. Anton and {Algaba}, Juan Carlos and {Allardi}, Alexander and {Bach}, Uwe and {Berthold}, Ryan and {Bintley}, Dan and {Byun}, Do-Young and {Casadio}, Carolina and {Chang}, Shu-Hao and {Chang}, Chih-Cheng and {Chang}, Song-Chu and {Chen}, Chung-Chen and {Chen}, Ming-Tang and {Chilson}, Ryan and {Chuter}, Tim C. and {Conway}, John and {Crew}, Geoffrey B. and {Dempsey}, Jessica T. and {Dornbusch}, Sven and {Faber}, Aaron and {Friberg}, Per and {Garc{\'\i}a}, Javier Gonz{\'a}lez and {Garrido}, Miguel G{\'o}mez and {Han}, Chih-Chiang and {Han}, Kuo-Chang and {Hasegawa}, Yutaka and {Herrero-Illana}, Ruben and {Huang}, Yau-De and {Huang}, Chih-Wei L. and {Impellizzeri}, Violette and {Jiang}, Homin and {Jinchi}, Hao and {Jung}, Taehyun and {Kallunki}, Juha and {Kirves}, Petri and {Kimura}, Kimihiro and {Koay}, Jun Yi and {Koch}, Patrick M. and {Kramer}, Carsten and {Kraus}, Alex and {Kubo}, Derek and {Kuo}, Cheng-Yu and {Li}, Chao-Te and {Lin}, Lupin Chun-Che and {Liu}, Ching-Tang and {Liu}, Kuan-Yu and {Lo}, Wen-Ping and {Lu}, Li-Ming and {MacDonald}, Nicholas and {Martin-Cocher}, Pierre and {Messias}, Hugo and {Meyer-Zhao}, Zheng and {Minter}, Anthony and {Nair}, Dhanya G. and {Nishioka}, Hiroaki and {Norton}, Timothy J. and {Nystrom}, George and {Ogawa}, Hideo and {Oshiro}, Peter and {Patel}, Nimesh A. and {Pen}, Ue-Li and {Pidopryhora}, Yurii and {Pradel}, Nicolas and {Raffin}, Philippe A. and {Rao}, Ramprasad and {Ruiz}, Ignacio and {Sanchez}, Salvador and {Shaw}, Paul and {Snow}, William and {Sridharan}, T.~K. and {Srinivasan}, Ranjani and {Tercero}, Bel{\'e}n and {Torne}, Pablo and {Traianou}, Efthalia and {Wagner}, Jan and {Walther}, Craig and {Wei}, Ta-Shun and {Yang}, Jun and {Yu}, Chen-Yu},
 doi = {10.1038/s41586-023-05843-w},
 eprint = {2304.13252},
 journal = {\nat},
 month = {April},
 number = {7958},
 pages = {686-690},
 primaryclass = {astro-ph.HE},
 title = {{A ring-like accretion structure in M87 connecting its black hole and jet}},
 volume = {616},
 year = {2023}
}

@article{2023Natur.621..711C,
 adsurl = {https://ui.adsabs.harvard.edu/abs/2023Natur.621..711C},
 archiveprefix = {arXiv},
 author = {{Cui}, Yuzhu and {Hada}, Kazuhiro and {Kawashima}, Tomohisa and {Kino}, Motoki and {Lin}, Weikang and {Mizuno}, Yosuke and {Ro}, Hyunwook and {Honma}, Mareki and {Yi}, Kunwoo and {Yu}, Jintao and {Park}, Jongho and {Jiang}, Wu and {Shen}, Zhiqiang and {Kravchenko}, Evgeniya and {Algaba}, Juan-Carlos and {Cheng}, Xiaopeng and {Cho}, Ilje and {Giovannini}, Gabriele and {Giroletti}, Marcello and {Jung}, Taehyun and {Lu}, Ru-Sen and {Niinuma}, Kotaro and {Oh}, Junghwan and {Ohsuga}, Ken and {Sawada-Satoh}, Satoko and {Sohn}, Bong Won and {Takahashi}, Hiroyuki R. and {Takamura}, Mieko and {Tazaki}, Fumie and {Trippe}, Sascha and {Wajima}, Kiyoaki and {Akiyama}, Kazunori and {An}, Tao and {Asada}, Keiichi and {Buttaccio}, Salvatore and {Byun}, Do-Young and {Cui}, Lang and {Hagiwara}, Yoshiaki and {Hirota}, Tomoya and {Hodgson}, Jeffrey and {Kawaguchi}, Noriyuki and {Kim}, Jae-Young and {Lee}, Sang-Sung and {Lee}, Jee Won and {Lee}, Jeong Ae and {Maccaferri}, Giuseppe and {Melis}, Andrea and {Melnikov}, Alexey and {Migoni}, Carlo and {Oh}, Se-Jin and {Sugiyama}, Koichiro and {Wang}, Xuezheng and {Zhang}, Yingkang and {Chen}, Zhong and {Hwang}, Ju-Yeon and {Jung}, Dong-Kyu and {Kim}, Hyo-Ryoung and {Kim}, Jeong-Sook and {Kobayashi}, Hideyuki and {Li}, Bin and {Li}, Guanghui and {Li}, Xiaofei and {Liu}, Zhiyong and {Liu}, Qinghui and {Liu}, Xiang and {Oh}, Chung-Sik and {Oyama}, Tomoaki and {Roh}, Duk-Gyoo and {Wang}, Jinqing and {Wang}, Na and {Wang}, Shiqiang and {Xia}, Bo and {Yan}, Hao and {Yeom}, Jae-Hwan and {Yonekura}, Yoshinori and {Yuan}, Jianping and {Zhang}, Hua and {Zhao}, Rongbing and {Zhong}, Weiye},
 doi = {10.1038/s41586-023-06479-6},
 eprint = {2310.09015},
 journal = {\nat},
 month = {September},
 number = {7980},
 pages = {711-715},
 primaryclass = {astro-ph.HE},
 title = {{Precessing jet nozzle connecting to a spinning black hole in M87}},
 volume = {621},
 year = {2023}
}

@article{2024MNRAS.528.6046B,
 adsurl = {https://ui.adsabs.harvard.edu/abs/2024MNRAS.528.6046B},
 archiveprefix = {arXiv},
 author = {{Beskin}, V.~S. and {Khalilov}, T.~I. and {Nokhrina}, E.~E. and {Pashchenko}, I.~N. and {Kravchenko}, E.~V.},
 doi = {10.1093/mnras/stae447},
 eprint = {2403.18366},
 journal = {\mnras},
 month = {March},
 number = {4},
 pages = {6046-6055},
 primaryclass = {astro-ph.GA},
 title = {{On the M87 jet structure near the central engine}},
 volume = {528},
 year = {2024}
}

@article{2025ApJ...988...28W,
 adsurl = {https://ui.adsabs.harvard.edu/abs/2025ApJ...988...28W},
 archiveprefix = {arXiv},
 author = {{Wu}, Linhui and {Xie}, Fu-Guo and {Zheng}, Qian and {Guo}, Quan and {Shan}, Huanyuan and {Hu}, Dan and {Duchesne}, Stefan W. and {Seymour}, Nick and {Wang}, Jingying and {Gu}, Junhua and {Wu}, Qingwen and {Zhu}, Zhenghao and {Johnston-Hollitt}, Melanie and {Riseley}, Christopher J. and {Fan}, Xu-Liang},
 doi = {10.3847/1538-4357/adddb7},
 eid = {28},
 eprint = {2505.21929},
 journal = {\apj},
 month = {July},
 number = {1},
 pages = {28},
 primaryclass = {astro-ph.GA},
 title = {{MWA and VLA Observations of Diffuse Radio Lobes in M87}},
 volume = {988},
 year = {2025}
}

@article{2026A&A...706A..27S,
 adsurl = {https://ui.adsabs.harvard.edu/abs/2026A&A...706A..27S},
 archiveprefix = {arXiv},
 author = {{Saurabh} and {M{\"u}ller}, Hendrik and {von Fellenberg}, Sebastiano D. and {Tiede}, Paul and {Janssen}, Michael and {Blackburn}, Lindy and {Broderick}, Avery E. and {Chavez}, Erandi and {Georgiev}, Boris and {Krichbaum}, Thomas P. and {Moriyama}, Kotaro and {Nair}, Dhanya G. and {Natarajan}, Iniyan and {Park}, Jongho and {West}, Andrew Thomas and {Wielgus}, Maciek and {Akiyama}, Kazunori and {Albentosa-Ru{\'\i}z}, Ezequiel and {Alberdi}, Antxon and {Alef}, Walter and {Algaba}, Juan Carlos and {Anantua}, Richard and {Asada}, Keiichi and {Azulay}, Rebecca and {Bach}, Uwe and {Baczko}, Anne-Kathrin and {Ball}, David and {Balokovi{\'c}}, Mislav and {Bandyopadhyay}, Bidisha and {Barrett}, John and {Baub{\"o}ck}, Michi and {Benson}, Bradford A. and {Bintley}, Dan and {Blundell}, Raymond and {Bouman}, Katherine L. and {Bower}, Geoffrey C. and {Bremer}, Michael and {Brissenden}, Roger and {Britzen}, Silke and {Broguiere}, Dominique and {Bronzwaer}, Thomas and {Bustamante}, Sandra and {Carlos}, Douglas F. and {Carlstrom}, John E. and {Chael}, Andrew and {Chan}, Chi-kwan and {Chang}, Dominic O. and {Chatterjee}, Koushik and {Chatterjee}, Shami and {Chen}, Ming-Tang and {Chen}, Yongjun and {Cheng}, Xiaopeng and {Chichura}, Paul and {Cho}, Ilje and {Christian}, Pierre and {Conroy}, Nicholas S. and {Conway}, John E. and {Crawford}, Thomas M. and {Crew}, Geoffrey B. and {Cruz-Osorio}, Alejandro and {Cui}, Yuzhu and {Curd}, Brandon and {Dahale}, Rohan and {Davelaar}, Jordy and {De Laurentis}, Mariafelicia and {Deane}, Roger and {Desvignes}, Gregory and {Dexter}, Jason and {Dhruv}, Vedant and {Dihingia}, Indu K. and {Doeleman}, Sheperd S. and {Dzib}, Sergio A. and {Eatough}, Ralph P. and {Emami}, Razieh and {Falcke}, Heino and {Farah}, Joseph and {Fish}, Vincent L. and {Fomalont}, Edward and {Alyson Ford}, H. and {Foschi}, Marianna and {Fraga-Encinas}, Raquel and {Freeman}, William T. and {Friberg}, Per and {Fromm}, Christian M. and {Fuentes}, Antonio and {Galison}, Peter and {Gammie}, Charles F. and {Garc{\'\i}a}, Roberto and {Gentaz}, Olivier and {Goddi}, Ciriaco and {Gold}, Roman and {G{\'o}mez-Ruiz}, Arturo I. and {G{\'o}mez}, Jos{\'e} L. and {Gu}, Minfeng and {Gurwell}, Mark and {Hada}, Kazuhiro and {Haggard}, Daryl and {Hesper}, Ronald and {Heumann}, Dirk and {Ho}, Luis C. and {Ho}, Paul and {Honma}, Mareki and {Huang}, Chih-Wei L. and {Huang}, Lei and {Hughes}, David H. and {Ikeda}, Shiro and {Violette Impellizzeri}, C.~M. and {Inoue}, Makoto and {Issaoun}, Sara and {James}, David J. and {Jannuzi}, Buell T. and {Jeter}, Britton and {Jiang}, Wu and {Jim{\'e}nez-Rosales}, Alejandra and {Johnson}, Michael D. and {Jorstad}, Svetlana and {Jones}, Adam C. and {Joshi}, Abhishek V. and {Jung}, Taehyun and {Karuppusamy}, Ramesh and {Kawashima}, Tomohisa and {Keating}, Garrett K. and {Kettenis}, Mark and {Kim}, Dong-Jin and {Kim}, Jae-Young and {Kim}, Jongsoo and {Kim}, Junhan and {Kino}, Motoki and {Koay}, Jun Yi and {Kocherlakota}, Prashant and {Kofuji}, Yutaro and {Koch}, Patrick M. and {Koyama}, Shoko and {Kramer}, Carsten and {Kramer}, Joana A. and {Kramer}, Michael and {Kuo}, Cheng-Yu and {La Bella}, Noemi and {Lee}, Deokhyeong and {Lee}, Sang-Sung and {Levis}, Aviad and {Li}, Shaoliang and {Li}, Zhiyuan and {Lico}, Rocco and {Lindahl}, Greg and {Lindqvist}, Michael and {Lisakov}, Mikhail and {Liu}, Jun and {Liu}, Kuo and {Liuzzo}, Elisabetta and {Lo}, Wen-Ping and {Lobanov}, Andrei P. and {Loinard}, Laurent and {Lonsdale}, Colin J. and {Lowitz}, Amy E. and {Lu}, Ru-Sen and {MacDonald}, Nicholas R. and {Mao}, Jirong and {Marchili}, Nicola and {Markoff}, Sera and {Marrone}, Daniel P. and {Marscher}, Alan P. and {Mart{\'\i}-Vidal}, Iv{\'a}n and {Matsushita}, Satoki and {Matthews}, Lynn D. and {Medeiros}, Lia and {Menten}, Karl M. and {Messias}, Hugo and {Mizuno}, Izumi and {Mizuno}, Yosuke and {Montgomery}, Joshua and {Moran}, James M. and {Moscibrodzka}, Monika and {Mulaudzi}, Wanga and {M{\"u}ller}, Cornelia and {Mus}, Alejandro and {Musoke}, Gibwa and {Myserlis}, Ioannis and {Nagai}, Hiroshi and {Nagar}, Neil M. and {Nakamura}, Masanori and {Narayanan}, Gopal and {Nathanail}, Antonios and {Fuentes}, Santiago Navarro and {Neilsen}, Joey and {Ni}, Chunchong and {Nowak}, Michael A. and {Oh}, Junghwan and {Okino}, Hiroki and {S{\'a}nchez}, H{\'e}ctor Ra{\'u}l Olivares and {Oyama}, Tomoaki and {{\"O}zel}, Feryal and {Palumbo}, Daniel C.~M. and {Paraschos}, Georgios Filippos and {Parsons}, Harriet and {Patel}, Nimesh and {Pen}, Ue-Li and {Pesce}, Dominic W. and {Pi{\'e}tu}, Vincent and {Plavin}, Alexander},
 doi = {10.1051/0004-6361/202557022},
 eid = {A27},
 eprint = {2512.08970},
 journal = {\aap},
 month = {January},
 pages = {A27},
 primaryclass = {astro-ph.HE},
 title = {{Probing jet base emission of M87* with the 2021 Event Horizon Telescope observations}},
 volume = {706},
 year = {2026}
}

@article{2026ApJ...999..169R,
 adsurl = {https://ui.adsabs.harvard.edu/abs/2026ApJ...999..169R},
 archiveprefix = {arXiv},
 author = {{Ro}, Hyunwook and {Kino}, Motoki and {Hada}, Kazuhiro and {Mizuno}, Yosuke and {Cui}, Yuzhu and {Yi}, Kunwoo and {Kawashima}, Tomohisa and {Park}, Jongho and {Sohn}, Bong Won},
 doi = {10.3847/1538-4357/ae355b},
 eid = {169},
 eprint = {2603.02660},
 journal = {\apj},
 month = {March},
 number = {2},
 pages = {169},
 primaryclass = {astro-ph.HE},
 title = {{Transverse Oscillations and Wave Propagation in the Magnetically Dominated M87 Jet}},
 volume = {999},
 year = {2026}
}

@inproceedings{aips,
 adsurl = {https://ui.adsabs.harvard.edu/abs/2003ASSL..285..109G},
 author = {{Greisen}, E.~W.},
 booktitle = {Information Handling in Astronomy - Historical Vistas},
 doi = {10.1007/0-306-48080-8_7},
 editor = {{Heck}, Andr{\'e}},
 month = {March},
 pages = {109},
 series = {Astrophysics and Space Science Library},
 title = {{AIPS, the VLA, and the VLBA}},
 volume = {285},
 year = {2003}
}

@article{Beskin06,
 adsurl = {http://adsabs.harvard.edu/abs/2006MNRAS.367..375B},
 author = {{Beskin}, V.~S. and {Nokhrina}, E.~E.},
 doi = {10.1111/j.1365-2966.2006.09957.x},
 journal = {\mnras},
 month = {March},
 pages = {375-386},
 title = {{The effective acceleration of plasma outflow in the paraboloidal magnetic field}},
 volume = {367},
 year = {2006}
}

@article{CLEAN,
 adsurl = {http://adsabs.harvard.edu/abs/1974A%26AS...15..417H},
 author = {{H{\"o}gbom}, J.~A.},
 journal = {\apjs},
 month = {June},
 pages = {417},
 title = {{Aperture Synthesis with a Non-Regular Distribution of Interferometer Baselines}},
 volume = {15},
 year = {1974}
}

@article{Komissarov_2009,
 adsurl = {http://adsabs.harvard.edu/abs/2009MNRAS.394.1182K},
 archiveprefix = {arXiv},
 author = {{Komissarov}, S.~S. and {Vlahakis}, N. and {K{\"o}nigl}, A. and 
{Barkov}, M.~V.},
 doi = {10.1111/j.1365-2966.2009.14410.x},
 eprint = {0811.1467},
 journal = {\mnras},
 month = {April},
 pages = {1182-1212},
 title = {{Magnetic acceleration of ultrarelativistic jets in gamma-ray burst sources}},
 volume = {394},
 year = {2009}
}

@article{Mizuno12,
 adsurl = {https://ui.adsabs.harvard.edu/abs/2012ApJ...757...16M/abstract},
 author = {{Mizuno}, Y. and {Lyubarsky}, Y. and {Nishikawa}, K.-I. and {Hardee}, P.~E.},
 doi = {10.1088/0004-637X/757/1/16 },
 journal = {\apj},
 month = {September},
 pages = {16},
 title = {{Three-dimensional Relativistic Magnetohydrodynamic Simulations of Current-driven Instability. III. Rotating Relativistic Jets}},
 volume = {757},
 year = {2012}
}

@article{Musoke+2023MNRAS,
 adsurl = {https://ui.adsabs.harvard.edu/abs/2023MNRAS.518.1656M},
 archiveprefix = {arXiv},
 author = {{Musoke}, G. and {Liska}, M. and {Porth}, O. and {van der Klis}, Michiel and {Ingram}, Adam},
 doi = {10.1093/mnras/stac2754},
 eprint = {2201.03085},
 journal = {\mnras},
 month = {January},
 number = {2},
 pages = {1656-1671},
 primaryclass = {astro-ph.HE},
 title = {{Disc tearing leads to low and high frequency quasi-periodic oscillations in a GRMHD simulation of a thin accretion disc}},
 volume = {518},
 year = {2023}
}

@article{N24b,
 adsurl = {https://ui.adsabs.harvard.edu/abs/2024MNRAS.535.2687N/abstract},
 author = {{Nokhrina}, E.~E.},
 doi = {10.1093/mnras/stae2515},
 journal = {\mnras},
 month = {December},
 pages = {2687-2696},
 title = {{On the possible core shift break in relativistic jets}},
 volume = {535},
 year = {2024}
}

@article{Nok17A,
 adsurl = {http://adsabs.harvard.edu/abs/2017MNRAS.468.2372N},
 author = {{Nokhrina}, E.E.},
 doi = {10.1093/mnras/stx521},
 journal = {\mnras},
 month = {June},
 pages = {2372},
 title = {{Brightness temperature --- obtaining physical properties of non-equipartition plasma}},
 volume = {468},
 year = {2017}
}

@article{Nokhrina+2019MNRAS,
 adsurl = {https://ui.adsabs.harvard.edu/abs/2019MNRAS.489.1197N},
 archiveprefix = {arXiv},
 author = {{Nokhrina}, E.~E. and {Gurvits}, L.~I. and {Beskin}, V.~S. and {Nakamura}, M. and {Asada}, K. and {Hada}, K.},
 doi = {10.1093/mnras/stz2116},
 eprint = {1904.05665},
 journal = {\mnras},
 month = {October},
 number = {1},
 pages = {1197-1205},
 primaryclass = {astro-ph.HE},
 title = {{M87 black hole mass and spin estimate through the position of the jet boundary shape break}},
 volume = {489},
 year = {2019}
}

@article{Nokhrina_Pushkarev_2024,
 adsurl = {https://ui.adsabs.harvard.edu/abs/2024MNRAS.528.2523N/abstract},
 author = {{Nokhrina}, E.~E. and {Pushkarev}, A.~B.},
 doi = {10.1093/mnras/stae179},
 journal = {\mnras},
 month = {February},
 pages = {2523-2532},
 title = {{Core shift in parabolic accelerating jets}},
 volume = {528},
 year = {2024}
}

@article{Steinle+2024PhRvD,
 adsurl = {https://ui.adsabs.harvard.edu/abs/2024PhRvD.110l3034S},
 archiveprefix = {arXiv},
 author = {{Steinle}, Nathan and {Gerosa}, Davide and {Krause}, Martin G.~H.},
 doi = {10.1103/PhysRevD.110.123034},
 eid = {123034},
 eprint = {2403.00066},
 journal = {\prd},
 month = {December},
 number = {12},
 pages = {123034},
 primaryclass = {astro-ph.HE},
 title = {{Probing AGN jet precession with LISA}},
 volume = {110},
 year = {2024}
}

\appendix
\section{Resolution and sensitivity estimates}
\label{apdx:snr_maps}

Since the limiting resolution depends on the signal-to-noise ratio (SNR) of detection, we estimate the minimum resolvable size of a component in an image following the convention of \citet{Lobanov_reslims_2005}:
\begin{equation}
    \theta_{\rm lim} = 2^{1-\beta/2} b_{\psi}\big[ \frac{{\rm ln}2}{\pi}{\rm ln}\frac{SNR}{SNR-1}\big]^{1/2},
\label{eq:theta_lim}
\end{equation}
\noindent where $\beta$ denotes the weighting function, which is 0 for natural weighting or 2 for uniform weighting, and $b_{\psi}$ is full width at half-maximum size of the elliptical synthesised beam measured along the direction $\psi$ of the axis for which the limit is calculated.  
The resulting maps of SNR and $\theta_{\rm lim}$ are illustrated in Fig.~\ref{fig:snr_su}.
\begin{figure}
\centering
\includegraphics[width=\columnwidth]{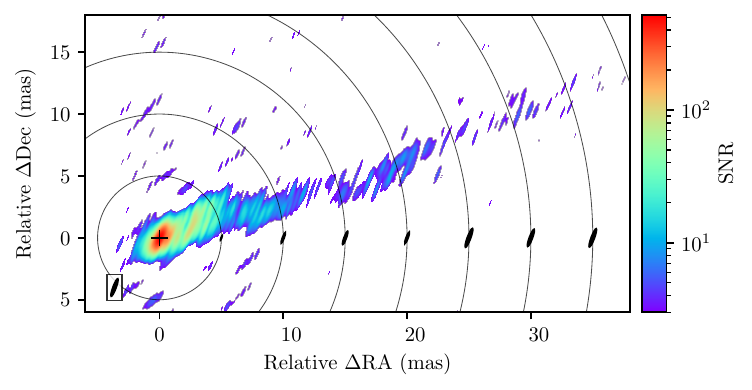}
\caption{The maps of signal-to-noise ratio (color) and of the limiting resolution (ellipses) calculated for the superuniform \textit{RadioAstron} image using eq.~\ref{eq:theta_lim}. Each ellipse size is calculated based on the maximum signal-to-noise in a giving radius.}
\label{fig:snr_su}
\end{figure}

\section{Double-peak substructure 20\,mas from the VLBI core}
\label{apdx:ehtim}

\begin{figure}
    \centering
    \includegraphics[width=\linewidth]{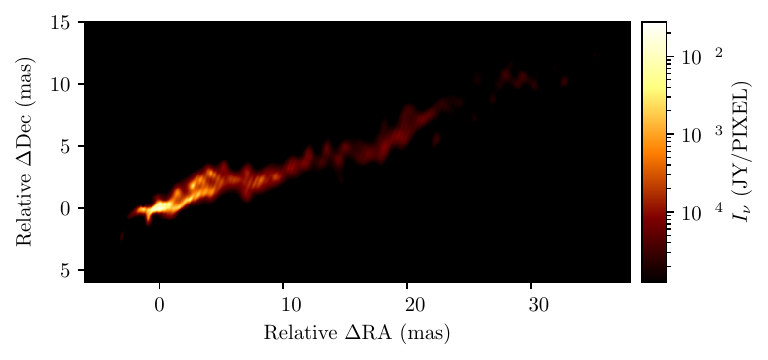}
    \caption{The RML reconstruction obtained with eht-imaging. The image is shown at its native reconstruction resolution, without subsequent Gaussian blurring or restoring-beam convolution. The colour scale represents the specific intensity in Jy pixel\(^{-1}\), while the axes show relative offsets in right ascension and declination in milliarcseconds. The cut-off noise level was adopted from the super-uniform \texttt{CLEAN} image.}
    \label{fig:ehtim-image}
\end{figure}

\begin{figure*}
    \centering
    \includegraphics[width=0.95\textwidth]{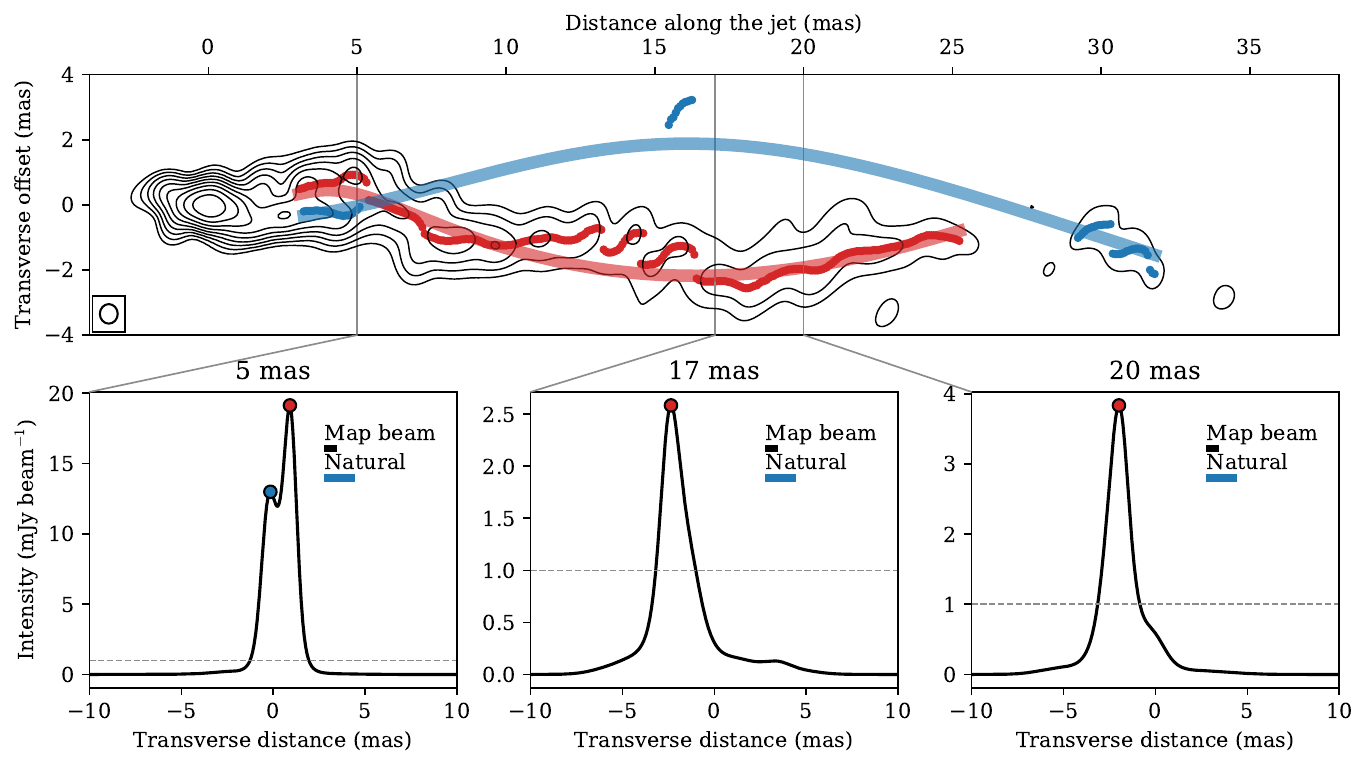}
    \caption{\textit{Top panel:} Regularized maximum-likelihood (RML) reconstruction of the \ra{} Stokes $I$ emission at 4.8~GHz, obtained with \texttt{eht-imaging} from the same visibility data as the \texttt{CLEAN} image shown in the main text. The black contours show the RML intensity distribution, with contour levels increasing by successive factors of $2$ and the lowest contour set to $3\sigma = 1$~mJy~beam$^{-1}$, where $\sigma$ is the rms noise of the corresponding superuniform-weighted \texttt{CLEAN} image. The thick red and blue curves show the best-fitting three-dimensional helical model projected on to the plane of the sky, assuming the M87 jet viewing angle of $\theta = 17^{\circ}$. The coloured markers indicate the intensity peaks identified in the transverse profiles and used to trace the corresponding Kelvin--Helmholtz instability threads. The $0.7$~mas circular beam adopted in the \texttt{CLEAN} analysis is shown in the lower-left corner for reference. As in the \texttt{CLEAN} image, the map has been rotated clockwise by $18^{\circ}$ for presentation. The grey vertical lines mark the transverse slices taken perpendicular to the jet axis at projected distances of 5, 17, and 20~mas from the core.
    \textit{Bottom panels:} Transverse intensity profiles through the RML reconstruction at the three slice positions marked in the upper panel. The black curves show the measured profiles, and the coloured circles mark the local intensity peaks used in the helical-model analysis. The horizontal gray dashed lines indicate the $3\sigma$ threshold derived from the superuniform-weighted \texttt{CLEAN} image. The black and blue horizontal bars labelled ``Map beam'' and ``Natural'', respectively, show the same reference angular scales as in the corresponding \texttt{CLEAN} analysis: the $0.7$~mas circular map beam and the projected transverse size of the natural beam ($1.6$~mas). The RML reconstruction provides an independent check of the transverse jet structure obtained with \texttt{CLEAN}, in particular in the region around 20~mas from the VLBI core. }
    \label{fig:ehtim}
\end{figure*}

The transverse-profile analysis of the \texttt{CLEAN} image reveals a double-peaked structure at a projected distance of approximately 20 mas from the VLBI core (see Sect.~\ref{sec:instability}). This feature, however, lies close to the effective angular-resolution limit, which is $\thicksim0.73$~mas along the direction of the transverse slice at this location - comparable to the 0.7~mas circular restoring beam adopted for the \texttt{CLEAN} image. Furthermore, the characteristic widths of the individual peaks are themselves comparable to these angular scales. The interpretation of the double structure therefore requires caution

The original superuniform-weighted \texttt{CLEAN} beam is highly elongated ($1.34\times0.33$~mas) owing to the long RadioAstron baselines. For the jet-structure analysis presented in the main text, the \texttt{CLEAN} model was restored with an equivalent-area circular 0.7 mas beam. While this choice offers a convenient representation of the high-resolution structure, it necessarily entails stronger super-resolution along the poorly resolved direction of the synthesized beam, thereby upweighting \texttt{CLEAN} reconstruction error in that direction \citep{2023MNRAS.523.1247P}. Furthermore, because \texttt{CLEAN} models the source brightness distribution as a collection of point components, weak sub-beam-scale structures may be particularly susceptible to deconvolution artefacts in regions of moderate signal-to-noise ratio.

To test whether the double-peaked structure near 20~mas is robust against the choice of imaging algorithm, we independently reconstructed the source using a regularized maximum-likelihood (RML) approach implemented in \texttt{eht-imaging} \citep{Chael_ehtim_2018}, applied to the same self-calibrated visibility data as used for the CLEAN reconstruction. To assess the stability of the recovered structure, we performed a parameter search over a grid of imaging settings, adopting a single circular Gaussian component with a radius of 40~mas as the prior image. The resulting reconstructions were generally very similar across the explored parameter range.

For the reconstruction adopted here, we used data terms based on complex visibilities, closure phases, and logarithmic closure amplitudes, along with a total-flux constraint, and without introducing additional image-domain regularization terms. For the final minimization, the \texttt{CLEAN} reconstruction convolved with a 3~mas circular beam served as the prior image. This relatively broad convolution beam preserves the overall source morphology and spatial localization while suppressing small-scale structure in the \texttt{CLEAN} reconstruction, thereby facilitating convergence of the minimization and reducing the possibility of imprinting fine-scale \texttt{CLEAN} features on the RML reconstruction. The final RML image, obtained with a field of view of 160~mas and a pixel size of 0.08~mas, together with the corresponding 3~mas-convolved \texttt{CLEAN} prior, is shown in Fig.~\ref{fig:ehtim-image}.

For a direct comparison with the \texttt{CLEAN} results, the reconstructed RML image was convolved with the same circular 0.7~mas beam used in the main analysis, and the transverse-profile measurements were repeated at the same projected distances of 5, 17, and 20~mas from the core. The corresponding profiles are shown in Fig.~\ref{fig:ehtim}. The RML reconstruction reproduces the main transverse structures at 5 and 17~mas. At 20~mas, however, the reconstructed profile is dominated by a single peak rather than the closely separated double structure found in the \texttt{CLEAN} reconstruction. This comparison suggests that the detailed two-component morphology at this location may not be robust with respect to the choice of imaging method, possibly being an artifact of the \texttt{CLEAN} algorithm \citep{2023MNRAS.523.1247P}, and should not be interpreted independently of the local resolution limit.

We additionally repeated the determination of the transverse intensity maxima using the RML image. Since the transverse structure is relatively simple in this reconstruction, the ridge positions were taken directly from the local maxima of the profiles rather than from a multi-component Gaussian decomposition. These positions are marked in Fig.~\ref{fig:ehtim}. Fitting them with the same three-dimensional helical model as used for the \texttt{CLEAN} reconstruction yields a qualitatively similar projected pattern of the Kelvin–Helmholtz instability threads. Therefore, the large-scale behaviour of the inferred filamentary structure does not depend strongly on the choice between \texttt{CLEAN} and RML imaging, even though some of the finer transverse substructure is less stable.

Moreover, the comparison reveals tentative agreement between the two reconstructions in regions containing features that we previously regarded as too weak to distinguish reliably from noise. One example occurs at a projected distance of approximately (15--17)~mas north of the core, close to where one of the modelled Kelvin–Helmholtz threads is expected to pass. In the \texttt{CLEAN} image, no significant emission was identified at this position, although a small isolated feature is visible near the 3$\sigma$ contour. We therefore excluded it from the original ridge analysis, as its significance was insufficient for a robust identification. The RML reconstruction, however, shows additional low-level emission at approximately the same location. Although this emission lies below the nominal 3$\sigma$ noise level adopted for the \texttt{CLEAN} image and should therefore not be regarded as an independent robust detection, its spatial coincidence is suggestive. It may indicate that the underlying plasma structure continues through this region but becomes too faint to be recovered reliably at the available sensitivity.

Estimating a noise level for an RML reconstruction is less straightforward than for a conventional \texttt{CLEAN} image. We explored an empirical estimate by subtracting the model visibilities corresponding to the final RML reconstruction from the observed visibilities and Fourier transforming the resulting residual visibilities. Interpreting the rms of such a residual image, however, requires a choice of image weighting and an associated dirty-beam normalization. For natural or uniform weighting, this normalization is set by the corresponding dirty beam, whose angular scale and structure generally differ from the effective resolution adopted for displaying the RML reconstruction. Consequently, no unique beam normalization allows the residual-image rms to be directly interpreted as the noise level of the RML image at its adopted resolution. In our tests, the residual-based rms was also substantially lower than the noise measured in the superuniform-weighted \texttt{CLEAN} image, suggesting that it provides an overly optimistic estimate of the effective image uncertainty.

For these reasons, we do not quote a separate RML noise level and instead adopt the rms measured from the superuniform-weighted \texttt{CLEAN} image as a conservative reference for the significance levels reported in this work. A rigorous definition of image-domain noise and significance for RML reconstructions, including the relation between visibility residuals, weighting, beam normalization, and the effective resolution of the reconstructed image, warrants a dedicated methodological investigation and lies beyond the scope of the present work.

\bsp 
\label{lastpage}
\end{document}